\documentclass[aip, rsi, amsmath,amssymb, reprint]{revtex4-1}

\usepackage{graphicx}
\usepackage{dcolumn}
\usepackage{bm}
\usepackage{hyperref}

\usepackage{color}
\usepackage[dvipsnames]{xcolor}
\usepackage{multirow}
\usepackage{rotating}
\usepackage{threeparttable}

\newcommand*{\vI}{\boldsymbol{I}}

\newcommand*{\vF}{\boldsymbol{F}}
\newcommand*{\vG}{\boldsymbol{G}}

\newcommand*{\vS}{\boldsymbol{S}}

\renewcommand*{\Pi}{{\varPi}}

\newcommand*{\vV}{\boldsymbol{V}}
\newcommand*{\vO}{\boldsymbol{O}}
\newcommand*{\vP}{\boldsymbol{P}}
\newcommand*{\vU}{\boldsymbol{U}}

\newcommand*{\vJ}{\boldsymbol{J}}
\newcommand*{\vK}{\boldsymbol{K}}

\newcommand*{\vD}{\boldsymbol{D}}

\graphicspath{{fig_F/}}

\usepackage{siunitx}
\usepackage{upgreek}
\usepackage{chemmacros}
\chemsetup{
	formula = mhchem,
	modules = thermodynamics,
	modules = redox,
	modules = reactions%
}
\usepackage{booktabs}

\begin{document}

\preprint{*}

\title{Environment effects on X-ray absorption spectra with quantum embedded real-time Time-dependent density functional theory approaches} 

\author{Matteo De Santis}
\author{Val\'erie Vallet}
\author{Andr\'e Severo Pereira Gomes}
\affiliation{Univ. Lille, CNRS, UMR 8523 -- PhLAM -- Physique des Lasers, Atomes et Molécules, F-59000 Lille,  France}
\email{andre.gomes@univ-lille.fr}

\date{\today}

\begin{abstract}
In this work we implement the real-time time-dependent block-orthogonalized Manby-Miller embedding (rt-BOMME) approach alongside our previously developed real-time frozen density embedding time-dependent density functional theory (rt-TDDFT-in-DFT FDE)  code, and investigate these methods' performance in reproducing X-ray absorption spectra (XAS) obtained with standard rt-TDDFT simulations, for model systems comprised of solvated fluoride and chloride ions ([X@\ce{(H2O)8]^-}, X = F, Cl). We observe that, for ground-state quantities such as core orbital energies, the BOMME approach shows significantly better agreement with supermolecular results than FDE for the strongly interacting fluoride system, while for chloride the two embedding approaches show more similar results. For the excited states, we see that while FDE (constrained not to have the environment densities relaxed in the ground state) is in good agreement with the reference calculations for the region around the K and L$_1$ edge, and is capable of reproducing the splitting of the $\mathrm{1s^{1} (n+1)p^1}$ final states ($n+1$ being the lowest virtual p orbital of the halides), it by and large fails to properly reproduce the $\mathrm{1s^{1} (n+2)p{^1}}$ states and misses the electronic states arising from excitation to orbitals with important contributions from the solvent. The BOMME results, on the other hand, provide a faithful qualitative representation of the spectra in all energy regions considered, though its intrinsic approximation of employing a lower-accuracy exchange-correlation functional for the environment induces non-negligible shifts in peak positions for the excitations from the halide to the environment. Our results thus confirm that QM/QM embedding approaches are viable alternatives to standard real-time simulations of X-ray absorption spectra of species in complex or confined environments.
\end{abstract}

\maketitle

\section{Introduction}

X-ray absorption spectroscopy (XAS) is a powerful technique to probe the structural and electronic properties of molecules from an atomistic picture, since  the absorbing photons in the X-ray energy range promote excitations of the core electrons to unoccupied or continuum states. The resulting absorption peaks are called edges in XAS and are labelled according to the origin of the core state, for instance K-edge for 1s, \ce{L1} for 2s, \ce{L2}-edge edge for $\mathrm{2p_{1/2}}$ and \ce{L3}-edge for the $\mathrm{2p_{3/2}}$. The spectra features near these edges are called X-ray absorption near-edge structures (XANES). Both the energy range and the spectral shapes directly provide information on the oxidation state, local symmetry, and coordination environment of a selected analyte in the gas, liquid or solid phase~\cite{xray-Zimmermann-CCR2020-423-213466,xray-Bokhoven-2016-Book,xray-Bunker-2009-Book,xray-Groot-CR2001-101-1779}. For instance, K-edges correspond to $\mathrm{1s \rightarrow (n+1)p}$ dipole transitions, $\mathrm{n+1}$, being the first virtual p~level, implying that, in a simple picture, the edge position is a direct measure of molecular valence states, thus allowing to monitor the effect of the local environment on a given atom within an analyte. The interpretation of such environmental interplay calls for electronic structure calculations that allow to access the atomic and molecular energy levels. More specifically, the theoretical modeling of XAS spectra implies the calculations of core-valence excitations. 

Within quantum chemical methods, density functional theory (DFT) based approaches such as time-dependent density functional theory (TDDFT) in its linear response (LR) (frequency domain) formulation, currently offers the best compromise between cost and accuracy for calculating electronic excitations~\cite{Norman2018:chemrev,Besley2021}. While a brute-force application of TDDFT to XAS would be prohibitively expensive as a large number of states (valence excitations, resonance, etc.) need to be determined before arriving at the energy regions pertaining to the core excitations, the introduction of restricted-excitation window TD-DFT (in which one can restrict the calculation to access only configurations in which particular core electrons are excited~\cite{dft-Stener-CPL2003-373-115,Besley2010,dft-Zhang-JCP2012-137-194306}) or the complex polarization propagator approach~\cite{Ekstrom2006a,Ekstrom2006b,Jiemchooroj2007,Villaume2010,Pedersen2014,Fahleson2016,Rinkevicius2016} (from which one can obtain the spectral profiles for a given range of frequencies of the external perturbations from the imaginary part of the dipole polarizability), has allowed TDDFT to be routinely used in simulating XAS. 

An alternative to the frequency domain approaches above that is gaining attention in recent years is that of the real-time formulation of TDDFT (rt-TDDFT)~\cite{dft-Goings-WCMS2017-8-e1341,li_rev2020}, in which time-dependent properties (such as electronically excited states) are obtained based on integrating the time-dependent Kohn-Sham (TDKS) equations in time. While the theoretical underpinnings, strengths, and limitations in respect to accuracy are similar to traditional linear response (LR) TDDFT methods for obtaining electronic spectra, rt-TDDFT provides fully time-resolved solutions that can potentially incorporate non-linear effects, and also allows for the case of strong external perturbations. With that, rt-TDDFT is used to compute not only spectroscopic properties including XAS~\cite{lopata2012,Kadek2015,li_rev2020} but also the time and space-resolved electronic response to arbitrary external stimuli (e.g., electron charge dynamics after laser excitation)~\cite{ati,Keldysh_2017,voohris,MOKKATH2020137905}.

However, as soon as one wishes to treat molecules surrounded by an environment (e.g.,\ species in solution or in otherwise confined spaces), the structural model for the system of interest might become too large to be treated with plain DFT approaches. In such cases, subsystem or embedding approaches~\cite{env-Gomes-ARPCSPC2012-108-222} appear as a computationally efficient strategy: they allow for combinations of different levels of theory for the subsystems of interest and their surroundings, thereby reducing the overall computational cost. Furthermore, embedding approaches allow for selectively switching on or off interactions between different subsystems, and thus can offer a powerful way to understand the physics of chemistry of a particular process, when the analysis of a full (supermolecular) calculation may prove much more cumbersome to analyze. This is the case when analyzing electronically excited states of confined systems, which can involve transitions within particular subsystems as well as between the different subsystems--and in core states in particular since core states can be embedded into states representing the continuum. 

There are been several propositions to couple rt-TDDFT methods to embedding approaches, perhaps the most widely used ones involving the coupling between a quantum subsystem and a classical environment~\cite{Lipparini2021}, described by continuum models~\cite{Pipolo2016,Gil2019} or classical force fields (QM/MM)~\cite{env-Marques-PRL2003-90-258101,env-Morzan-JCP2014-140-164105,Wu2017,Parise2018}. Although the obvious advantages are in cost reduction, these approaches may not properly describe specific interactions such as hydrogen bonds (for the continuum models) or rely on the availability of an appropriate classical force fields (for QM/MM). Classical approaches will in any case will be limited in their ability to properly describe phenomena in which a quantum description of the environment is important (such as charge delocalization, coupled excitations, or excitations across many parts of the systems not confined to a small fragment). The alternative in this case is to use quantum embedding theories (QM/QM)\cite{env-Gomes-ARPCSPC2012-108-222,env-Jacob-WCMS2014-4-325,wesolowski_frozen-density_2015,sun2016,Goez2018}, and among the fully quantum mechanical approaches to include environmental effects in the molecular response property we note the family of subsystem DFT approaches~\cite{env-Jacob-WCMS2014-4-325,wesolowski_frozen-density_2015}, to which the frozen density embedding (FDE) scheme is a member. It corresponds to a partitioning of a given system into a set of subsystems that can be, for instance, all represented within the Kohn-Sham framework, which interact through a local embedding potential. A subsystem DFT formulation of the real-time methodology (rt-TDDFT-in-DFT) has been presented in a seminal work by~\citeauthor{env-Krishtal-JCP2015-142-154116} together with its formulation within the FDE framework~\cite{env-Krishtal-JCP2015-142-154116,Krishtal2016,P2017,Genova2017}. 

This initial formulation, based upon plane-wave basis representations for the different subsystems, has been shown to properly capture the coupling in the response of the different subsystems, through the dependency of the time-dependent embedding potential on the time-dependent electron densities of all (or a subset) of subsystems, whenever such coupling is of importance. It should be noted that such couplings between the response of subsystems to external perturbation can also be taken into account in a frequency-domain formulation, but at the expense of determining second (or higher) order derivatives of the interaction energy~\cite{neugebauer2007,neugebauer2009,neugebauer2009b,visscher12,Pavanello2013,konig2013}. That said, applications of linear-response or real-time TDDFT-in-DFT showed that in many cases, the coupling between the response of different subsystems can be ignored and a so-called ``uncoupled'' TDDFT-in-DFT approach can yield accurate results~\cite{actinide-Gomes-PCCP2008-10-5353,actinide-Gomes-PCCP2013-15-15153,olejniczak2017,rt_fde2020}, provided the coupling between subsystems in the ground state is well described by the embedding potential representing the subsystems' interaction. FDE-based calculation has been shown to perform well for situations in which there are no strong interactions between subsystems, such as covalent bonds. This makes it possible in general to describe interactions such as hydrogen bonds, though in certain cases the approximations intrinsic to FDE, due to the use of approximate kinetic energy density functionals (KEDFs)~\cite{Beyhan2010,Grimmel2019} in the description of the embedding potential, prevent it from accurately describing stronger non-covalent interactions~\cite{fux_exemb,halides-water-Bouchafra-PRL2018-121-266001}. While in such cases, a pragmatic solution is to enlarge the active subsystem, that can be potentially problematic in respect to the increase of computational costs, especially if one is interested in replacing DFT by higher-level approaches such as coupled cluster to describe the subsystem of interest.

There, another QM/QM family of embedding approaches closely connected to the subsystem DFT approaches mentioned above involve the use of projection operator techniques~\cite{toelle_1,toelle_2,Scholz2020,Niemeyer2020,manby2012,goodpaster_exemb,goodpaster_exemb2,ding2017,Lee2019,Graham2020}, and by foregoing the use of the approximate KEDFs, show a better performance in describing strong interactions. These approaches, in some variants also referred to as Manby-Miller embedding (MME), have shown to be particularly adept at allowing the fragmentation of a particular system through covalent bonds. More recently, the block-orthogonalized MME (BOMME)~\cite{ding2017} has been introduced to alleviate issues that plagued prior MME variants~\cite{manby2012}. BOMME allows one to treat the target system with high-level Fock matrix and the remaining degrees of freedom with a less expensive Fock matrix by reducing the quality of basis set and exchange. A combination of BOMME with rt-TDDFT has been recently proposed by~\citeauthor{parkhill_2017}~\cite{parkhill_2017}. They demonstrated that rt-BOMME can capture both intermolecular and intramolecular couplings and their induced effects, namely solvent shifts, on spectra of chromophores. 

However, in that work only processes involving valence electrons have been considered. Given the interest of core spectra as a means to characterize species in complex environments, it is of great interest to explore the behavior of rt-BOMME for describing XAS. We note that the same is also the case for FDE or TDDFT-in-DFT since these have also, to the best of our knowledge, not yet been explored for core excitations.

Thus, the main goal of this work is to describe a first investigation o the performance of rt-TDDFT-in-DFT and rt-BOMME for core excitations. To do so, we have extended our recently developed Psi4Numpy-based rt-TDDFT-in-DFT to implement the BOMME and rt-BOMME methods. As discussed below, in this manuscript we shall focus on the K and \ce{L1} edges of hydrated halides, since halogenated species and their interaction with species in solution or at interfaces are of particular interest in atmospheric sciences~\cite{FinlaysonPitts2013,PillarLittle2013,Simpson2015,Kong2017,FinlaysonPitts2019,Yu2021,BartelsRausch2021}.
Here, however, in order to simplify our discussion we have considered relatively simple model systems representing the first hydration shell of the halides (\ce{[F(H2O)8]-} and \ce{[Cl(H2O)8]-}) that nevertheless can gauge of the ability of the different embedding methods to describe interactions of varying strengths between halides and their environment. Also, due to the scarce experimental data for XAS on such systems, we shall restrict ourselves to comparisons to two limiting cases: the free ions and the rt-TDDFT calculations on the supermolecular system (which then serve as our benchmarks).

\section{Materials and Methods}

\subsection{Theoretical background}
\label{sec:methods}
The Frozen Density (FDE)~\cite{wesolowski93,env-Gomes-ARPCSPC2012-108-222,env-Jacob-WCMS2014-4-325,wesolowski_frozen-density_2015} and Block-Orthogonalized Manby-Miller embedding (BOMME) approaches~\cite{ding2017}, and their 
extension to the rt-TDDFT framework has been described in previous works~\cite{rt_fde2020,parkhill_2017}. 
In this Section, after brief recapitulation of the rt-TDDFT
method, we will outline analogies and differences of rt-FDE and rt-BOMME approaches.

In rt-TDDFT, the one-electron density matrix $\vD(t)$ 
representing in the algebraic approximation the time-dependent electron density evolves in time
according to the 
\begin{equation}\label{eq:Uop}
\vD(t)= \vU(t,t_0)\,\vD(t_0)\,\vU(t,t_0)^{\dagger},
\end{equation}
where $\vU(t,t_0)$ is the matrix representation of the time-evolution operator:
\begin{equation}\label{eq:lvn}
	\vU(t,t_0) = \hat{T} \exp \bigg(-i \int_{t_0}^{t} \vF(t') dt' \bigg).
\end{equation}
The real-time approach is based on the repeated application of Eq.~\ref{eq:Uop} over a
discretized time-domain. Time discretization allows to devise advantageous representations of
 $\vU$ to be employed in real computer codes.
In this work we employ the exponential midpoint ansatz, which has been successfully employed
in the study of valence and core excitations~\cite{lopata2012,nwchem}
Extensive discussion on the computational strategies employed to carry-out the time-evolution
propagation can be found in the seminal work by Castro and co-workers~\cite{castro}.
In rt-TDDFT the Fock matrix is defined as 
\begin{equation}\label{eq:td_fock}
\vF(t) = \bm{h}_0 + \vG[\vD(t)]+v_{\text{ext}}(t),
\end{equation}
 where $\bm{h}_0$ represents the one-electron operator while $\vG$ is the two-electron term
\begin{equation}
\vG[\vD(t)] = \vJ[\vD(t)] + c_x \vK[\vD(t)] + c_x\vV_{\text{xc}}[\vD(t)],
\end{equation}
$c_x$ being the fraction of Hartree-Fock exchange in the exchange correlation potential $V_{xc}$.
It is worth noting the the Fock matrix appearing in Eq.~\ref{eq:td_fock} has an implicit time-dependence
due to a time dependent density matrix, and the explicit time-dependence due to the external potential
$v_{\text{ext}}(t)$.

An embedding mean-field approach  is based on the mapping of two different domains within the total system, 
into two different-quality levels of theory to be employed in each domain. This can be realized by assigning a high-level Fock matrix to the subsystem to be treated accurately, while letting the remaining part to be described by a low-level Fock matrix in a reduced basis set.
The block-orthogonalized (BO) partitioning scheme proposed by Ding and co-workers\cite{ding2017} relies on a projected basis in place of the conventional
atomic-orbital (AO) partitioning to define the high- and low-level components of the system. Such scheme proved to
be suitable to remove the artifacts related to the embedding scheme while keeping the expression of the low-level Fock as simple as:
\begin{equation}\label{eq:bomme_terms}
\tilde{\bm{h}}_0 = \vO^T \bm{h}_0 \vO, \quad  \tilde{\vG}^{\text{Low}}[\tilde{\vD}] = \vO^T \vG^{\text{Low}}\vO, \quad \tilde{\vD}=\vO \vD \vO^T.
\end{equation} 
In Eq.~\ref{eq:bomme_terms} quantities expressed in the block-orthogonalized basis are denoted by tildes, and $\vO$
is the transformation matrix from the non-orthogonal AO basis set to the BO basis set: 
\begin{equation}
    \vO = \begin{pmatrix} \vI^{\text{AA}} & -\vP^{\text{AB}}\\ \bm{0} & \vI^{\text{BB}}  \end{pmatrix}.
\end{equation}
The sub-blocks appearing in the transformation matrix are the identity matrices $\vI^{\text{AA}}$ and $\vI^{\text{BB}}$ having dimensions of $n_a$ and $n_b$, mapping subsystem A and B basis sets respectively, and 
the projection matrix  $\vP^{AB}=(\vS^{\text{AA}})^{-1}\vS^{\text{AB}}$, in which $\vS^{AB}$ is the AO overlap between the subsystems. Here and hereafter   the AA  (BB) block denote the subsystem with  high - (low-)level theory. The Fock matrix in the BO basis reads as:
\begin{equation}
\vF =\tilde{\bm{h}}_0 + \tilde{\vG}^{\text{Low}}[\tilde{\vD}]  +(\tilde{\vG}^{\text{High}}[\tilde{\vD}^{\text{AA}}] -\tilde{\vG}^{\text{Low}}[\tilde{\vD}^{\text{AA}}]).
\end{equation}
In this context different schemes for the calculation of the exchange term (in $\vG^{\text{High}}$ ) are available. Following~\citeauthor{parkhill_2017}~\cite{parkhill_2017} we adopted the simplest scheme for $E_{\text{EX}}[\tilde{\vD}^{\text{AA}}]$ which takes into account only the exact exchange interaction within the AA block:
\begin{equation}
E_{\text{EX}0}= -\frac{1}{4}\sum_{\mu\kappa\nu\lambda}(\mu\kappa|\nu\lambda)\vD^{\text{AA}}_{\mu \nu}\vD^{\text{AA}}_{\kappa \lambda}.   
\end{equation}

In the Frozen Density formulation of DFT the entire system is partitioned into N subsystems,
and the total density $\rho_\text{tot}(\bm{r})$ is represented as the sum of electron densities of the 
various subsystems [i.e., $\rho_a(\bm{r})$ ($a = 1,..,N$)].  In this work we restrict our consideration to 
a simplified model in which 
the total density is partitioned in only two 
contributions as
\begin{equation}
	\rho_\text{tot}(\bm{r}) =\rho_\text{I}(\bm{r}) + \rho_\text{II}(\bm{r}).
\end{equation}
The total energy of the system can then be written as 
\begin{equation}
	E_\text{tot}[\rho_\text{I},\rho_\text{II}] = E_\text{I}[\rho_\text{I}] + E_\text{II}[\rho_\text{II}] + E_\text{int}[\rho_\text{I},\rho_\text{II}],
\label{eq:etot}
\end{equation}
with the energy of each subsystem ($E_i[\rho_i]$, with $i=\text{I},\text{II}$) given according to the usual 
definition in DFT as
\begin{equation}
\begin{aligned}
E_i[\rho_i]   &= \int\rho_i(\bm{r})v_\text{nuc}^{i}(\bm{r}) {\rm d}^3r 
                          + \frac{1}{2}\iint\frac{\rho_i(\bm{r})\rho_{i}(\bm{r}')}{|\bm{r}-\bm{r}'|}{\rm d}^3r {\rm d}^3r' \\
              &+ E_\text{xc}[\rho_i] + T_s[\rho_i] + E_\text{nuc}^{i}.
\end{aligned}
\end{equation}
In the above expression, $v_\text{nuc}^{i}(\bm{r})$ is the nuclear potential due to the set of atoms which defines the 
subsystem and $E_\text{nuc}^{i}$ is the related nuclear repulsion energy.
$T_s[\rho_i]$ is the kinetic energy of the auxiliary non-interacting system,
which is, within the Kohn-Sham (KS) approach, commonly evaluated using the KS orbitals.
The interaction energy is given by the expression:
\begin{equation}
\begin{aligned}
  \label{eq:eint}
	E_\text{int}[\rho_\text{I},\rho_\text{II}] &= \int\rho_\text{I}(\bm{r})v_\text{nuc}^\text{II}(\bm{r}){\rm d}^3r 
	            +\int\rho_\text{II}(\bm{r})v_\text{nuc}^\text{I}(\bm{r}) {\rm d}^3r \\
        &+ E^\text{I,II}_\text{nuc} +\iint\frac{\rho_\text{I}(\bm{r})\rho_\text{II}(\bm{r}')}{|\bm{r}-\bm{r}'|} {\rm d}^3r {\rm d}^3r' \\
	&+E^\text{nadd}_\text{xc}[\rho_\text{I},\rho_\text{II}] + T^\text{nadd}_s[\rho_\text{I},\rho_\text{II}],
\end{aligned}
\end{equation}
with $v_\text{nuc}^\text{I}$ and $v_\text{nuc}^\text{II}$ the nuclear potentials due to the set of atoms associated with the subsystem $\text{I}$ 
and $\text{II}$,
respectively. The repulsion energy for nuclei belonging to different subsystems is described by the $E^\text{I,II}_\text{nuc}$ term.
The non-additive contributions are defined as:
\begin{equation}
	X^\text{nadd}[\rho_\text{I},\rho_\text{II}] = X[\rho_\text{I}+\rho_\text{II}] - X[\rho_\text{I}] - X[\rho_\text{II}],
\end{equation}
with $X=E_\text{xc}, T_s$. These terms arise because both 
exchange-correlation and kinetic energy, in contrast to the Coulomb interaction, 
are not linear functionals of the density.

The electron density of a given fragment ($\rho_\text{I}$ or $\rho_\text{II}$ in this case) 
can be determined by minimizing the total energy
functional (Eq.~\ref{eq:etot}) with respect to the density of the fragment while
keeping the density of the other subsystem frozen. This procedure is the essence of the
FDE scheme and leads to a set of Kohn-Sham-like equations (one for  each subsystem)
\begin{equation}\label{eq:act_opt}
        \Big[ -\frac{\nabla^2}{2} + v^\text{KS}_\text{eff}[\rho_\text{I}](\bm{r}) 
                + v_\text{emb}^\text{I}[\rho_\text{I},\rho_\text{II}](\bm{r})\Big]\phi_k^\text{I}(\bm{r}) = \varepsilon_k^\text{I}\phi_k^\text{I}(\bm{r})
\end{equation}
which are coupled by the embedding potential term $v^\text{I}_\text{emb}(\bm{r})$, 
that carries all dependence on the other fragment's density.
In the framework of FDE theory, $v^\text{I}_\text{emb}(\bm{r})$ is explicitly given by 
\begin{equation}
\begin{aligned}
	\label{eq:vemb}
	v^\text{I}_\text{emb}[\rho_\text{I},\rho_\text{II}](\bm{r}) 
      = \frac{\delta E_\text{int}[\rho_\text{I},\rho_\text{II}]}{\delta\rho_\text{I}(\bm{r})} 
       = v_\text{nuc}^\text{II}(\bm{r})  &+ \int\frac{\rho_\text{II}(\bm{r}')}{|\bm{r}-\bm{r}'|} {\rm d}^3r' \\
       +\int\frac{\rho_\text{II}(\bm{r}')}{|\bm{r}-\bm{r}'|} {\rm d}^3r' 
        + \frac{\delta E_\text{xc}^\text{nadd}[\rho_\text{I}, \rho_\text{II}]}{\delta\rho_\text{I}(\bm{r})} 
        &+ \frac{\delta T^\text{nadd}_s[\rho_\text{I}, \rho_\text{II}]}{\delta\rho_\text{I}(\bm{r})},
\end{aligned}
\end{equation}
where the non-additive exchange-correlation and kinetic energy contributions 
are defined as the difference between the associated exchange-correlation and kinetic
potentials defined using $\rho_\text{tot}(\bm{r})$ and $\rho_\text{I}(\bm{r})$. 
It is worth noting that only the density for the total system is available so that potentials requiring 
KS orbitals as input are excluded.

For the exchange-correlation potential, one may make use of accurate density functional 
approximations and its quality is therefore similar to that of ordinary KS. The potential for the non-additive kinetic term 
($\delta T^\text{nadd}_s[\rho]/\delta\rho_\text{I}(\bm{r})$,
in Eq.~\ref{eq:vemb}) is more problematic as it relies on less accurate orbital-free kinetic energy density functionals (KEDFs).

In this context, the Thomas-Fermi (TF) 
kinetic energy functional~\cite{thomas_1927} or 
the GGA functional  PW91k~\cite{PhysRevA.50.5328}, are customarily employed. 
The reader interested in applicability and shortcomings of the functionals associated to $T^\text{nadd}_s[\rho_\text{I}, \rho_\text{II}]$ term can refer to
Ref.~\citenum{fux_exemb} and references therein.

In general, FDE scheme provides a set of coupled equations for the subsystems have to be solved iteratively.
Typically, the ``freeze-and-thaw'' (FnT) procedure is employed, meaning that the
electron density of the active subsystem is determined keeping frozen
the electron density of the other subsystems, which is then frozen
when the electron density of the other subsystems is worked out.
The subsystems' densities are converged repeatedly applying the procedure.

We conclude this section highlighting the main differences between FDE and BOMME approaches.
FDE approach employing an explicit embedding potential allows to optimize the subsystem of interest limiting the basis set to the sole 'active' basis subset. We have already mentioned that the embedding potential relies on the KEDFs, which are in general less accurate than the exchange-correlation counterpart. On the contrary in BOMME formulation, the self-consistent calculation is carried out in the supermolecular basis, and the embedding is handled implicitly in the calculation.
As far as the coupling between subsystem is concerned in the FDE approach is trivial to estimate the interaction energy of the subsystem and eventually evaluate the net effect of the environment polarization performing an unrelaxed calculation (keeping the environment frozen). It is important to note that since the total density is obtained as the sum of subsystem densities, the partitioning reflects on mean values of observables. In the BOMME approach
the high-level system (AA block) and its environment (BB block) are optimized on the same footing, thus disentangling them could result in a cumbersome procedure. 
Nevertheless it could be possible to investigate the contribution of the different domains to the overall value of an observable using localization techniques.
\subsection{Computational Details}
\label{sec:computdetails}

In the ADF~\cite{ADF2001} calculations, all of which performed with version \texttt{2019.307}~\cite{ADF2019.3}, we have employed the AUG/ATZP basis sets for the halogens, and the single-z without polarization (SZ) basis set for the water molecules~\cite{basis-Van-Lenthe-JCC2003-24-1142}.  In supermolecular calculations, we employed the B3LYP functional. The ADF FDE and FnT calculations were performed via the PyADF scripting framework~\cite{Jacob2011}. The halogen subsystem has been calculations with the B3LYP functional and the water molecules with BLYP. Since the use of different density functionals for different subsystems in a FnT calculation is currently not possible from within the ADF implementation~\cite{Jacob2008}, a PyADF script to carry out such calculations in provided as part of the dataset accompanying this manuscript~\cite{DeSantis-dataset-xas}, and in this case we employed a convergence criteria on the energy of $1\times 10^{-6}$. In all case, the Thomas-Fermi and BLYP functionals have been used to calculate the non-additive kinetic energy and exchange-correlation contributions to the embedding potentials, respectively. We employed supermolecular integration grids of normal (6.0) accuracy in all calculations.

In the (rt-)BOMME and (rt-)FDE calculations in the Psi4Numpy~\cite{Smith2018} framework, we employed the version \texttt{1.3.2} of the Psi4 code~\cite{Smith2020} as computational backend. We have employed the equivalent functionals as for the ADF calculations for the halogen and water subsystems (B3LYP and BLYP, respectively). As for basis sets, we employed aug-cc-pVTZ~\cite{basis-Dunning-JCP1989-90-1007,basis-Kendall-JCP1992-96-6796,basis-Woon-JCP1994-100-2975} and STO-3G~\cite{basis-Hehre-JCP1969-51-2657} basis for the halogen and the water cluster respectively.

For the real-time simulations, the electronic ground state of the halogen-water complex calculated in absence of an external electric field, was perturbed by an analytic $\delta$-function pulse with a strength of $\kappa =$ 5.0$\times10^{-4}$ a.u. along the three directions, $x,y,z$. The induced dipole moment has been collected for 56000 time steps with a length of 0.025 a.u. per time step, corresponding to \SI{33.9}{\fs} of simulation. The choice of such fine-grained time grid ensures in principle an observable frequency up to 3419.5 eV in the power spectrum distribution. In the case of the fluorine-water complex, the near-edge structure is located in the range of 665-700 eV, thus the time-dependent dipole moment was down-sampled halving the amount of sampling points. The use of Pad\'e approximant-based Fourier Transform allowed to further reduce the length of the signal to be sampled corresponding to an ``effective'' dipole moment of 24 and 29 fs for the fluorine- and chlorine-water complex respectively. In both cases prior to Fourier transformation an exponential damping $e^{-\lambda\cdot t}$ with $\lambda=3.0\times 10^{-4}$ was applied.

The code implementing the rt-FDE in the Psi4Numpy framework used in this work is part of the PyBertha package~\cite{rt_pybertha2020,rt_fde2020,pybertha-github} (revision~\texttt{3c752072}). The code implementing the (rt-)BOMME approach is under version control (Git) but does not yet have a public release version (one is  envisaged for 2022). The simulations described in the paper have been carried out with revision~\texttt{3c4c334b}.

\begin{figure}[htbp]
\centering
\begin{minipage}{0.48\linewidth}
\includegraphics[width=0.95\linewidth]{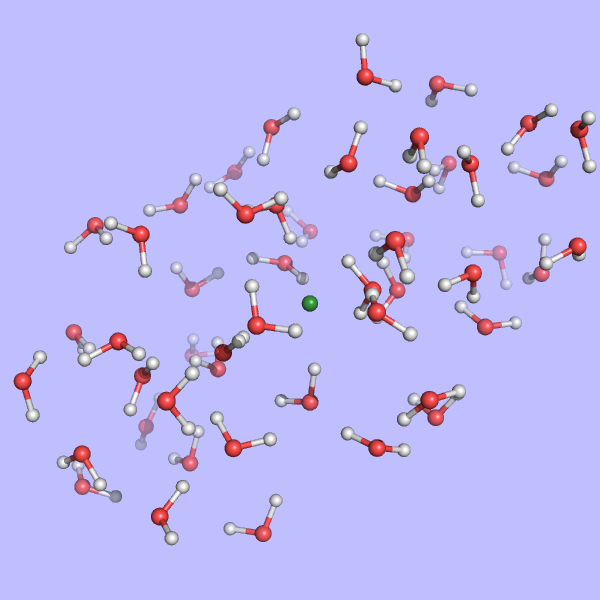}
\end{minipage}
\begin{minipage}{0.48\linewidth}
\includegraphics[width=0.95\linewidth]{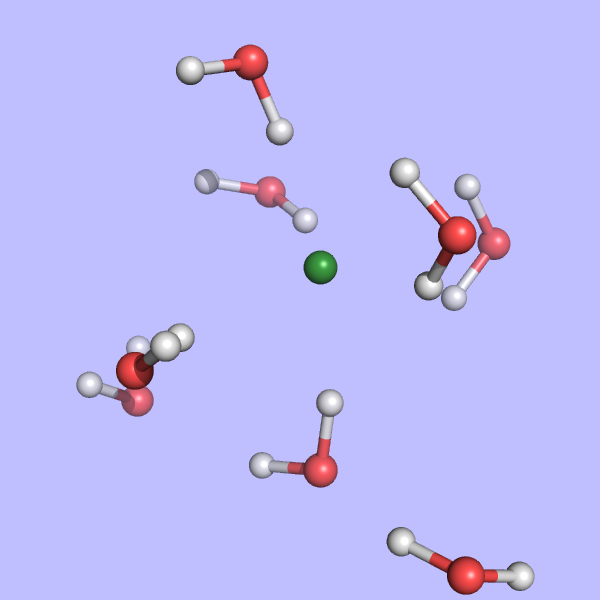}
\end{minipage}
\caption{Structure for the~\texttt{snapshot 1} of the fluoride-water droplet system taken from ~\citeauthor{halides-water-Bouchafra-PRL2018-121-266001}~\cite{halides-water-Bouchafra-PRL2018-121-266001,halides-water-Bouchafra-PRL2018-121-266001-Zenodo} (left) and the model used in this work (right), in which the halide and the 8 nearest water molecules making up its first solvation shell were extracted from the 50-water droplet.}
\label{fig:structural_model}
\end{figure}

The structures employed in the calculations were taken from the structures generated by~\citeauthor{halides-water-Bouchafra-PRL2018-121-266001}\cite{halides-water-Bouchafra-PRL2018-121-266001,halides-water-Bouchafra-PRL2018-121-266001-Zenodo} for halogens in 50-water droplets--in particular,~\texttt{snapshot 619} for chloride and~\texttt{snapshot 1} for fluoride--and, for reasons of computational cost, we have only kept the nearest 8 water molecules that correspond to the first solvation shell. This setup is exemplified in figure~\ref{fig:structural_model}.
 
\section{Results and Discussion}

We proceed now to the presentation and discussion of our results. Before doing so, we recall that since we are interested on the relative performance of the embedding methods with respect to a calculation on the whole system, and not on a comparison to experiment, we have opted to disregard both relativistic effects and statistical sampling of different solute configurations (e.g.,\ by considering different snapshots from molecular dynamics simulations) as done by some of us in~\citeauthor{halides-water-Bouchafra-PRL2018-121-266001}~\cite{halides-water-Bouchafra-PRL2018-121-266001}, which we aim to consider in subsequent work. Second, we chose to focus only on transitions from the core s orbitals of the halogens, that is, the 1s (K edges) for \ce{F-} and \ce{Cl-}, and the 2s (\ce{L1} edge) of \ce{Cl-}, since they provide sufficient information for our method comparisons.

\subsection{Ground states}

Before investigating the outcome of the real-time propagation of the wavefunctions, it is instructive to analyze the differences between the different models: isolated atoms, embedding approaches (FDE and BOMME) and standard (supermolecular) DFT calculations. To this end, we shall focus on the comparison of core orbital energies, on the one hand, since there is a direct connection between them and how environment effects are incorporated (see discussion on theoretical approaches in Section~\ref{sec:methods}), and, on the other hand, their values provide an approximation to the ionization potentials--though still missing will be the very important effects of wavefunction relaxation due to the creation of the core hole. 

While core orbitals are naturally rather localized, they are nevertheless quite sensitive to changes in the surroundings of the atom due to the presence of the solvent molecules, as we can see from the comparison of values for the isolation anions and the supermolecular systems. With the embedding approaches we expect the orbital energies to be much closer to the supermolecular values, since they introduce the different interactions (electrostatic, kinetic energy and exchange-correlation) between the halides and the water molecules, albeit in more or less approximated manners. Consequently, the closer an embedding approach yields orbital energies to supermolecular ones, the better suited it can be considered to replace the supermolecular calculation.

Before we can proceed to a comparison between the FDE and BOMME results shown in Table~\ref{tab:fnt_adf}, we should note that our Psi4Numpy-based code, in which both embedding approaches are implemented, does not yet implement the "freeze-and-thaw" (FnT) procedure for the FDE case. While that posed no problem in its first application to the rt-TDDFT-in-DFT simulation of a neutral system~\cite{rt_fde2020}, prior work by some of us~\cite{actinide-Gomes-PCCP2008-10-5353,actinide-Gomes-PCCP2013-15-15153,halides-water-Bouchafra-PRL2018-121-266001} has shown that for charged systems such as those considered here, the manner in which the environment density has been constructed is important, and that a relaxation of both active subsystem and the environment densities via FnT can improve the results over a pure FDE calculation, in which the environment's density and electrostatic potential have been obtained in the absence of the halides.

To estimate the effect of relaxing the environment on the orbital energies, and indirectly part of its influence on the simulation of core spectra (the other part coming from the effect on the halide virtual orbitals), we also present in Table~\ref{tab:fnt_adf}, results obtained with the ADF code, in which both FDE and FnT calculations have been carried out. We see that the FDE calculations tend to overestimate the effect of the environment, via overall more attractive embedding potentials, reflected in lower orbital energies than the supermolecular case, whereas FnT reverses this trend but overcorrects somewhat and yields energies which are slightly higher than the supermolecular ones. For fluoride FDE and FnT differ by roughly 1.2 eV, whereas for chloride there is a much less important difference, of around 0.4 eV. In addition to being smaller in magnitude, the shift for chloride is roughly the same for both 1s and 2s orbitals, an observation that is consistent with prior work~\cite{actinide-Gomes-PCCP2013-15-15153} in which we observed that the embedding potential shifted orbital energies in a nearly constant manner across different occupied orbitals.

\begin{center}
\begin{table*}[tb]
	\caption{Orbital energies ($\epsilon$, in eV) of the core s orbitals the halogen atom in the \ce{[X(H2O)n]-} clusters, (X = \ce{F-}, \ce{Cl-}), obtained with different models: (0) isolated halogen atoms; (1) DFT-in-DFT without the relaxation of the solvent (FDE); (2) DFT-in-DFT with the relaxation of the solvent environment (FnT); and (3) Block-orthogonalized Manby-Miller embedding (BOMME). In addition to the energies obtained with embedding, we provide energy differences with respect to reference supermolecular DFT calculations ($\Delta \epsilon = \epsilon_\text{sup} - \epsilon_\text{model}$), represented by $\Delta \epsilon_0, \Delta \epsilon_1, \Delta \epsilon_2$ and $\Delta \epsilon_3$, respectively. Due to the fact that for technical reasons, in our FDE implementation based on the Psi4Numpy framework, we are currently not able to perform embedding scheme (2), we provide results for (1) and (2) obtained with the ADF code.}
\vspace{1mm}
\label{tab:fnt_adf}
\centering
\begin{tabular}{lll*{8}r}
\hline
\hline
            &       &           & \multicolumn{4}{c}{Orbital energies (\si{\electronvolt})} & & & & \\
\cline{4-7}
framework	&	X	&	orbital	& {iso. (0)} & {FDE (1)}	&	{FnT (2)}	&	{BOMME (3)}	&	{$\Delta \epsilon_0$} & {$\Delta \epsilon_1$}	&	{$\Delta \epsilon_2$}	& {$\Delta \epsilon_3$}	\\
\hline
ADF     & \ce{F-}  &    1s   &  -659.67 &  -661.16  &  -661.59  &            & -2.57  & -0.67   &   0.46    &          \\
        & \ce{Cl-} &    1s   & -2753.62 & -2755.05  & -2755.48  &            & -1.55  & -0.12   &   0.32    &          \\
        &          &    2s   &  -248.46 &  -249.92  &  -250.36  &            & -1.59  & -0.12   &   0.31    &          \\
        &          &         &          &           &           &            &        &         &           &          \\
Psi4Numpy    
        & \ce{F-}  &    1s   &  -659.88 &   -661.96 &           &  -662.13   & -2.42  & -0.34   &           &  -0.17    \\      
        & \ce{Cl-} &    1s   & -2754.83 &  -2756.28 &           & -2756.25   & -1.50  & -0.05   &           &   -0.08   \\
        &          &    2s   &  -248.78 &   -250.27 &           &  -250.25   & -1.55  & -0.06   &           &   -0.08   \\\bottomrule
\hline
\hline
\end{tabular}
\end{table*}
\end{center}

Comparing the differences between FDE and supermolecule results between ADF and Psi4Numpy, we see a similar trend in that FDE overestimates the effect of the environment. From the comparison of $\Delta \epsilon_0$ for the two codes, we see that discrepancies of around 0.15 eV (for fluoride) and 0.05 eV (for chloride) can be attributed to differences inherent to the two sets of calculations (Slater vs. Gaussian basis sets, etc.), with values calculated with ADF showing larger discrepancies between isolated and supermolecular calculations than ADF. If we correct for these differences, we see that for chloride the $\Delta \epsilon_1$ values are consistent between codes, though for fluoride even taking into account such corrections, non-negligible differences between codes, of around 0.15 eV, remain. From this comparison, we believe that we can conclude that, if we were able to carry out such calculations, the Psi4Numpy FnT $\Delta \epsilon_2$ would likely be of around 0.2-0.3 eV for chloride, and 0.4-0.5 eV for fluoride. 

The BOMME results show a similar trend as the FDE in overestimating the effect of the environment with respect to supermolecular results, and that such overestimation is larger for fluoride than for chloride. The magnitude of such effect, however, is about half of that of FDE for fluoride (-0.17 eV vs -0.34 eV), and roughly equivalent to that of FDE (less than -0.1 eV) for chloride. The differences between BOMME and FDE are consistent with what is known in the literature between the more reliable behavior of projection-based embedding (such as BOMME) in describing cases in which there are strong interactions between the different subsystems with respect to FDE, which suffers from the limited accuracy of the non-additive kinetic energy density functionals used to calculate the non-additive kinetic energy contribution to the embedding potential~\cite{env-Gomes-ARPCSPC2012-108-222}. 

From the discussion above, and assuming that the dominant effects in the electronic spectra would come mainly from the energy differences between the core and low-lying virtuals either on the halogen (for both BOMME and FDE) or on the environment (for BOMME), we can expect to see that BOMME excitation energies would be consistently closer to the supermolecular results than FDE, but that such a difference would decrease for chloride. In the following we shall see to what extent this picture holds true. In any case, for the core orbital energies, the FDE results seem to provide a fortuitous error cancellation that places the relatively simple FDE model at par with the much more sophisticated BOMME.

\subsection{Core excited states}

Before discussing the behavior of the different approaches for the core states, we note that in the following we shall focus on the edge region for the K edges of both systems, considering energies spanning a somewhat broad window (20–30 eV higher than the first peaks), in order to have a wider region in which to compare the different models) and \ce{L1} edge of chloride. Furthermore, in the discussion below we shall focus on combined contributions from the $x, y, z$ components of the perturbing field. We present a breakdown of these by component of the perturbing field, along with the spectra for the whole
regions under consideration (Figures~\ref{fig:fluoride_K_general},~\ref{fig:chloride_K_general} and~\ref{fig:chloride_L1_general}) in the supplemental information. 

\subsubsection{K edge}

The spectra for the K edges of fluoride and chloride are shown in Figure~\ref{fig:F_Cl_K_edge_sections}. Starting with the simplest systems, the isolated anions, we note that, as expected, the K edge spectra corresponds to transitions from the 1s orbitals to the first virtual halide $\mathrm{p}$ orbitals ($\mathrm{(n+1)p}$). The second peak in the energy range considered, on the other hand, corresponds to transitions from the 1s to a next, higher-lying halide $\mathrm{p}$-type orbitals ($\mathrm{(n+2)p}$). 

Second, at the other extreme we have the supermolecular calculations on the microsolvated anions (which here serve as a benchmark to which the embedding approaches will be compared). There, the first remarkable difference from the free ions is that there is an environment-induced shift in the first region (panes \textbf{a} and \textbf{b} in Figure~\ref{fig:F_Cl_K_edge_sections}), which at around 1 eV is fairly similar between systems, but about half of what would be expected from the difference in orbital energies between the free ions and the supermolecular systems ($\Delta\varepsilon_0$). This is a first indication that core orbital energies are useful from understanding the K-edge absorption spectra from a semi-quantitative viewpoint at best. 

The asymmetrical first hydration shell environment also breaks the atomic symmetry, which has as consequence the lifting of the selection rules for the atom, and introduces differential interactions with the different $p$ orbital that become occupied in the excited states. As a result of that, for both fluoride and chloride, in the supermolecular calculations we observe four transitions within roughly a 1 eV interval, with spacings of around 0.5 eV between the first three peaks (with the fourth being much closer to the third).

For the region corresponding to the second free ion transition (panes \textbf{e} and \textbf{f} in Figure~\ref{fig:F_Cl_K_edge_sections}), we observe a similar situation to that of the first, with the environment inducing a symmetry breaking of the higher-lying $\mathrm{p}$ orbitals. In the case of fluoride, for the supermolecular calculations we observe five peaks, one rather close to that of the free ion (around 691.5 eV), followed by two other peaks around 2 eV higher (at 694 eV), and two more peaks between 697 and 699 eV. 
In the case of chloride, we also observe five peaks, one nearest to that of the free ion, another peak around 2775.5 eV, and then three additional peaks between 2778 and 2780 eV. Finally, in between the two free ion peaks (panes \textbf{c} and \textbf{d} in Figure~\ref{fig:F_Cl_K_edge_sections}), in the supermolecular calculation we have a region that contains several peaks.

Considering now the FDE calculations--and recalling that these correspond to a situation in which the density of the environment has not been relaxed in the presence of the anions--we see, for fluoride, a semi-quantitative agreement with the supermolecular calculation; for the lower energy region (panes \textbf{a} and \textbf{b} in Figure~\ref{fig:F_Cl_K_edge_sections}), the first peak appears in a slightly (around 0.2 ev) lower energies than the supermolecular ones, while the second, third and fourth peaks appear at slightly higher energies.
For chloride, the peak positions are overall closer (0.1 eV or less) to the supermolecular one than for fluoride, but now the energies of the first three peaks are slightly overestimated with respect to supermolecule, while the fourth we see a slight underestimation. We note that this is in line with the better agreement between FDE and supermolecular 1s orbital energies than for fluoride. Such a tendency was also observed by~\citeauthor{halides-water-Bouchafra-PRL2018-121-266001}~\cite{halides-water-Bouchafra-PRL2018-121-266001}, though for valence ionizations. There, it was shown that a FDE model in which only the halide belonged to the active subsystem was not a good representation for the solvated ion, due to the strong water-fluoride interactions in the valence regions, whereas for chloride (and other heavier halides) this simple FDE model containing no explicit halide-water interactions was quite good. In the higher energy region (panes \textbf{e} and \textbf{f} in figure~\ref{fig:F_Cl_K_edge_sections}) we see that the peaks from FDE calculations are also close to that of the free ion and of supermolecule (around 2774 eV), though for FDE we observe another two other peaks, just over 2774 eV (that show a very small splitting), and four others between 2777 and 2778 eV. The behavior of FDE for this higher energy range is, therefore, in stark contrast to the lowest energy range considered, since there is not even qualitative agreement with the supermolecular results.

Having in mind that in the supermolecular case the complete system is allowed to respond to the external perturbation, but that by construction the response of the environment is absent in the FDE case, this discrepancy provides a first indication of the importance of the response of the environment for higher energies. This is further underscored by the fact that the FDE calculations show no peaks in the intermediate energy range considered. Consequently, we can safely say that the intermediate energy range is, in effect, dominated by excitations from the halide to virtuals with strong (if not dominant) contributions from the environment. Furthermore, while the behavior of FDE is in line with the difference between the isolated and FDE orbital energies, in particular for the lower energy part of the spectra, the situation is less clear-cut with respect to a comparison to the supermolecule. We consider the discrepancies in this case to partly arise from the lack of relaxation for fluoride virtual orbitals and partly from the lack of coupling between the response of the subsystems, as discussed below.

Now comparing supermolecular and BOMME calculations, we see that for both systems the latter provides an overall improvement over FDE--already in qualitative terms, with BOMME we are able to capture the contributions to the solvent to the different electronic states, and, furthermore, in all energy ranges BOMME systematically approaches (underestimates) the excitation energies. In quantitative terms, for fluoride, BOMME clearly perform better than FDE; this can be already seen from the orbital energy differences, and in the low-energy range, the differences in absolute are not very large but BOMME does show smaller differences. For the larger energy range, where FDE is not even qualitatively correct, BOMME shows discrepancies of around 0.1 eV or less. 

For the intermediate energy range, on the other hand, we see more significant differences between the BOMME and supermolecular energies. We attribute this to the use of a GGA functional for the environment, since GGAs tend to underestimate excitation energies with respect to hybrid functionals)~\cite{Besley2021}, and it is precisely in this region that contributions from the environment have a prominent role. While the goal of BOMME is to replace a high-level description of the environment for a lower-level one, our results are a first indication that, for core excited states in which the environment plays an important role, the quality of the low-level of theory may matter much more than for valence states.

The trends outlined above for fluoride are also seen for chloride; if for the low-energy region BOMME does not bring about as significant an improvement over FDE as for fluoride, for the other two energy ranges BOMME provides a semi-quantitative agreement due to its systematic behavior. We see, however, that for the higher energy range, we have often energy differences between BOMME and supermolecule in the order of 1.0 eV, and nearly so for many of the states in the intermediate region. This further underscores the importance of the quality of the density functionals employed for core energies, and in particular deeper cores such as the chloride K edge.

We can now compare these results to what one could expect the simple argument put forward above that the halide orbital energies would provide a dominant contribution to the excited states. From the orbital energies alone, we would expect that both BOMME and FDE would yield excitation energies larger than supermolecular ones; considering only the low-energy range, for both halides this is at odds with the BOMME results, which always shown lower energies than supermolecule, but the simple orbital picture is more consistent with the FDE results, since for three peaks out of four they appear at slightly higher excitation energies (for fluoride, the first FDE peaks appears at a lower energy than supermolecule, whereas for chloride that happens for the fourth peak). We consider this is yet another evidence of the importance of the virtual orbitals from the environment to characterize the excitation energies obtained with BOMME, since such contributions are absent by construction in the case of FDE.

To deepen the discussion on the origin of differences between embeddding and supermolecular results, it is useful to analyse the molecular orbitals involved in the K-edges.
In the real-time framework, similarly to LR,
the absorption cross-section can be interpreted in terms of occupied-virtual
molecular orbital (MO) pairs. This approach has been proposed originally by Repisky and coworkers \cite{repisky2015excitation} and implemented in the relativistic code ReSpect~\cite{Repisky2020}. 
The scheme has been slightly reworked by~\citeauthor{bruner2016accelerated}~\cite{bruner2016accelerated} accelerating the methodology applying the Pad\'e approximants to the Fourier transform of the deconvolution of the induced dipole into molecular orbital pairs. The method gives the MO contribution to the dipole strength function at all frequency. The relative areas under each peak correspond to MO  contribution to an excitation at that frequency, which gives a representation of transitions consistent with the linear response.

In order to have a more visual interpretation of transitions, one can recur to transition density plots. We recall that he simplest direct approach~\cite{kuemmel2012} is to evaluate the transition density at a preset frequency according to:
\begin{equation}
\rho(\bm r,\omega)\propto -\mathrm{Im}\{\delta\tilde{\rho}(\bm{r}, \omega)\},
\end{equation}
where $\delta\tilde{\rho}(\bm{r}, \omega)$ is the Fourier Transform of the time-dependent (TD) induced density $\delta \rho(\boldsymbol{r},t)= \rho(\boldsymbol{r},t)-\rho(\boldsymbol{r},0)$.
Many research groups have recently contributed to this topic~\cite{erhart2017,weissker2018,kuemmel2018,samoza2019}, focusing on low-frequency excitations. To the best of our knowledge, these methods have not been applied yet in the high-frequency range.

In the supplemental information we can observe from the isosurface plot,  $\delta\tilde{\rho}(\bm{r}, \omega_i)$ for $\omega_i =$ 668.586 eV, that the surrounding waters contribute to the K-edge. At the same time, we should point out that the Fourier Transform of the TD induced dipole (using GNU Octave~\cite{octave} routines) is of exceptionally poor quality in the frequency range of fluorine K-edge. It can be argued reasonably that also $\delta\tilde{\rho}(\bm{r}, \omega_i)$ is not of the best quality possible. With that, we have been unable to apply the methodology developed by Schelter and co-workers, in which the accurate value of the oscillator strength is extracted from the refined dipole strength function (DSF) of the TD induced dipole moment. The lack high of quality DSF make difficult to provide an accurate estimation of both the transition dipole moment and transition density. This further prevents from applying more elaborate methods, namely Natural Transition Orbitals. These difficulties should be addressed in dedicated future works.

In the meantime, in the present work we provide in the supplemental information an analysis of a LR-TDDFT calculation on the supermolecular system, performed with the NWChem code~\cite{Valiev2010} (using the same basis sets and density functionals as the Psi4 calculations), from which we can determine that lower-energy K edge transitions of \ce{[F(H2O)8]-} indeed involve virtual orbital with contributions from orbitals centered on the oxygen atoms, in line with the qualitative picture we have managed to extract from the isosurface plots of $\delta\tilde{\rho}(\bm{r}, \omega_i)$.

\subsubsection{\ce{L1} edge}

\begin{figure*}
    \centering
    \includegraphics[width=\linewidth]{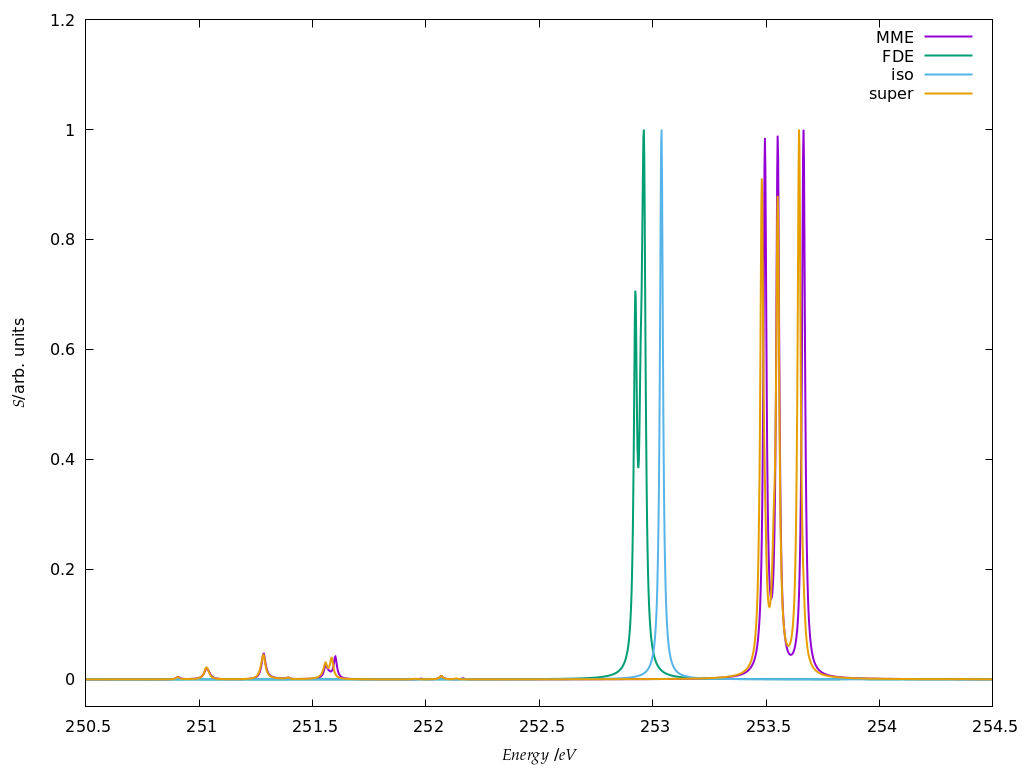}
    \caption{Simulated L$_1$-edge spectra for the fluoride model system in edge region (contrary to the K edge, no peaks of appreciable intensity have been observed at higher energies). It should be noted that here the peak heights (in arbitrary units) for each family of models: free ion (=iso), FDE, BOMME and supermolecule (super) have been scaled, with a height of 1 assigned to the most intense transition.}
    \label{fig:chloride_L1_general}
\end{figure*}

The spectra for the chloride \ce{L1} edge region are shown in figure~\ref{fig:chloride_L1_general}. Unlike the case for the K edge, here we do not show a larger energy range since, for energies between 245 eV and 260 eV, there are no other peaks with appreciable intensity other than those in the picture. A comparison between the free ion at around 253 eV and the supermolecular peaks here show that the energy shift due to the environment is not as marked (around 0.5 eV higher) but nevertheless sufficient to clearly characterize the interaction with the waters through the splitting of the peaks. In this case, we have a near perfect agreement between the supermolecular and BOMME results in terms of peak positions, at around 253.5 eV but also for low intensity transitions around 251.5 eV. We also observe that, unlike for the K edge, here the BOMME results slightly overestimate the supermolecular ones, in agreement with the orbital energy differences ($\Delta \epsilon_3$) in table~\ref{tab:fnt_adf}. We consider that this point, and the absence of other peaks as in the K edge that would indicate more or less important contributions from the environment, we have the L$_1$ edge are almost exclusively dominated by halide to halide transitions. The FDE results, on the other hand, underestimate the effect of environment and show almost no difference to the free ion results, apart from the fact that a splitting of the peak, much less significant than seen for BOMME or supermolecule, is also seen. This would be a further indication that the FDE approach has not properly captured the perturbations to the virtual orbitals of chloride induced by the solvent.

\begin{figure*}
    \centering
    \includegraphics[width=\linewidth]{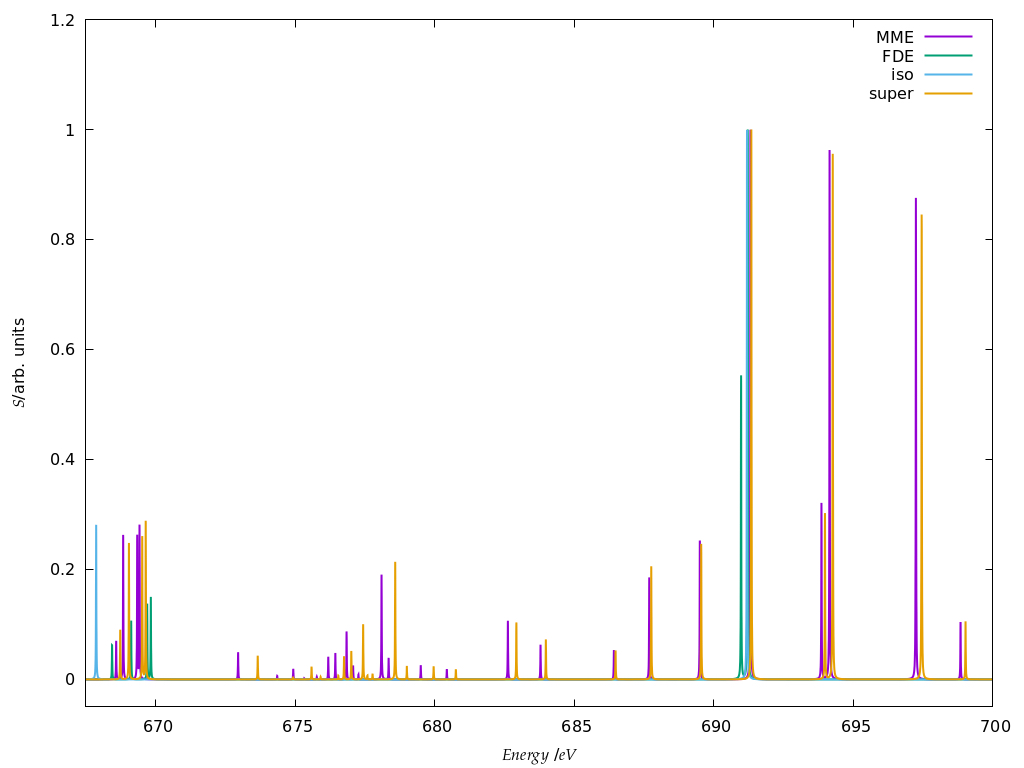}
    \caption{Simulated K-edge spectra for the fluoride model system, over the roughly 30 eV interval starting at the free ion edge peak. It should be noted that here the peak heights (in arbitrary units) for each family of models: free ion (=iso), FDE, BOMME and supermolecule (super) have been scaled, with a height of 1 assigned to the most intense transition.}
    \label{fig:fluoride_K_general}
\end{figure*}

\begin{figure*}
    \centering
    \includegraphics[width=\linewidth]{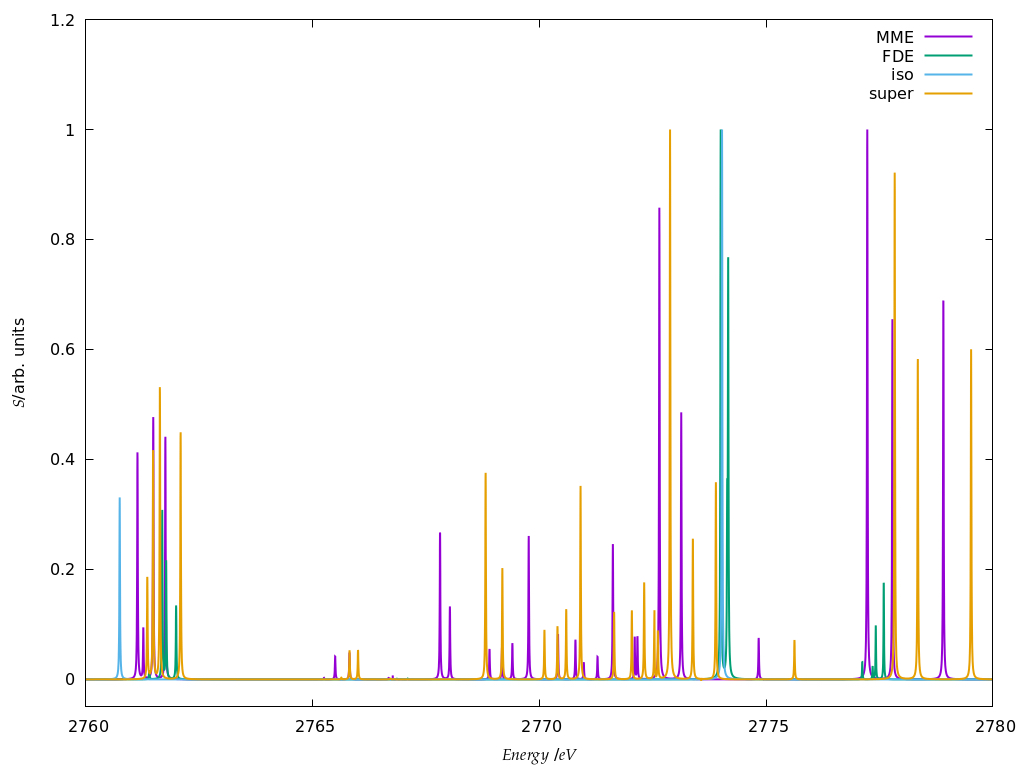}
    \caption{Simulated K-edge spectra for the chloride model system, over the roughly 30 eV interval starting at the free ion edge peak. It should be noted that here the peak heights (in arbitrary units) for each family of models: free ion (=iso), FDE, BOMME and supermolecule (=super) have been scaled, with a height of 1 assigned to the most intense transition.}
    \label{fig:chloride_K_general}
\end{figure*}

\begin{figure*}[htbp]
\centering
\begin{minipage}{0.48\linewidth}
\includegraphics[width=0.95\linewidth]{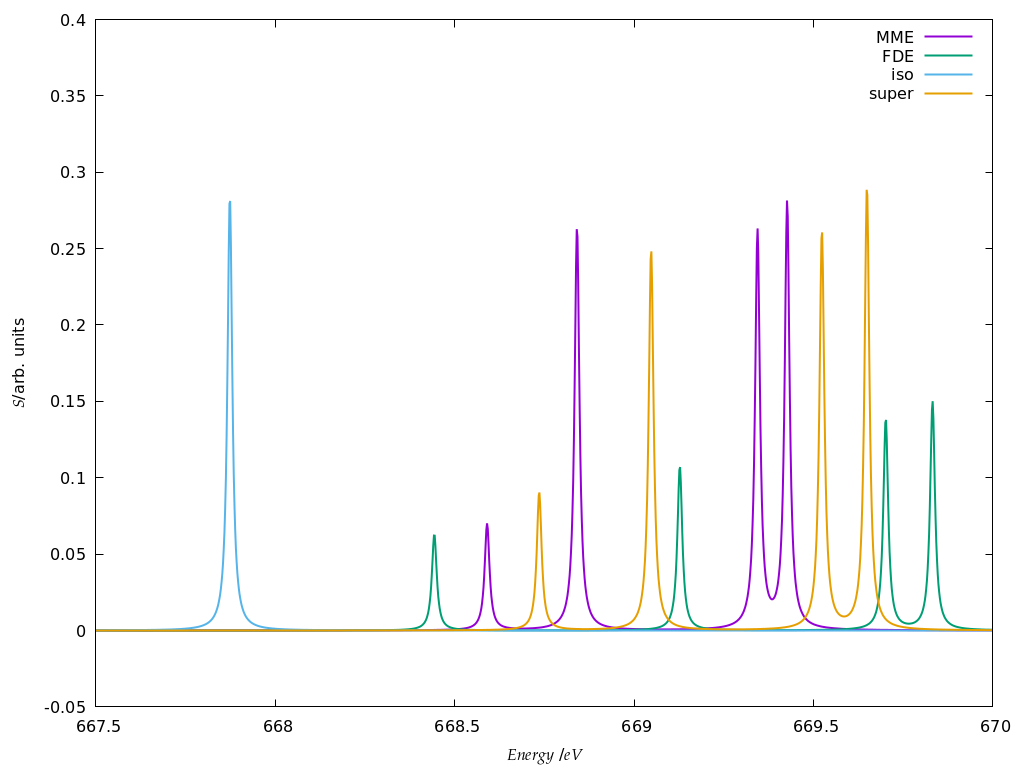}
\end{minipage}
\begin{minipage}{0.48\linewidth}
\includegraphics[width=0.95\linewidth]{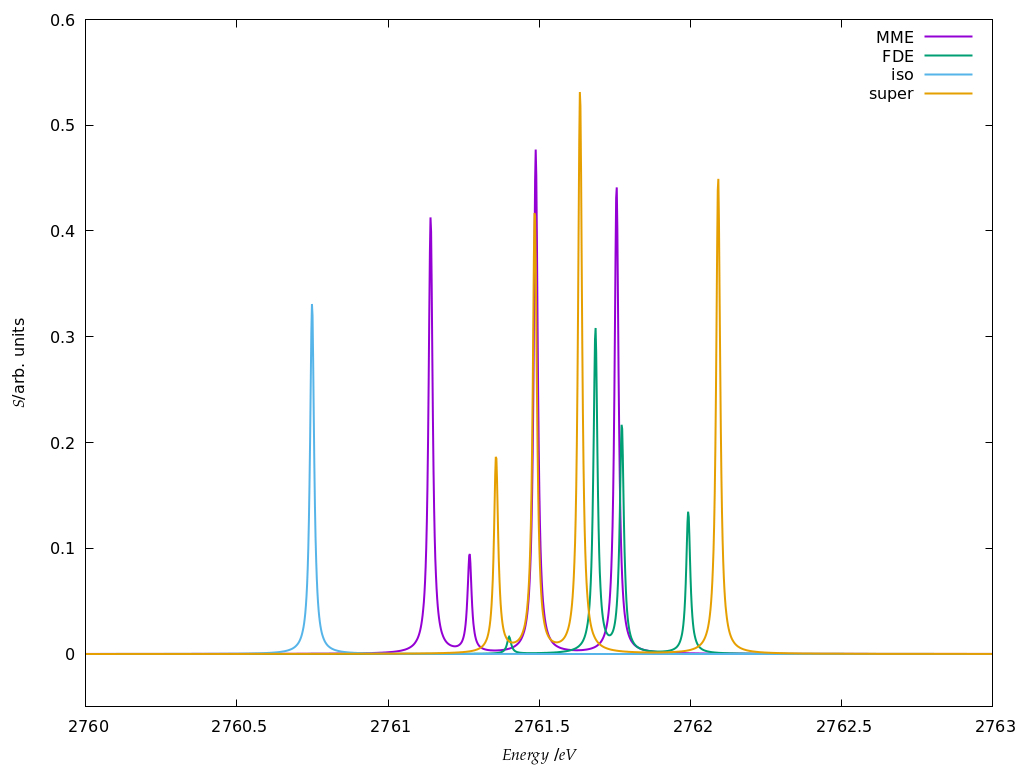}
\end{minipage}
\begin{minipage}{0.48\linewidth}
\includegraphics[width=0.95\linewidth]{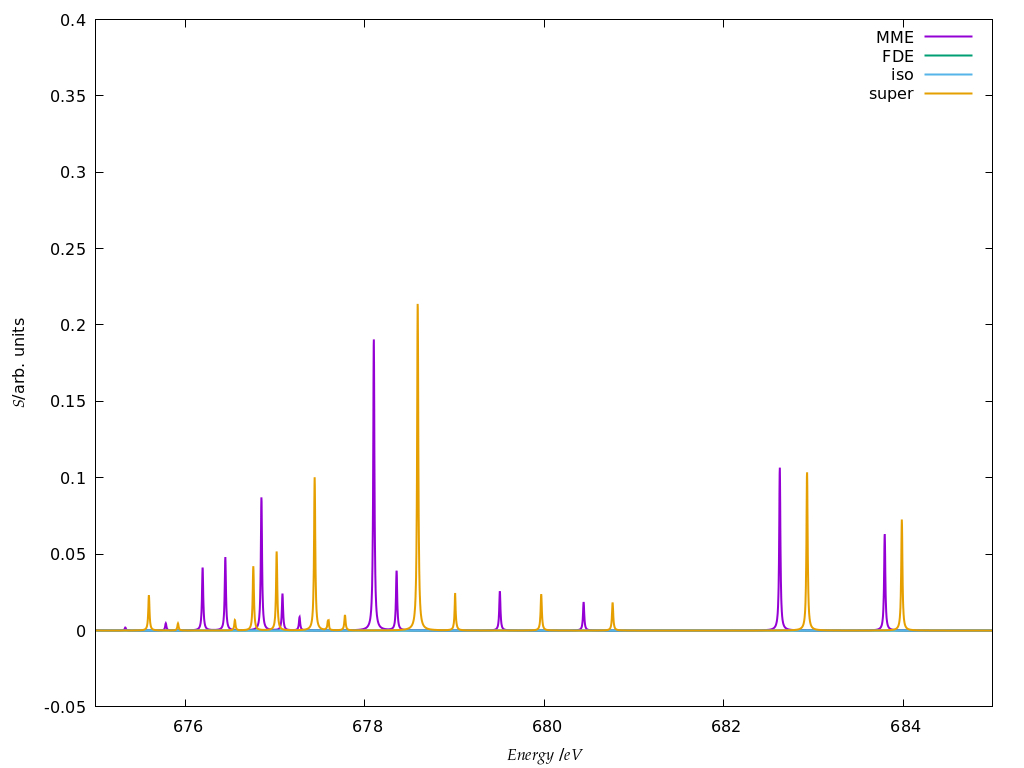}
\end{minipage}
\begin{minipage}{0.48\linewidth}
\includegraphics[width=0.95\linewidth]{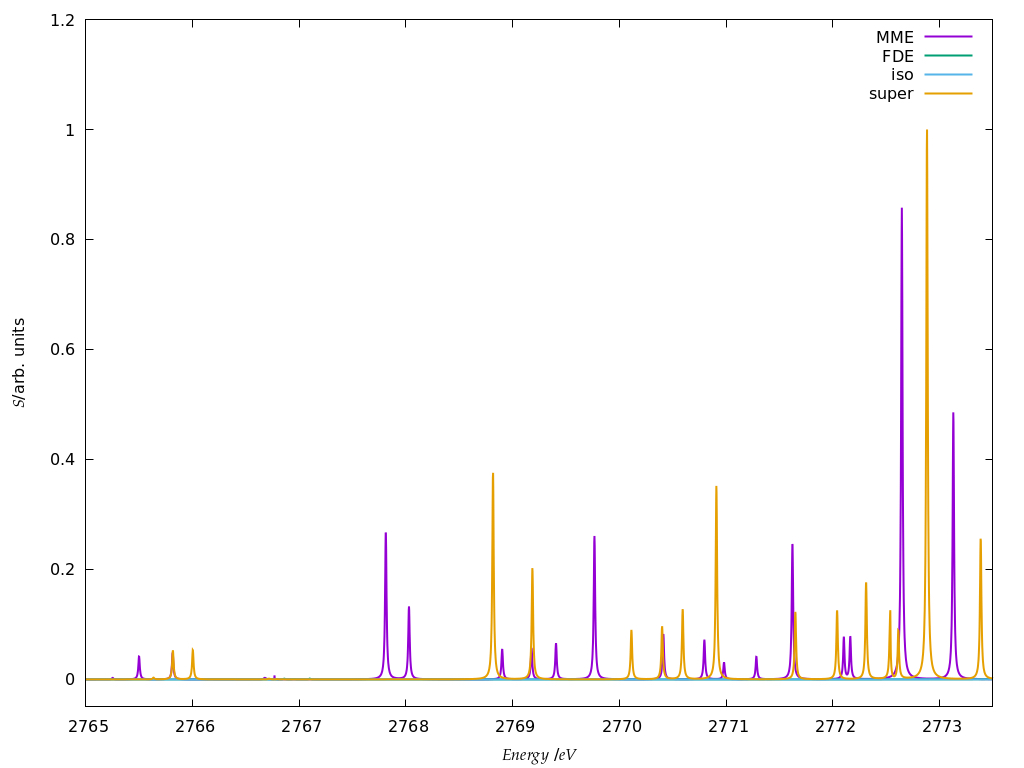}
\end{minipage}
\begin{minipage}{0.48\linewidth}
\includegraphics[width=0.95\linewidth]{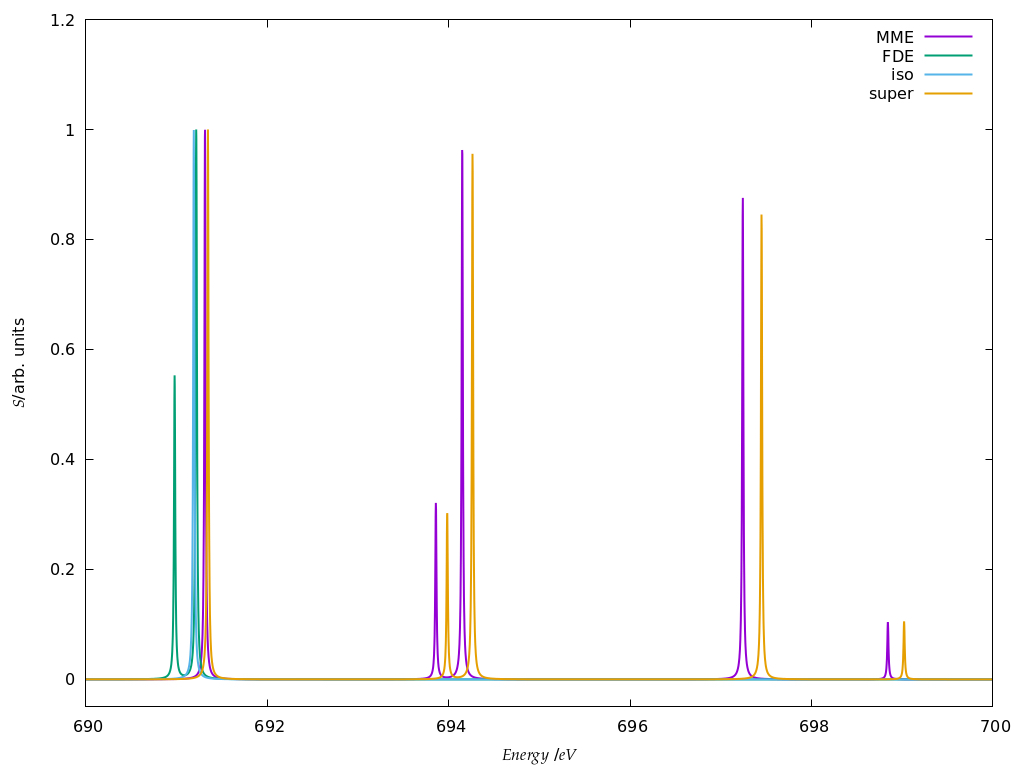}
\end{minipage}
\begin{minipage}{0.48\linewidth}
\includegraphics[width=0.95\linewidth]{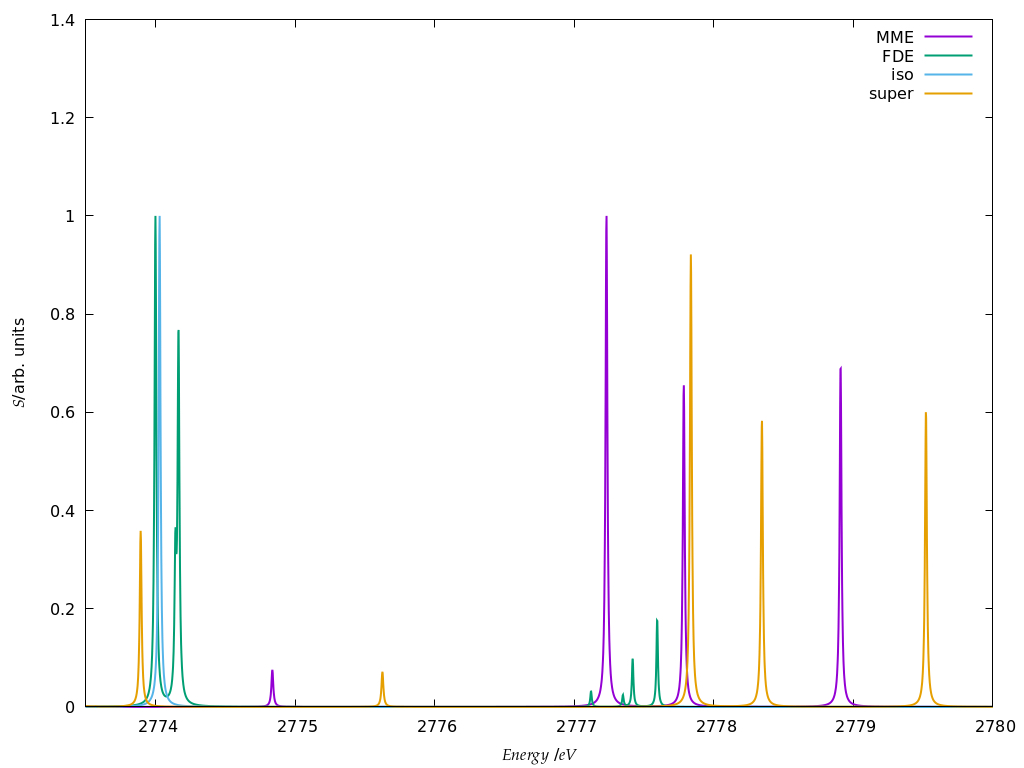}
\end{minipage}
\caption{Details on the three main energy ranges for the K-edge spectra of \ce{F-} (left) and \ce{Cl-} (right) model systems selected from the spectra shown in figures~\ref{fig:fluoride_K_general} and~\ref{fig:chloride_K_general}.}
\label{fig:F_Cl_K_edge_sections}
\end{figure*}

\section{Conclusion}

In this manuscript we have carried out an investigation, to the best of our knowledge for the first time, of the accuracy of fully quantum mechanical (QM/QM), DFT-based embedding approaches--namely, the frozen density embedding (FDE) and block-orthogonalized Manby-Miller embedding (BOMME) approaches--in the description of core excitation spectra (XAS), obtained by the real-time propagation of the electron density, for model systems representing the hydration of halide ions, comprising the halide ions (fluoride and chloride) as active subsystems, and the 8 water molecules in the first solvation shells as the environment.

We note that the BOMME approach and its real-time variant have been implemented within the Psi4Numpy framework in which some of us had previously implemented the real-time TDDFT-in-DFT FDE method, thereby facilitating a rigorous one-to-one comparison between approaches.

From our comparison of the two embedding methods to reference DFT calculations on the whole model systems, we observe first that the BOMME approach can better describe the fluoride core orbital energies in the fluoride-water system than FDE, due to its better handling of the stronger interactions between subsystems, while for chloride both BOMME and FDE perform rather similarly.

In the case of real-time simulations, we have found that the rt-BOMME approach follows the behavior of the reference rt-TDDFT calculations in a very systematic manner across all energy ranges investigated, and as such can potentially become very useful in the investigation of core spectra of species in confined or complex environments. 

We observe that rt-BOMME tends to slightly underestimates the supermolecular results around the edge region for both K and \ce{L1} edges, where excitations mostly take place between orbitals belonging to the halides. On the other hand, we observe much more important discrepancies in higher energy regions for which, it turns out, the environment plays a more important role. 

We attribute these discrepancies to the fact that in rt-BOMME the environment is described with a lower-accuracy GGA functional, a functional class which tends to underestimate core excitation energies due to larger self-interaction errors in comparison to the hybrid functionals which were employed for the active subsystem, and for the whole system in the supermolecular calculations. Our results call for particular attention, in the case of core spectra, in choosing the density functional for the environment, in order to minimize artifacts in the simulations.

The rt-TDDFT-in-DFT simulations carried out under the constraint that the density of the environment has not been relaxed, have nevertheless shown performances similar to the rt-BOMME and reference rt-TDDFT simulations for the (pre-)edge regions. However, since in our implementation the response of the environment is also lacking, large parts of the spectra are either inaccessible or are not correctly described. We intend to address this issue, and introduce the coupling between the response of the different subsystems, in subsequent work.

\section{Data Availability Statement} 

The data presented in this paper are available at the Zenodo repository~\cite{DeSantis-dataset-xas}.

\section{Author contributions}

All authors listed have made a substantial, direct, and intellectual contribution to the work and approved it for publication.

\section{Funding}

We acknowledge funding from projects Labex CaPPA (ANR-11-LABX-0005-01) and CompRIXS (ANR-19-CE29-0019, DFG JA 2329/6-1), the I-SITE ULNE project OVERSEE and MESONM International Associated Laboratory (LAI) (ANR-16-IDEX-0004), and support from the French national supercomputing facilities (grant DARI A0090801859).

\bibliography{ms.bib}

\begin{thebibliography}{102}%
\makeatletter
\providecommand \@ifxundefined [1]{%
 \@ifx{#1\undefined}
}%
\providecommand \@ifnum [1]{%
 \ifnum #1\expandafter \@firstoftwo
 \else \expandafter \@secondoftwo
 \fi
}%
\providecommand \@ifx [1]{%
 \ifx #1\expandafter \@firstoftwo
 \else \expandafter \@secondoftwo
 \fi
}%
\providecommand \natexlab [1]{#1}%
\providecommand \enquote  [1]{``#1''}%
\providecommand \bibnamefont  [1]{#1}%
\providecommand \bibfnamefont [1]{#1}%
\providecommand \citenamefont [1]{#1}%
\providecommand \href@noop [0]{\@secondoftwo}%
\providecommand \href [0]{\begingroup \@sanitize@url \@href}%
\providecommand \@href[1]{\@@startlink{#1}\@@href}%
\providecommand \@@href[1]{\endgroup#1\@@endlink}%
\providecommand \@sanitize@url [0]{\catcode `\\12\catcode `\$12\catcode
  `\&12\catcode `\#12\catcode `\^12\catcode `\_12\catcode `\%12\relax}%
\providecommand \@@startlink[1]{}%
\providecommand \@@endlink[0]{}%
\providecommand \url  [0]{\begingroup\@sanitize@url \@url }%
\providecommand \@url [1]{\endgroup\@href {#1}{\urlprefix }}%
\providecommand \urlprefix  [0]{URL }%
\providecommand \Eprint [0]{\href }%
\providecommand \doibase [0]{http://dx.doi.org/}%
\providecommand \selectlanguage [0]{\@gobble}%
\providecommand \bibinfo  [0]{\@secondoftwo}%
\providecommand \bibfield  [0]{\@secondoftwo}%
\providecommand \translation [1]{[#1]}%
\providecommand \BibitemOpen [0]{}%
\providecommand \bibitemStop [0]{}%
\providecommand \bibitemNoStop [0]{.\EOS\space}%
\providecommand \EOS [0]{\spacefactor3000\relax}%
\providecommand \BibitemShut  [1]{\csname bibitem#1\endcsname}%
\let\auto@bib@innerbib\@empty
\bibitem [{\citenamefont {Zimmermann}\ \emph {et~al.}(2020)\citenamefont
  {Zimmermann}, \citenamefont {Peredkov}, \citenamefont {Abdala}, \citenamefont
  {DeBeer}, \citenamefont {Tromp}, \citenamefont {M\"{u}ller},\ and\
  \citenamefont {van Bokhoven}}]{xray-Zimmermann-CCR2020-423-213466}%
  \BibitemOpen
  \bibfield  {author} {\bibinfo {author} {\bibfnamefont {P.}~\bibnamefont
  {Zimmermann}}, \bibinfo {author} {\bibfnamefont {S.}~\bibnamefont
  {Peredkov}}, \bibinfo {author} {\bibfnamefont {P.~M.}\ \bibnamefont
  {Abdala}}, \bibinfo {author} {\bibfnamefont {S.}~\bibnamefont {DeBeer}},
  \bibinfo {author} {\bibfnamefont {M.}~\bibnamefont {Tromp}}, \bibinfo
  {author} {\bibfnamefont {C.}~\bibnamefont {M\"{u}ller}}, \ and\ \bibinfo
  {author} {\bibfnamefont {J.~A.}\ \bibnamefont {van Bokhoven}},\ }\href
  {\doibase 10.1016/j.ccr.2020.213466} {\bibfield  {journal} {\bibinfo
  {journal} {Coord. Chem. Rev.}\ }\textbf {\bibinfo {volume} {423}},\ \bibinfo
  {pages} {213466} (\bibinfo {year} {2020})}\BibitemShut {NoStop}%
\bibitem [{\citenamefont {Bokhoven}\ and\ \citenamefont
  {Lamberti}(2016)}]{xray-Bokhoven-2016-Book}%
  \BibitemOpen
  \bibinfo {editor} {\bibfnamefont {J.~A.~V.}\ \bibnamefont {Bokhoven}}\ and\
  \bibinfo {editor} {\bibfnamefont {C.}~\bibnamefont {Lamberti}},\ eds.,\ \href
  {\doibase 10.1002/9781118844243} {\emph {\bibinfo {title} {X-Ray Absorption
  and X-Ray Emission Spectroscopy}}}\ (\bibinfo  {publisher} {John Wiley {\&}
  Sons, Ltd},\ \bibinfo {year} {2016})\BibitemShut {NoStop}%
\bibitem [{\citenamefont {Bunker}(2009)}]{xray-Bunker-2009-Book}%
  \BibitemOpen
  \bibfield  {author} {\bibinfo {author} {\bibfnamefont {G.}~\bibnamefont
  {Bunker}},\ }\href {\doibase 10.1017/cbo9780511809194} {\emph {\bibinfo
  {title} {Introduction to {XAFS}}}}\ (\bibinfo  {publisher} {Cambridge
  University Press},\ \bibinfo {year} {2009})\BibitemShut {NoStop}%
\bibitem [{\citenamefont {de~Groot}(2001)}]{xray-Groot-CR2001-101-1779}%
  \BibitemOpen
  \bibfield  {author} {\bibinfo {author} {\bibfnamefont {F.}~\bibnamefont
  {de~Groot}},\ }\href {\doibase 10.1021/cr9900681} {\bibfield  {journal}
  {\bibinfo  {journal} {Chem. Rev.}\ }\textbf {\bibinfo {volume} {101}},\
  \bibinfo {pages} {1779} (\bibinfo {year} {2001})}\BibitemShut {NoStop}%
\bibitem [{\citenamefont {Norman}\ and\ \citenamefont
  {Dreuw}(2018)}]{Norman2018:chemrev}%
  \BibitemOpen
  \bibfield  {author} {\bibinfo {author} {\bibfnamefont {P.}~\bibnamefont
  {Norman}}\ and\ \bibinfo {author} {\bibfnamefont {A.}~\bibnamefont {Dreuw}},\
  }\href {\doibase 10.1021/acs.chemrev.8b00156} {\bibfield  {journal} {\bibinfo
   {journal} {Chem. Rev.}\ }\textbf {\bibinfo {volume} {118}},\ \bibinfo
  {pages} {7208} (\bibinfo {year} {2018})},\ \Eprint
  {http://arxiv.org/abs/https://doi.org/10.1021/acs.chemrev.8b00156}
  {https://doi.org/10.1021/acs.chemrev.8b00156} \BibitemShut {NoStop}%
\bibitem [{\citenamefont {Besley}(2021)}]{Besley2021}%
  \BibitemOpen
  \bibfield  {author} {\bibinfo {author} {\bibfnamefont {N.~A.}\ \bibnamefont
  {Besley}},\ }\href {\doibase 10.1002/wcms.1527} {\bibfield  {journal}
  {\bibinfo  {journal} {{WIREs} Comput. Mol. Sci.}\ }\textbf {\bibinfo {volume}
  {11}} (\bibinfo {year} {2021}),\ 10.1002/wcms.1527}\BibitemShut {NoStop}%
\bibitem [{\citenamefont {Stener}, \citenamefont {Fronzoni},\ and\
  \citenamefont {de~Simone}(2003)}]{dft-Stener-CPL2003-373-115}%
  \BibitemOpen
  \bibfield  {author} {\bibinfo {author} {\bibfnamefont {M.}~\bibnamefont
  {Stener}}, \bibinfo {author} {\bibfnamefont {G.}~\bibnamefont {Fronzoni}}, \
  and\ \bibinfo {author} {\bibfnamefont {M.}~\bibnamefont {de~Simone}},\ }\href
  {\doibase 10.1016/s0009-2614(03)00543-8} {\bibfield  {journal} {\bibinfo
  {journal} {Chem. Phys. Lett.}\ }\textbf {\bibinfo {volume} {373}},\ \bibinfo
  {pages} {115} (\bibinfo {year} {2003})}\BibitemShut {NoStop}%
\bibitem [{\citenamefont {Besley}\ and\ \citenamefont
  {Asmuruf}(2010)}]{Besley2010}%
  \BibitemOpen
  \bibfield  {author} {\bibinfo {author} {\bibfnamefont {N.~A.}\ \bibnamefont
  {Besley}}\ and\ \bibinfo {author} {\bibfnamefont {F.~A.}\ \bibnamefont
  {Asmuruf}},\ }\href {\doibase 10.1039/C002207A} {\bibfield  {journal}
  {\bibinfo  {journal} {Phys. Chem. Chem. Phys.}\ }\textbf {\bibinfo {volume}
  {12}},\ \bibinfo {pages} {12024} (\bibinfo {year} {2010})}\BibitemShut
  {NoStop}%
\bibitem [{\citenamefont {Zhang}\ \emph {et~al.}(2012)\citenamefont {Zhang},
  \citenamefont {Biggs}, \citenamefont {Healion}, \citenamefont {Govind},\ and\
  \citenamefont {Mukamel}}]{dft-Zhang-JCP2012-137-194306}%
  \BibitemOpen
  \bibfield  {author} {\bibinfo {author} {\bibfnamefont {Y.}~\bibnamefont
  {Zhang}}, \bibinfo {author} {\bibfnamefont {J.~D.}\ \bibnamefont {Biggs}},
  \bibinfo {author} {\bibfnamefont {D.}~\bibnamefont {Healion}}, \bibinfo
  {author} {\bibfnamefont {N.}~\bibnamefont {Govind}}, \ and\ \bibinfo {author}
  {\bibfnamefont {S.}~\bibnamefont {Mukamel}},\ }\href {\doibase
  10.1063/1.4766356} {\bibfield  {journal} {\bibinfo  {journal} {J. Chem.
  Phys.}\ }\textbf {\bibinfo {volume} {137}},\ \bibinfo {pages} {194306}
  (\bibinfo {year} {2012})}\BibitemShut {NoStop}%
\bibitem [{\citenamefont {Ekstr\"om}\ \emph {et~al.}(2006)\citenamefont
  {Ekstr\"om}, \citenamefont {Norman}, \citenamefont {Carravetta},\ and\
  \citenamefont {\AA{}gren}}]{Ekstrom2006a}%
  \BibitemOpen
  \bibfield  {author} {\bibinfo {author} {\bibfnamefont {U.}~\bibnamefont
  {Ekstr\"om}}, \bibinfo {author} {\bibfnamefont {P.}~\bibnamefont {Norman}},
  \bibinfo {author} {\bibfnamefont {V.}~\bibnamefont {Carravetta}}, \ and\
  \bibinfo {author} {\bibfnamefont {H.}~\bibnamefont {\AA{}gren}},\ }\href
  {\doibase 10.1103/PhysRevLett.97.143001} {\bibfield  {journal} {\bibinfo
  {journal} {Phys. Rev. Lett.}\ }\textbf {\bibinfo {volume} {97}},\ \bibinfo
  {pages} {143001} (\bibinfo {year} {2006})}\BibitemShut {NoStop}%
\bibitem [{\citenamefont {Ekstr\"om}\ and\ \citenamefont
  {Norman}(2006)}]{Ekstrom2006b}%
  \BibitemOpen
  \bibfield  {author} {\bibinfo {author} {\bibfnamefont {U.}~\bibnamefont
  {Ekstr\"om}}\ and\ \bibinfo {author} {\bibfnamefont {P.}~\bibnamefont
  {Norman}},\ }\href {\doibase 10.1103/PhysRevA.74.042722} {\bibfield
  {journal} {\bibinfo  {journal} {Phys. Rev. A}\ }\textbf {\bibinfo {volume}
  {74}},\ \bibinfo {pages} {042722} (\bibinfo {year} {2006})}\BibitemShut
  {NoStop}%
\bibitem [{\citenamefont {Jiemchooroj}, \citenamefont {Ekström},\ and\
  \citenamefont {Norman}(2007)}]{Jiemchooroj2007}%
  \BibitemOpen
  \bibfield  {author} {\bibinfo {author} {\bibfnamefont {A.}~\bibnamefont
  {Jiemchooroj}}, \bibinfo {author} {\bibfnamefont {U.}~\bibnamefont
  {Ekström}}, \ and\ \bibinfo {author} {\bibfnamefont {P.}~\bibnamefont
  {Norman}},\ }\href {\doibase 10.1063/1.2800024} {\bibfield  {journal}
  {\bibinfo  {journal} {J. Chem. Phys.}\ }\textbf {\bibinfo {volume} {127}},\
  \bibinfo {pages} {165104} (\bibinfo {year} {2007})},\ \Eprint
  {http://arxiv.org/abs/https://doi.org/10.1063/1.2800024}
  {https://doi.org/10.1063/1.2800024} \BibitemShut {NoStop}%
\bibitem [{\citenamefont {Villaume}, \citenamefont {Saue},\ and\ \citenamefont
  {Norman}(2010)}]{Villaume2010}%
  \BibitemOpen
  \bibfield  {author} {\bibinfo {author} {\bibfnamefont {S.}~\bibnamefont
  {Villaume}}, \bibinfo {author} {\bibfnamefont {T.}~\bibnamefont {Saue}}, \
  and\ \bibinfo {author} {\bibfnamefont {P.}~\bibnamefont {Norman}},\ }\href
  {\doibase 10.1063/1.3461163} {\bibfield  {journal} {\bibinfo  {journal} {J.
  Chem. Phys.}\ }\textbf {\bibinfo {volume} {133}},\ \bibinfo {pages} {064105}
  (\bibinfo {year} {2010})},\ \Eprint
  {http://arxiv.org/abs/https://doi.org/10.1063/1.3461163}
  {https://doi.org/10.1063/1.3461163} \BibitemShut {NoStop}%
\bibitem [{\citenamefont {Pedersen}\ \emph {et~al.}(2014)\citenamefont
  {Pedersen}, \citenamefont {Hedegård}, \citenamefont {Olsen}, \citenamefont
  {Kauczor}, \citenamefont {Norman},\ and\ \citenamefont
  {Kongsted}}]{Pedersen2014}%
  \BibitemOpen
  \bibfield  {author} {\bibinfo {author} {\bibfnamefont {M.~N.}\ \bibnamefont
  {Pedersen}}, \bibinfo {author} {\bibfnamefont {E.~D.}\ \bibnamefont
  {Hedegård}}, \bibinfo {author} {\bibfnamefont {J.~M.~H.}\ \bibnamefont
  {Olsen}}, \bibinfo {author} {\bibfnamefont {J.}~\bibnamefont {Kauczor}},
  \bibinfo {author} {\bibfnamefont {P.}~\bibnamefont {Norman}}, \ and\ \bibinfo
  {author} {\bibfnamefont {J.}~\bibnamefont {Kongsted}},\ }\href {\doibase
  10.1021/ct400946k} {\bibfield  {journal} {\bibinfo  {journal} {J. Chem.
  Theory Comput.}\ }\textbf {\bibinfo {volume} {10}},\ \bibinfo {pages} {1164}
  (\bibinfo {year} {2014})},\ \Eprint
  {http://arxiv.org/abs/https://doi.org/10.1021/ct400946k}
  {https://doi.org/10.1021/ct400946k} \BibitemShut {NoStop}%
\bibitem [{\citenamefont {Fahleson}, \citenamefont {Ågren},\ and\
  \citenamefont {Norman}(2016)}]{Fahleson2016}%
  \BibitemOpen
  \bibfield  {author} {\bibinfo {author} {\bibfnamefont {T.}~\bibnamefont
  {Fahleson}}, \bibinfo {author} {\bibfnamefont {H.}~\bibnamefont {Ågren}}, \
  and\ \bibinfo {author} {\bibfnamefont {P.}~\bibnamefont {Norman}},\ }\href
  {\doibase 10.1021/acs.jpclett.6b00750} {\bibfield  {journal} {\bibinfo
  {journal} {J. Phys. Chem. Lett.}\ }\textbf {\bibinfo {volume} {7}},\ \bibinfo
  {pages} {1991} (\bibinfo {year} {2016})},\ \Eprint
  {http://arxiv.org/abs/https://doi.org/10.1021/acs.jpclett.6b00750}
  {https://doi.org/10.1021/acs.jpclett.6b00750} \BibitemShut {NoStop}%
\bibitem [{\citenamefont {Rinkevicius}\ \emph {et~al.}(2016)\citenamefont
  {Rinkevicius}, \citenamefont {Sandberg}, \citenamefont {Li}, \citenamefont
  {Linares}, \citenamefont {Norman},\ and\ \citenamefont
  {Ågren}}]{Rinkevicius2016}%
  \BibitemOpen
  \bibfield  {author} {\bibinfo {author} {\bibfnamefont {Z.}~\bibnamefont
  {Rinkevicius}}, \bibinfo {author} {\bibfnamefont {J.~A.~R.}\ \bibnamefont
  {Sandberg}}, \bibinfo {author} {\bibfnamefont {X.}~\bibnamefont {Li}},
  \bibinfo {author} {\bibfnamefont {M.}~\bibnamefont {Linares}}, \bibinfo
  {author} {\bibfnamefont {P.}~\bibnamefont {Norman}}, \ and\ \bibinfo {author}
  {\bibfnamefont {H.}~\bibnamefont {Ågren}},\ }\href {\doibase
  10.1021/acs.jctc.6b00255} {\bibfield  {journal} {\bibinfo  {journal} {J.
  Chem. Theory Comput.}\ }\textbf {\bibinfo {volume} {12}},\ \bibinfo {pages}
  {2661} (\bibinfo {year} {2016})},\ \Eprint
  {http://arxiv.org/abs/https://doi.org/10.1021/acs.jctc.6b00255}
  {https://doi.org/10.1021/acs.jctc.6b00255} \BibitemShut {NoStop}%
\bibitem [{\citenamefont {Goings}, \citenamefont {Lestrange},\ and\
  \citenamefont {Li}(2017)}]{dft-Goings-WCMS2017-8-e1341}%
  \BibitemOpen
  \bibfield  {author} {\bibinfo {author} {\bibfnamefont {J.~J.}\ \bibnamefont
  {Goings}}, \bibinfo {author} {\bibfnamefont {P.~J.}\ \bibnamefont
  {Lestrange}}, \ and\ \bibinfo {author} {\bibfnamefont {X.}~\bibnamefont
  {Li}},\ }\href {\doibase 10.1002/wcms.1341} {\bibfield  {journal} {\bibinfo
  {journal} {{WIREs} Comput. Mol. Sci.}\ }\textbf {\bibinfo {volume} {8}},\
  \bibinfo {pages} {e1341} (\bibinfo {year} {2017})}\BibitemShut {NoStop}%
\bibitem [{\citenamefont {Li}\ \emph {et~al.}(2020)\citenamefont {Li},
  \citenamefont {Govind}, \citenamefont {Isborn}, \citenamefont {DePrince},\
  and\ \citenamefont {Lopata}}]{li_rev2020}%
  \BibitemOpen
  \bibfield  {author} {\bibinfo {author} {\bibfnamefont {X.}~\bibnamefont
  {Li}}, \bibinfo {author} {\bibfnamefont {N.}~\bibnamefont {Govind}}, \bibinfo
  {author} {\bibfnamefont {C.}~\bibnamefont {Isborn}}, \bibinfo {author}
  {\bibfnamefont {A.~E.}\ \bibnamefont {DePrince}}, \ and\ \bibinfo {author}
  {\bibfnamefont {K.}~\bibnamefont {Lopata}},\ }\href {\doibase
  10.1021/acs.chemrev.0c00223} {\bibfield  {journal} {\bibinfo  {journal}
  {Chem. Rev.}\ }\textbf {\bibinfo {volume} {120}},\ \bibinfo {pages} {9951}
  (\bibinfo {year} {2020})},\ \Eprint
  {http://arxiv.org/abs/https://doi.org/10.1021/acs.chemrev.0c00223}
  {https://doi.org/10.1021/acs.chemrev.0c00223} \BibitemShut {NoStop}%
\bibitem [{\citenamefont {Lopata}\ \emph {et~al.}(2012)\citenamefont {Lopata},
  \citenamefont {Van~Kuiken}, \citenamefont {Khalil},\ and\ \citenamefont
  {Govind}}]{lopata2012}%
  \BibitemOpen
  \bibfield  {author} {\bibinfo {author} {\bibfnamefont {K.}~\bibnamefont
  {Lopata}}, \bibinfo {author} {\bibfnamefont {B.~E.}\ \bibnamefont
  {Van~Kuiken}}, \bibinfo {author} {\bibfnamefont {M.}~\bibnamefont {Khalil}},
  \ and\ \bibinfo {author} {\bibfnamefont {N.}~\bibnamefont {Govind}},\ }\href
  {\doibase 10.1021/ct3005613} {\bibfield  {journal} {\bibinfo  {journal} {J.
  Chem. Theory Comput.}\ }\textbf {\bibinfo {volume} {8}},\ \bibinfo {pages}
  {3284} (\bibinfo {year} {2012})},\ \Eprint
  {http://arxiv.org/abs/https://doi.org/10.1021/ct3005613}
  {https://doi.org/10.1021/ct3005613} \BibitemShut {NoStop}%
\bibitem [{\citenamefont {Kadek}\ \emph {et~al.}(2015)\citenamefont {Kadek},
  \citenamefont {Konecny}, \citenamefont {Gao}, \citenamefont {Repisky},\ and\
  \citenamefont {Ruud}}]{Kadek2015}%
  \BibitemOpen
  \bibfield  {author} {\bibinfo {author} {\bibfnamefont {M.}~\bibnamefont
  {Kadek}}, \bibinfo {author} {\bibfnamefont {L.}~\bibnamefont {Konecny}},
  \bibinfo {author} {\bibfnamefont {B.}~\bibnamefont {Gao}}, \bibinfo {author}
  {\bibfnamefont {M.}~\bibnamefont {Repisky}}, \ and\ \bibinfo {author}
  {\bibfnamefont {K.}~\bibnamefont {Ruud}},\ }\href {\doibase
  10.1039/C5CP03712C} {\bibfield  {journal} {\bibinfo  {journal} {Phys. Chem.
  Chem. Phys.}\ }\textbf {\bibinfo {volume} {17}},\ \bibinfo {pages} {22566}
  (\bibinfo {year} {2015})}\BibitemShut {NoStop}%
\bibitem [{\citenamefont {Eberly}, \citenamefont {Javanainen},\ and\
  \citenamefont {Rza{\.z}ewski}(1991)}]{ati}%
  \BibitemOpen
  \bibfield  {author} {\bibinfo {author} {\bibfnamefont {J.}~\bibnamefont
  {Eberly}}, \bibinfo {author} {\bibfnamefont {J.}~\bibnamefont {Javanainen}},
  \ and\ \bibinfo {author} {\bibfnamefont {K.}~\bibnamefont {Rza{\.z}ewski}},\
  }\href {\doibase 10.1016/0370-1573(91)90131-5} {\bibfield  {journal}
  {\bibinfo  {journal} {Phys. Rep.}\ }\textbf {\bibinfo {volume} {204}},\
  \bibinfo {pages} {331 } (\bibinfo {year} {1991})}\BibitemShut {NoStop}%
\bibitem [{\citenamefont {Keldysh}(2017)}]{Keldysh_2017}%
  \BibitemOpen
  \bibfield  {author} {\bibinfo {author} {\bibfnamefont {L.~V.}\ \bibnamefont
  {Keldysh}},\ }\href {\doibase 10.3367/ufne.2017.10.038229} {\bibfield
  {journal} {\bibinfo  {journal} {Phys.-Uspekhi}\ }\textbf {\bibinfo {volume}
  {60}},\ \bibinfo {pages} {1187} (\bibinfo {year} {2017})}\BibitemShut
  {NoStop}%
\bibitem [{\citenamefont {Cheng}, \citenamefont {Evans},\ and\ \citenamefont
  {Van~Voorhis}(2006)}]{voohris}%
  \BibitemOpen
  \bibfield  {author} {\bibinfo {author} {\bibfnamefont {C.-L.}\ \bibnamefont
  {Cheng}}, \bibinfo {author} {\bibfnamefont {J.~S.}\ \bibnamefont {Evans}}, \
  and\ \bibinfo {author} {\bibfnamefont {T.}~\bibnamefont {Van~Voorhis}},\
  }\href {\doibase 10.1103/PhysRevB.74.155112} {\bibfield  {journal} {\bibinfo
  {journal} {Phys. Rev. B}\ }\textbf {\bibinfo {volume} {74}},\ \bibinfo
  {pages} {155112} (\bibinfo {year} {2006})}\BibitemShut {NoStop}%
\bibitem [{\citenamefont {Mokkath}(2020)}]{MOKKATH2020137905}%
  \BibitemOpen
  \bibfield  {author} {\bibinfo {author} {\bibfnamefont {J.~H.}\ \bibnamefont
  {Mokkath}},\ }\href {\doibase 10.1016/j.cplett.2020.137905} {\bibfield
  {journal} {\bibinfo  {journal} {Chem. Phys. Lett.}\ }\textbf {\bibinfo
  {volume} {758}},\ \bibinfo {pages} {137905} (\bibinfo {year}
  {2020})}\BibitemShut {NoStop}%
\bibitem [{\citenamefont {Gomes}\ and\ \citenamefont
  {Jacob}(2012)}]{env-Gomes-ARPCSPC2012-108-222}%
  \BibitemOpen
  \bibfield  {author} {\bibinfo {author} {\bibfnamefont {A.~S.~P.}\
  \bibnamefont {Gomes}}\ and\ \bibinfo {author} {\bibfnamefont {C.~R.}\
  \bibnamefont {Jacob}},\ }\href {\doibase 10.1039/C2PC90007F} {\bibfield
  {journal} {\bibinfo  {journal} {Annu. Rep. Prog. Chem.{,} Sect. C: Phys.
  Chem.}\ }\textbf {\bibinfo {volume} {108}},\ \bibinfo {pages} {222} (\bibinfo
  {year} {2012})}\BibitemShut {NoStop}%
\bibitem [{\citenamefont {Lipparini}\ and\ \citenamefont
  {Mennucci}(2021)}]{Lipparini2021}%
  \BibitemOpen
  \bibfield  {author} {\bibinfo {author} {\bibfnamefont {F.}~\bibnamefont
  {Lipparini}}\ and\ \bibinfo {author} {\bibfnamefont {B.}~\bibnamefont
  {Mennucci}},\ }\href {\doibase 10.1063/5.0064075} {\bibfield  {journal}
  {\bibinfo  {journal} {Chem. Phys. Rev.}\ }\textbf {\bibinfo {volume} {2}},\
  \bibinfo {pages} {041303} (\bibinfo {year} {2021})},\ \Eprint
  {http://arxiv.org/abs/https://doi.org/10.1063/5.0064075}
  {https://doi.org/10.1063/5.0064075} \BibitemShut {NoStop}%
\bibitem [{\citenamefont {Pipolo}\ and\ \citenamefont
  {Corni}(2016)}]{Pipolo2016}%
  \BibitemOpen
  \bibfield  {author} {\bibinfo {author} {\bibfnamefont {S.}~\bibnamefont
  {Pipolo}}\ and\ \bibinfo {author} {\bibfnamefont {S.}~\bibnamefont {Corni}},\
  }\href {\doibase 10.1021/acs.jpcc.6b11084} {\bibfield  {journal} {\bibinfo
  {journal} {J. Phys. Chem. C}\ }\textbf {\bibinfo {volume} {120}},\ \bibinfo
  {pages} {28774} (\bibinfo {year} {2016})},\ \Eprint
  {http://arxiv.org/abs/https://doi.org/10.1021/acs.jpcc.6b11084}
  {https://doi.org/10.1021/acs.jpcc.6b11084} \BibitemShut {NoStop}%
\bibitem [{\citenamefont {Gil}\ \emph {et~al.}(2019)\citenamefont {Gil},
  \citenamefont {Pipolo}, \citenamefont {Delgado}, \citenamefont {Rozzi},\ and\
  \citenamefont {Corni}}]{Gil2019}%
  \BibitemOpen
  \bibfield  {author} {\bibinfo {author} {\bibfnamefont {G.}~\bibnamefont
  {Gil}}, \bibinfo {author} {\bibfnamefont {S.}~\bibnamefont {Pipolo}},
  \bibinfo {author} {\bibfnamefont {A.}~\bibnamefont {Delgado}}, \bibinfo
  {author} {\bibfnamefont {C.~A.}\ \bibnamefont {Rozzi}}, \ and\ \bibinfo
  {author} {\bibfnamefont {S.}~\bibnamefont {Corni}},\ }\href {\doibase
  10.1021/acs.jctc.9b00010} {\bibfield  {journal} {\bibinfo  {journal} {J.
  Chem. Theory Comput.}\ }\textbf {\bibinfo {volume} {15}},\ \bibinfo {pages}
  {2306} (\bibinfo {year} {2019})},\ \Eprint
  {http://arxiv.org/abs/https://doi.org/10.1021/acs.jctc.9b00010}
  {https://doi.org/10.1021/acs.jctc.9b00010} \BibitemShut {NoStop}%
\bibitem [{\citenamefont {Marques}\ \emph {et~al.}(2003)\citenamefont
  {Marques}, \citenamefont {L{\'{o}}pez}, \citenamefont {Varsano},
  \citenamefont {Castro},\ and\ \citenamefont
  {Rubio}}]{env-Marques-PRL2003-90-258101}%
  \BibitemOpen
  \bibfield  {author} {\bibinfo {author} {\bibfnamefont {M.~A.~L.}\
  \bibnamefont {Marques}}, \bibinfo {author} {\bibfnamefont {X.}~\bibnamefont
  {L{\'{o}}pez}}, \bibinfo {author} {\bibfnamefont {D.}~\bibnamefont
  {Varsano}}, \bibinfo {author} {\bibfnamefont {A.}~\bibnamefont {Castro}}, \
  and\ \bibinfo {author} {\bibfnamefont {A.}~\bibnamefont {Rubio}},\ }\href
  {\doibase 10.1103/physrevlett.90.258101} {\bibfield  {journal} {\bibinfo
  {journal} {Physical Review Letters}\ }\textbf {\bibinfo {volume} {90}},\
  \bibinfo {pages} {258101} (\bibinfo {year} {2003})}\BibitemShut {NoStop}%
\bibitem [{\citenamefont {Morzan}\ \emph {et~al.}(2014)\citenamefont {Morzan},
  \citenamefont {Ram{\'{\i}}rez}, \citenamefont {Oviedo}, \citenamefont
  {S{\'{a}}nchez}, \citenamefont {Scherlis},\ and\ \citenamefont
  {Lebrero}}]{env-Morzan-JCP2014-140-164105}%
  \BibitemOpen
  \bibfield  {author} {\bibinfo {author} {\bibfnamefont {U.~N.}\ \bibnamefont
  {Morzan}}, \bibinfo {author} {\bibfnamefont {F.~F.}\ \bibnamefont
  {Ram{\'{\i}}rez}}, \bibinfo {author} {\bibfnamefont {M.~B.}\ \bibnamefont
  {Oviedo}}, \bibinfo {author} {\bibfnamefont {C.~G.}\ \bibnamefont
  {S{\'{a}}nchez}}, \bibinfo {author} {\bibfnamefont {D.~A.}\ \bibnamefont
  {Scherlis}}, \ and\ \bibinfo {author} {\bibfnamefont {M.~C.~G.}\ \bibnamefont
  {Lebrero}},\ }\href {\doibase 10.1063/1.4871688} {\bibfield  {journal}
  {\bibinfo  {journal} {J. Chem. Phys.}\ }\textbf {\bibinfo {volume} {140}},\
  \bibinfo {pages} {164105} (\bibinfo {year} {2014})}\BibitemShut {NoStop}%
\bibitem [{\citenamefont {Wu}\ \emph {et~al.}(2017)\citenamefont {Wu},
  \citenamefont {Teuler}, \citenamefont {Cailliez}, \citenamefont
  {Clavaguéra}, \citenamefont {Salahub},\ and\ \citenamefont {de~la
  Lande}}]{Wu2017}%
  \BibitemOpen
  \bibfield  {author} {\bibinfo {author} {\bibfnamefont {X.}~\bibnamefont
  {Wu}}, \bibinfo {author} {\bibfnamefont {J.-M.}\ \bibnamefont {Teuler}},
  \bibinfo {author} {\bibfnamefont {F.}~\bibnamefont {Cailliez}}, \bibinfo
  {author} {\bibfnamefont {C.}~\bibnamefont {Clavaguéra}}, \bibinfo {author}
  {\bibfnamefont {D.~R.}\ \bibnamefont {Salahub}}, \ and\ \bibinfo {author}
  {\bibfnamefont {A.}~\bibnamefont {de~la Lande}},\ }\href {\doibase
  10.1021/acs.jctc.7b00251} {\bibfield  {journal} {\bibinfo  {journal} {J.
  Chem. Theory Comput.}\ }\textbf {\bibinfo {volume} {13}},\ \bibinfo {pages}
  {3985} (\bibinfo {year} {2017})},\ \Eprint
  {http://arxiv.org/abs/https://doi.org/10.1021/acs.jctc.7b00251}
  {https://doi.org/10.1021/acs.jctc.7b00251} \BibitemShut {NoStop}%
\bibitem [{\citenamefont {Parise}\ \emph {et~al.}(2018)\citenamefont {Parise},
  \citenamefont {Alvarez-Ibarra}, \citenamefont {Wu}, \citenamefont {Zhao},
  \citenamefont {Pilm{\'{e}}},\ and\ \citenamefont {de~la Lande}}]{Parise2018}%
  \BibitemOpen
  \bibfield  {author} {\bibinfo {author} {\bibfnamefont {A.}~\bibnamefont
  {Parise}}, \bibinfo {author} {\bibfnamefont {A.}~\bibnamefont
  {Alvarez-Ibarra}}, \bibinfo {author} {\bibfnamefont {X.}~\bibnamefont {Wu}},
  \bibinfo {author} {\bibfnamefont {X.}~\bibnamefont {Zhao}}, \bibinfo {author}
  {\bibfnamefont {J.}~\bibnamefont {Pilm{\'{e}}}}, \ and\ \bibinfo {author}
  {\bibfnamefont {A.}~\bibnamefont {de~la Lande}},\ }\href {\doibase
  10.1021/acs.jpclett.7b03379} {\bibfield  {journal} {\bibinfo  {journal} {J.
  Phys. Chem. Lett.}\ }\textbf {\bibinfo {volume} {9}},\ \bibinfo {pages} {844}
  (\bibinfo {year} {2018})}\BibitemShut {NoStop}%
\bibitem [{\citenamefont {Jacob}\ and\ \citenamefont
  {Neugebauer}(2014)}]{env-Jacob-WCMS2014-4-325}%
  \BibitemOpen
  \bibfield  {author} {\bibinfo {author} {\bibfnamefont {C.~R.}\ \bibnamefont
  {Jacob}}\ and\ \bibinfo {author} {\bibfnamefont {J.}~\bibnamefont
  {Neugebauer}},\ }\href {\doibase 10.1002/wcms.1175} {\bibfield  {journal}
  {\bibinfo  {journal} {{WIREs} Comput. Mol. Sci.}\ }\textbf {\bibinfo {volume}
  {4}},\ \bibinfo {pages} {325} (\bibinfo {year} {2014})},\ \Eprint
  {http://arxiv.org/abs/https://wires.onlinelibrary.wiley.com/doi/pdf/10.1002/wcms.1175}
  {https://wires.onlinelibrary.wiley.com/doi/pdf/10.1002/wcms.1175}
  \BibitemShut {NoStop}%
\bibitem [{\citenamefont {Wesolowski}, \citenamefont {Shedge},\ and\
  \citenamefont {Zhou}(2015)}]{wesolowski_frozen-density_2015}%
  \BibitemOpen
  \bibfield  {author} {\bibinfo {author} {\bibfnamefont {T.~A.}\ \bibnamefont
  {Wesolowski}}, \bibinfo {author} {\bibfnamefont {S.}~\bibnamefont {Shedge}},
  \ and\ \bibinfo {author} {\bibfnamefont {X.}~\bibnamefont {Zhou}},\
  }\href@noop {} {\bibfield  {journal} {\bibinfo  {journal} {Chem. Rev.}\
  }\textbf {\bibinfo {volume} {115}},\ \bibinfo {pages} {5891} (\bibinfo {year}
  {2015})}\BibitemShut {NoStop}%
\bibitem [{\citenamefont {Sun}\ and\ \citenamefont {Chan}(2016)}]{sun2016}%
  \BibitemOpen
  \bibfield  {author} {\bibinfo {author} {\bibfnamefont {Q.}~\bibnamefont
  {Sun}}\ and\ \bibinfo {author} {\bibfnamefont {G.~K.-L.}\ \bibnamefont
  {Chan}},\ }\href {\doibase 10.1021/acs.accounts.6b00356} {\bibfield
  {journal} {\bibinfo  {journal} {Acc. Chem. Res.}\ }\textbf {\bibinfo {volume}
  {49}},\ \bibinfo {pages} {2705} (\bibinfo {year} {2016})},\ \Eprint
  {http://arxiv.org/abs/https://doi.org/10.1021/acs.accounts.6b00356}
  {https://doi.org/10.1021/acs.accounts.6b00356} \BibitemShut {NoStop}%
\bibitem [{\citenamefont {Goez}\ and\ \citenamefont
  {Neugebauer}(2018)}]{Goez2018}%
  \BibitemOpen
  \bibfield  {author} {\bibinfo {author} {\bibfnamefont {A.}~\bibnamefont
  {Goez}}\ and\ \bibinfo {author} {\bibfnamefont {J.}~\bibnamefont
  {Neugebauer}},\ }in\ \href {\doibase 10.1007/978-981-10-5651-2_7} {\emph
  {\bibinfo {booktitle} {Frontiers of Quantum Chemistry}}},\ \bibinfo {editor}
  {edited by\ \bibinfo {editor} {\bibfnamefont {M.~J.}\ \bibnamefont
  {W{\'o}jcik}}, \bibinfo {editor} {\bibfnamefont {H.}~\bibnamefont
  {Nakatsuji}}, \bibinfo {editor} {\bibfnamefont {B.}~\bibnamefont {Kirtman}},
  \ and\ \bibinfo {editor} {\bibfnamefont {Y.}~\bibnamefont {Ozaki}}}\
  (\bibinfo  {publisher} {Springer Singapore},\ \bibinfo {address}
  {Singapore},\ \bibinfo {year} {2018})\ pp.\ \bibinfo {pages}
  {139--179}\BibitemShut {NoStop}%
\bibitem [{\citenamefont {Krishtal}, \citenamefont {Ceresoli},\ and\
  \citenamefont {Pavanello}(2015)}]{env-Krishtal-JCP2015-142-154116}%
  \BibitemOpen
  \bibfield  {author} {\bibinfo {author} {\bibfnamefont {A.}~\bibnamefont
  {Krishtal}}, \bibinfo {author} {\bibfnamefont {D.}~\bibnamefont {Ceresoli}},
  \ and\ \bibinfo {author} {\bibfnamefont {M.}~\bibnamefont {Pavanello}},\
  }\href {\doibase 10.1063/1.4918276} {\bibfield  {journal} {\bibinfo
  {journal} {J. Chem. Phys.}\ }\textbf {\bibinfo {volume} {142}},\ \bibinfo
  {pages} {154116} (\bibinfo {year} {2015})},\ \Eprint
  {http://arxiv.org/abs/https://doi.org/10.1063/1.4918276}
  {https://doi.org/10.1063/1.4918276} \BibitemShut {NoStop}%
\bibitem [{\citenamefont {Krishtal}\ and\ \citenamefont
  {Pavanello}(2016)}]{Krishtal2016}%
  \BibitemOpen
  \bibfield  {author} {\bibinfo {author} {\bibfnamefont {A.}~\bibnamefont
  {Krishtal}}\ and\ \bibinfo {author} {\bibfnamefont {M.}~\bibnamefont
  {Pavanello}},\ }\href {\doibase 10.1063/1.4944526} {\bibfield  {journal}
  {\bibinfo  {journal} {J. Chem. Phys.}\ }\textbf {\bibinfo {volume} {144}},\
  \bibinfo {pages} {124118} (\bibinfo {year} {2016})}\BibitemShut {NoStop}%
\bibitem [{\citenamefont {P.}, \citenamefont {Genova},\ and\ \citenamefont
  {Pavanello}(2017)}]{P2017}%
  \BibitemOpen
  \bibfield  {author} {\bibinfo {author} {\bibfnamefont {S.~K.}\ \bibnamefont
  {P.}}, \bibinfo {author} {\bibfnamefont {A.}~\bibnamefont {Genova}}, \ and\
  \bibinfo {author} {\bibfnamefont {M.}~\bibnamefont {Pavanello}},\ }\href
  {\doibase 10.1021/acs.jpclett.7b02212} {\bibfield  {journal} {\bibinfo
  {journal} {J. Phys. Chem. Lett.}\ }\textbf {\bibinfo {volume} {8}},\ \bibinfo
  {pages} {5077} (\bibinfo {year} {2017})}\BibitemShut {NoStop}%
\bibitem [{\citenamefont {Genova}\ \emph {et~al.}(2017)\citenamefont {Genova},
  \citenamefont {Ceresoli}, \citenamefont {Krishtal}, \citenamefont
  {Andreussi}, \citenamefont {DiStasio},\ and\ \citenamefont
  {Pavanello}}]{Genova2017}%
  \BibitemOpen
  \bibfield  {author} {\bibinfo {author} {\bibfnamefont {A.}~\bibnamefont
  {Genova}}, \bibinfo {author} {\bibfnamefont {D.}~\bibnamefont {Ceresoli}},
  \bibinfo {author} {\bibfnamefont {A.}~\bibnamefont {Krishtal}}, \bibinfo
  {author} {\bibfnamefont {O.}~\bibnamefont {Andreussi}}, \bibinfo {author}
  {\bibfnamefont {R.~A.}\ \bibnamefont {DiStasio}}, \ and\ \bibinfo {author}
  {\bibfnamefont {M.}~\bibnamefont {Pavanello}},\ }\href {\doibase
  10.1002/qua.25401} {\bibfield  {journal} {\bibinfo  {journal} {Int. J.
  Quantum Chem.}\ }\textbf {\bibinfo {volume} {117}} (\bibinfo {year} {2017}),\
  10.1002/qua.25401}\BibitemShut {NoStop}%
\bibitem [{\citenamefont {Neugebauer}(2007)}]{neugebauer2007}%
  \BibitemOpen
  \bibfield  {author} {\bibinfo {author} {\bibfnamefont {J.}~\bibnamefont
  {Neugebauer}},\ }\href {\doibase 10.1063/1.2713754} {\bibfield  {journal}
  {\bibinfo  {journal} {J. Chem. Phys.}\ }\textbf {\bibinfo {volume} {126}},\
  \bibinfo {pages} {134116} (\bibinfo {year} {2007})},\ \Eprint
  {http://arxiv.org/abs/https://doi.org/10.1063/1.2713754}
  {https://doi.org/10.1063/1.2713754} \BibitemShut {NoStop}%
\bibitem [{\citenamefont {Neugebauer}(2009{\natexlab{a}})}]{neugebauer2009}%
  \BibitemOpen
  \bibfield  {author} {\bibinfo {author} {\bibfnamefont {J.}~\bibnamefont
  {Neugebauer}},\ }\href {\doibase 10.1063/1.3212883} {\bibfield  {journal}
  {\bibinfo  {journal} {J. Chem. Phys.}\ }\textbf {\bibinfo {volume} {131}},\
  \bibinfo {pages} {084104} (\bibinfo {year} {2009}{\natexlab{a}})},\ \Eprint
  {http://arxiv.org/abs/https://doi.org/10.1063/1.3212883}
  {https://doi.org/10.1063/1.3212883} \BibitemShut {NoStop}%
\bibitem [{\citenamefont {Neugebauer}(2009{\natexlab{b}})}]{neugebauer2009b}%
  \BibitemOpen
  \bibfield  {author} {\bibinfo {author} {\bibfnamefont {J.}~\bibnamefont
  {Neugebauer}},\ }\href {\doibase 10.1002/cphc.200900538} {\bibfield
  {journal} {\bibinfo  {journal} {ChemPhysChem}\ }\textbf {\bibinfo {volume}
  {10}},\ \bibinfo {pages} {3148} (\bibinfo {year} {2009}{\natexlab{b}})},\
  \Eprint
  {http://arxiv.org/abs/https://chemistry-europe.onlinelibrary.wiley.com/doi/pdf/10.1002/cphc.200900538}
  {https://chemistry-europe.onlinelibrary.wiley.com/doi/pdf/10.1002/cphc.200900538}
  \BibitemShut {NoStop}%
\bibitem [{\citenamefont {H\"ofener}, \citenamefont {Severo Pereira~Gomes},\
  and\ \citenamefont {Visscher}(2012)}]{visscher12}%
  \BibitemOpen
  \bibfield  {author} {\bibinfo {author} {\bibfnamefont {S.}~\bibnamefont
  {H\"ofener}}, \bibinfo {author} {\bibfnamefont {A.}~\bibnamefont {Severo
  Pereira~Gomes}}, \ and\ \bibinfo {author} {\bibfnamefont {L.}~\bibnamefont
  {Visscher}},\ }\href {\doibase 10.1063/1.3675845} {\bibfield  {journal}
  {\bibinfo  {journal} {J. Comp. Phys.}\ }\textbf {\bibinfo {volume} {136}},\
  \bibinfo {pages} {044104} (\bibinfo {year} {2012})},\ \Eprint
  {http://arxiv.org/abs/https://doi.org/10.1063/1.3675845}
  {https://doi.org/10.1063/1.3675845} \BibitemShut {NoStop}%
\bibitem [{\citenamefont {Pavanello}(2013)}]{Pavanello2013}%
  \BibitemOpen
  \bibfield  {author} {\bibinfo {author} {\bibfnamefont {M.}~\bibnamefont
  {Pavanello}},\ }\href {\doibase 10.1063/1.4807059} {\bibfield  {journal}
  {\bibinfo  {journal} {J. Chem. Phys.}\ }\textbf {\bibinfo {volume} {138}},\
  \bibinfo {pages} {204118} (\bibinfo {year} {2013})},\ \Eprint
  {http://arxiv.org/abs/https://doi.org/10.1063/1.4807059}
  {https://doi.org/10.1063/1.4807059} \BibitemShut {NoStop}%
\bibitem [{\citenamefont {König}\ and\ \citenamefont
  {Neugebauer}(2013)}]{konig2013}%
  \BibitemOpen
  \bibfield  {author} {\bibinfo {author} {\bibfnamefont {C.}~\bibnamefont
  {König}}\ and\ \bibinfo {author} {\bibfnamefont {J.}~\bibnamefont
  {Neugebauer}},\ }\href {\doibase 10.1021/jp3105419} {\bibfield  {journal}
  {\bibinfo  {journal} {J. Phys. Chem. B}\ }\textbf {\bibinfo {volume} {117}},\
  \bibinfo {pages} {3480} (\bibinfo {year} {2013})},\ \bibinfo {note} {pMID:
  23528045},\ \Eprint {http://arxiv.org/abs/https://doi.org/10.1021/jp3105419}
  {https://doi.org/10.1021/jp3105419} \BibitemShut {NoStop}%
\bibitem [{\citenamefont {Gomes}, \citenamefont {Jacob},\ and\ \citenamefont
  {Visscher}(2008)}]{actinide-Gomes-PCCP2008-10-5353}%
  \BibitemOpen
  \bibfield  {author} {\bibinfo {author} {\bibfnamefont {A.~S.~P.}\
  \bibnamefont {Gomes}}, \bibinfo {author} {\bibfnamefont {C.~R.}\ \bibnamefont
  {Jacob}}, \ and\ \bibinfo {author} {\bibfnamefont {L.}~\bibnamefont
  {Visscher}},\ }\href {\doibase 10.1039/b805739g} {\bibfield  {journal}
  {\bibinfo  {journal} {Phys. Chem. Chem. Phys.}\ }\textbf {\bibinfo {volume}
  {10}},\ \bibinfo {pages} {5353} (\bibinfo {year} {2008})}\BibitemShut
  {NoStop}%
\bibitem [{\citenamefont {Gomes}\ \emph {et~al.}(2013)\citenamefont {Gomes},
  \citenamefont {Jacob}, \citenamefont {R{\'e}al}, \citenamefont {Visscher},\
  and\ \citenamefont {Vallet}}]{actinide-Gomes-PCCP2013-15-15153}%
  \BibitemOpen
  \bibfield  {author} {\bibinfo {author} {\bibfnamefont {A.~S.~P.}\
  \bibnamefont {Gomes}}, \bibinfo {author} {\bibfnamefont {C.~R.}\ \bibnamefont
  {Jacob}}, \bibinfo {author} {\bibfnamefont {F.}~\bibnamefont {R{\'e}al}},
  \bibinfo {author} {\bibfnamefont {L.}~\bibnamefont {Visscher}}, \ and\
  \bibinfo {author} {\bibfnamefont {V.}~\bibnamefont {Vallet}},\ }\href
  {\doibase 10.1039/C3CP52090K} {\bibfield  {journal} {\bibinfo  {journal}
  {Phys. Chem. Chem. Phys.}\ }\textbf {\bibinfo {volume} {15}},\ \bibinfo
  {pages} {15153} (\bibinfo {year} {2013})}\BibitemShut {NoStop}%
\bibitem [{\citenamefont {Olejniczak}, \citenamefont {Bast},\ and\
  \citenamefont {Pereira~Gomes}(2017)}]{olejniczak2017}%
  \BibitemOpen
  \bibfield  {author} {\bibinfo {author} {\bibfnamefont {M.}~\bibnamefont
  {Olejniczak}}, \bibinfo {author} {\bibfnamefont {R.}~\bibnamefont {Bast}}, \
  and\ \bibinfo {author} {\bibfnamefont {A.~S.}\ \bibnamefont
  {Pereira~Gomes}},\ }\href {\doibase 10.1039/C6CP08561J} {\bibfield  {journal}
  {\bibinfo  {journal} {Phys. Chem. Chem. Phys.}\ }\textbf {\bibinfo {volume}
  {19}},\ \bibinfo {pages} {8400} (\bibinfo {year} {2017})}\BibitemShut
  {NoStop}%
\bibitem [{\citenamefont {De~Santis}\ \emph
  {et~al.}(2020{\natexlab{a}})\citenamefont {De~Santis}, \citenamefont
  {Belpassi}, \citenamefont {Jacob}, \citenamefont {Severo Pereira~Gomes},
  \citenamefont {Tarantelli}, \citenamefont {Visscher},\ and\ \citenamefont
  {Storchi}}]{rt_fde2020}%
  \BibitemOpen
  \bibfield  {author} {\bibinfo {author} {\bibfnamefont {M.}~\bibnamefont
  {De~Santis}}, \bibinfo {author} {\bibfnamefont {L.}~\bibnamefont {Belpassi}},
  \bibinfo {author} {\bibfnamefont {C.~R.}\ \bibnamefont {Jacob}}, \bibinfo
  {author} {\bibfnamefont {A.}~\bibnamefont {Severo Pereira~Gomes}}, \bibinfo
  {author} {\bibfnamefont {F.}~\bibnamefont {Tarantelli}}, \bibinfo {author}
  {\bibfnamefont {L.}~\bibnamefont {Visscher}}, \ and\ \bibinfo {author}
  {\bibfnamefont {L.}~\bibnamefont {Storchi}},\ }\href {\doibase
  10.1021/acs.jctc.0c00603} {\bibfield  {journal} {\bibinfo  {journal} {J.
  Chem. Theory Comput.}\ }\textbf {\bibinfo {volume} {16}},\ \bibinfo {pages}
  {5695} (\bibinfo {year} {2020}{\natexlab{a}})},\ \Eprint
  {http://arxiv.org/abs/https://doi.org/10.1021/acs.jctc.0c00603}
  {https://doi.org/10.1021/acs.jctc.0c00603} \BibitemShut {NoStop}%
\bibitem [{\citenamefont {Beyhan}\ \emph {et~al.}(2010)\citenamefont {Beyhan},
  \citenamefont {G\"{o}tz}, \citenamefont {Jacob},\ and\ \citenamefont
  {Visscher}}]{Beyhan2010}%
  \BibitemOpen
  \bibfield  {author} {\bibinfo {author} {\bibfnamefont {S.~M.}\ \bibnamefont
  {Beyhan}}, \bibinfo {author} {\bibfnamefont {A.~W.}\ \bibnamefont
  {G\"{o}tz}}, \bibinfo {author} {\bibfnamefont {C.~R.}\ \bibnamefont {Jacob}},
  \ and\ \bibinfo {author} {\bibfnamefont {L.}~\bibnamefont {Visscher}},\
  }\href {\doibase 10.1063/1.3297886} {\bibfield  {journal} {\bibinfo
  {journal} {J. Chem. Phys.}\ }\textbf {\bibinfo {volume} {132}},\ \bibinfo
  {pages} {044114} (\bibinfo {year} {2010})}\BibitemShut {NoStop}%
\bibitem [{\citenamefont {Grimmel}, \citenamefont {Teodoro},\ and\
  \citenamefont {Visscher}(2019)}]{Grimmel2019}%
  \BibitemOpen
  \bibfield  {author} {\bibinfo {author} {\bibfnamefont {S.~A.}\ \bibnamefont
  {Grimmel}}, \bibinfo {author} {\bibfnamefont {T.~Q.}\ \bibnamefont
  {Teodoro}}, \ and\ \bibinfo {author} {\bibfnamefont {L.}~\bibnamefont
  {Visscher}},\ }\href {\doibase 10.1002/qua.26111} {\bibfield  {journal}
  {\bibinfo  {journal} {Int. J. Quantum Chem.}\ }\textbf {\bibinfo {volume}
  {120}} (\bibinfo {year} {2019}),\ 10.1002/qua.26111}\BibitemShut {NoStop}%
\bibitem [{\citenamefont {Fux}\ \emph {et~al.}(2010)\citenamefont {Fux},
  \citenamefont {Jacob}, \citenamefont {Neugebauer}, \citenamefont {Visscher},\
  and\ \citenamefont {Reiher}}]{fux_exemb}%
  \BibitemOpen
  \bibfield  {author} {\bibinfo {author} {\bibfnamefont {S.}~\bibnamefont
  {Fux}}, \bibinfo {author} {\bibfnamefont {C.~R.}\ \bibnamefont {Jacob}},
  \bibinfo {author} {\bibfnamefont {J.}~\bibnamefont {Neugebauer}}, \bibinfo
  {author} {\bibfnamefont {L.}~\bibnamefont {Visscher}}, \ and\ \bibinfo
  {author} {\bibfnamefont {M.}~\bibnamefont {Reiher}},\ }\href {\doibase
  10.1063/1.3376251} {\bibfield  {journal} {\bibinfo  {journal} {J. Comp.
  Phys.}\ }\textbf {\bibinfo {volume} {132}},\ \bibinfo {pages} {164101}
  (\bibinfo {year} {2010})},\ \Eprint
  {http://arxiv.org/abs/https://doi.org/10.1063/1.3376251}
  {https://doi.org/10.1063/1.3376251} \BibitemShut {NoStop}%
\bibitem [{\citenamefont {Bouchafra}\ \emph {et~al.}(2018)\citenamefont
  {Bouchafra}, \citenamefont {Shee}, \citenamefont {R{\'e}al}, \citenamefont
  {Vallet},\ and\ \citenamefont
  {Gomes}}]{halides-water-Bouchafra-PRL2018-121-266001}%
  \BibitemOpen
  \bibfield  {author} {\bibinfo {author} {\bibfnamefont {Y.}~\bibnamefont
  {Bouchafra}}, \bibinfo {author} {\bibfnamefont {A.}~\bibnamefont {Shee}},
  \bibinfo {author} {\bibfnamefont {F.}~\bibnamefont {R{\'e}al}}, \bibinfo
  {author} {\bibfnamefont {V.}~\bibnamefont {Vallet}}, \ and\ \bibinfo {author}
  {\bibfnamefont {A.~S.~P.}\ \bibnamefont {Gomes}},\ }\href {\doibase
  10.1103/PhysRevLett.121.266001} {\bibfield  {journal} {\bibinfo  {journal}
  {Phys. Rev. Lett.}\ }\textbf {\bibinfo {volume} {121}},\ \bibinfo {pages}
  {266001} (\bibinfo {year} {2018})}\BibitemShut {NoStop}%
\bibitem [{\citenamefont {T\"olle}, \citenamefont {B\"ockers},\ and\
  \citenamefont {Neugebauer}(2019)}]{toelle_1}%
  \BibitemOpen
  \bibfield  {author} {\bibinfo {author} {\bibfnamefont {J.}~\bibnamefont
  {T\"olle}}, \bibinfo {author} {\bibfnamefont {M.}~\bibnamefont {B\"ockers}},
  \ and\ \bibinfo {author} {\bibfnamefont {J.}~\bibnamefont {Neugebauer}},\
  }\href {\doibase 10.1063/1.5097124} {\bibfield  {journal} {\bibinfo
  {journal} {J. Comp. Phys.}\ }\textbf {\bibinfo {volume} {150}},\ \bibinfo
  {pages} {181101} (\bibinfo {year} {2019})},\ \Eprint
  {http://arxiv.org/abs/https://doi.org/10.1063/1.5097124}
  {https://doi.org/10.1063/1.5097124} \BibitemShut {NoStop}%
\bibitem [{\citenamefont {T\"olle}\ \emph {et~al.}(2019)\citenamefont
  {T\"olle}, \citenamefont {B\"ockers}, \citenamefont {Niemeyer},\ and\
  \citenamefont {Neugebauer}}]{toelle_2}%
  \BibitemOpen
  \bibfield  {author} {\bibinfo {author} {\bibfnamefont {J.}~\bibnamefont
  {T\"olle}}, \bibinfo {author} {\bibfnamefont {M.}~\bibnamefont {B\"ockers}},
  \bibinfo {author} {\bibfnamefont {N.}~\bibnamefont {Niemeyer}}, \ and\
  \bibinfo {author} {\bibfnamefont {J.}~\bibnamefont {Neugebauer}},\ }\href
  {\doibase 10.1063/1.5121908} {\bibfield  {journal} {\bibinfo  {journal} {J.
  Comp. Phys.}\ }\textbf {\bibinfo {volume} {151}},\ \bibinfo {pages} {174109}
  (\bibinfo {year} {2019})},\ \Eprint
  {http://arxiv.org/abs/https://doi.org/10.1063/1.5121908}
  {https://doi.org/10.1063/1.5121908} \BibitemShut {NoStop}%
\bibitem [{\citenamefont {Scholz}, \citenamefont {Tölle},\ and\ \citenamefont
  {Neugebauer}(2020)}]{Scholz2020}%
  \BibitemOpen
  \bibfield  {author} {\bibinfo {author} {\bibfnamefont {L.}~\bibnamefont
  {Scholz}}, \bibinfo {author} {\bibfnamefont {J.}~\bibnamefont {Tölle}}, \
  and\ \bibinfo {author} {\bibfnamefont {J.}~\bibnamefont {Neugebauer}},\
  }\href {\doibase https://doi.org/10.1002/qua.26213} {\bibfield  {journal}
  {\bibinfo  {journal} {Int. J. Quantum Chem.}\ }\textbf {\bibinfo {volume}
  {120}},\ \bibinfo {pages} {e26213} (\bibinfo {year} {2020})},\ \Eprint
  {http://arxiv.org/abs/https://onlinelibrary.wiley.com/doi/pdf/10.1002/qua.26213}
  {https://onlinelibrary.wiley.com/doi/pdf/10.1002/qua.26213} \BibitemShut
  {NoStop}%
\bibitem [{\citenamefont {Niemeyer}, \citenamefont {Tölle},\ and\
  \citenamefont {Neugebauer}(2020)}]{Niemeyer2020}%
  \BibitemOpen
  \bibfield  {author} {\bibinfo {author} {\bibfnamefont {N.}~\bibnamefont
  {Niemeyer}}, \bibinfo {author} {\bibfnamefont {J.}~\bibnamefont {Tölle}}, \
  and\ \bibinfo {author} {\bibfnamefont {J.}~\bibnamefont {Neugebauer}},\
  }\href {\doibase 10.1021/acs.jctc.0c00125} {\bibfield  {journal} {\bibinfo
  {journal} {J. Chem. Theory Comput.}\ }\textbf {\bibinfo {volume} {16}},\
  \bibinfo {pages} {3104} (\bibinfo {year} {2020})},\ \Eprint
  {http://arxiv.org/abs/https://doi.org/10.1021/acs.jctc.0c00125}
  {https://doi.org/10.1021/acs.jctc.0c00125} \BibitemShut {NoStop}%
\bibitem [{\citenamefont {Manby}\ \emph {et~al.}(2012)\citenamefont {Manby},
  \citenamefont {Stella}, \citenamefont {Goodpaster},\ and\ \citenamefont
  {Miller}}]{manby2012}%
  \BibitemOpen
  \bibfield  {author} {\bibinfo {author} {\bibfnamefont {F.~R.}\ \bibnamefont
  {Manby}}, \bibinfo {author} {\bibfnamefont {M.}~\bibnamefont {Stella}},
  \bibinfo {author} {\bibfnamefont {J.~D.}\ \bibnamefont {Goodpaster}}, \ and\
  \bibinfo {author} {\bibfnamefont {T.~F.}\ \bibnamefont {Miller}},\ }\href
  {\doibase 10.1021/ct300544e} {\bibfield  {journal} {\bibinfo  {journal} {J.
  Chem. Theory Comput.}\ }\textbf {\bibinfo {volume} {8}},\ \bibinfo {pages}
  {2564} (\bibinfo {year} {2012})},\ \Eprint
  {http://arxiv.org/abs/https://doi.org/10.1021/ct300544e}
  {https://doi.org/10.1021/ct300544e} \BibitemShut {NoStop}%
\bibitem [{\citenamefont {Goodpaster}\ \emph {et~al.}(2010)\citenamefont
  {Goodpaster}, \citenamefont {Ananth}, \citenamefont {Manby},\ and\
  \citenamefont {Miller}}]{goodpaster_exemb}%
  \BibitemOpen
  \bibfield  {author} {\bibinfo {author} {\bibfnamefont {J.~D.}\ \bibnamefont
  {Goodpaster}}, \bibinfo {author} {\bibfnamefont {N.}~\bibnamefont {Ananth}},
  \bibinfo {author} {\bibfnamefont {F.~R.}\ \bibnamefont {Manby}}, \ and\
  \bibinfo {author} {\bibfnamefont {T.~F.}\ \bibnamefont {Miller}},\ }\href
  {\doibase 10.1063/1.3474575} {\bibfield  {journal} {\bibinfo  {journal} {J.
  Comp. Phys.}\ }\textbf {\bibinfo {volume} {133}},\ \bibinfo {pages} {084103}
  (\bibinfo {year} {2010})},\ \Eprint
  {http://arxiv.org/abs/https://doi.org/10.1063/1.3474575}
  {https://doi.org/10.1063/1.3474575} \BibitemShut {NoStop}%
\bibitem [{\citenamefont {Goodpaster}, \citenamefont {Barnes},\ and\
  \citenamefont {Miller}(2011)}]{goodpaster_exemb2}%
  \BibitemOpen
  \bibfield  {author} {\bibinfo {author} {\bibfnamefont {J.~D.}\ \bibnamefont
  {Goodpaster}}, \bibinfo {author} {\bibfnamefont {T.~A.}\ \bibnamefont
  {Barnes}}, \ and\ \bibinfo {author} {\bibfnamefont {T.~F.}\ \bibnamefont
  {Miller}},\ }\href {\doibase 10.1063/1.3582913} {\bibfield  {journal}
  {\bibinfo  {journal} {J. Comp. Phys.}\ }\textbf {\bibinfo {volume} {134}},\
  \bibinfo {pages} {164108} (\bibinfo {year} {2011})},\ \Eprint
  {http://arxiv.org/abs/https://doi.org/10.1063/1.3582913}
  {https://doi.org/10.1063/1.3582913} \BibitemShut {NoStop}%
\bibitem [{\citenamefont {Ding}, \citenamefont {Manby},\ and\ \citenamefont
  {Miller}(2017)}]{ding2017}%
  \BibitemOpen
  \bibfield  {author} {\bibinfo {author} {\bibfnamefont {F.}~\bibnamefont
  {Ding}}, \bibinfo {author} {\bibfnamefont {F.~R.}\ \bibnamefont {Manby}}, \
  and\ \bibinfo {author} {\bibfnamefont {T.~F.}\ \bibnamefont {Miller}},\
  }\href {\doibase 10.1021/acs.jctc.6b01065} {\bibfield  {journal} {\bibinfo
  {journal} {J. Chem. Theory Comput.}\ }\textbf {\bibinfo {volume} {13}},\
  \bibinfo {pages} {1605} (\bibinfo {year} {2017})},\ \Eprint
  {http://arxiv.org/abs/https://doi.org/10.1021/acs.jctc.6b01065}
  {https://doi.org/10.1021/acs.jctc.6b01065} \BibitemShut {NoStop}%
\bibitem [{\citenamefont {Lee}\ \emph {et~al.}(2019)\citenamefont {Lee},
  \citenamefont {Welborn}, \citenamefont {Manby},\ and\ \citenamefont
  {Miller}}]{Lee2019}%
  \BibitemOpen
  \bibfield  {author} {\bibinfo {author} {\bibfnamefont {S.~J.~R.}\
  \bibnamefont {Lee}}, \bibinfo {author} {\bibfnamefont {M.}~\bibnamefont
  {Welborn}}, \bibinfo {author} {\bibfnamefont {F.~R.}\ \bibnamefont {Manby}},
  \ and\ \bibinfo {author} {\bibfnamefont {T.~F.}\ \bibnamefont {Miller}},\
  }\href {\doibase 10.1021/acs.accounts.8b00672} {\bibfield  {journal}
  {\bibinfo  {journal} {Acc. Chem. Res.}\ }\textbf {\bibinfo {volume} {52}},\
  \bibinfo {pages} {1359} (\bibinfo {year} {2019})}\BibitemShut {NoStop}%
\bibitem [{\citenamefont {Graham}\ \emph {et~al.}(2020)\citenamefont {Graham},
  \citenamefont {Wen}, \citenamefont {Chulhai},\ and\ \citenamefont
  {Goodpaster}}]{Graham2020}%
  \BibitemOpen
  \bibfield  {author} {\bibinfo {author} {\bibfnamefont {D.~S.}\ \bibnamefont
  {Graham}}, \bibinfo {author} {\bibfnamefont {X.}~\bibnamefont {Wen}},
  \bibinfo {author} {\bibfnamefont {D.~V.}\ \bibnamefont {Chulhai}}, \ and\
  \bibinfo {author} {\bibfnamefont {J.~D.}\ \bibnamefont {Goodpaster}},\ }\href
  {\doibase 10.1021/acs.jctc.9b01185} {\bibfield  {journal} {\bibinfo
  {journal} {J. Chem. Theory Comput.}\ }\textbf {\bibinfo {volume} {16}},\
  \bibinfo {pages} {2284} (\bibinfo {year} {2020})},\ \Eprint
  {http://arxiv.org/abs/https://doi.org/10.1021/acs.jctc.9b01185}
  {https://doi.org/10.1021/acs.jctc.9b01185} \BibitemShut {NoStop}%
\bibitem [{\citenamefont {Koh}, \citenamefont {Nguyen-Beck},\ and\
  \citenamefont {Parkhill}(2017)}]{parkhill_2017}%
  \BibitemOpen
  \bibfield  {author} {\bibinfo {author} {\bibfnamefont {K.~J.}\ \bibnamefont
  {Koh}}, \bibinfo {author} {\bibfnamefont {T.~S.}\ \bibnamefont
  {Nguyen-Beck}}, \ and\ \bibinfo {author} {\bibfnamefont {J.}~\bibnamefont
  {Parkhill}},\ }\href {\doibase 10.1021/acs.jctc.7b00494} {\bibfield
  {journal} {\bibinfo  {journal} {J. Chem. Theory Comput.}\ }\textbf {\bibinfo
  {volume} {13}},\ \bibinfo {pages} {4173} (\bibinfo {year} {2017})},\ \Eprint
  {http://arxiv.org/abs/https://doi.org/10.1021/acs.jctc.7b00494}
  {https://doi.org/10.1021/acs.jctc.7b00494} \BibitemShut {NoStop}%
\bibitem [{\citenamefont {Finlayson-Pitts}(2013)}]{FinlaysonPitts2013}%
  \BibitemOpen
  \bibfield  {author} {\bibinfo {author} {\bibfnamefont {B.~J.}\ \bibnamefont
  {Finlayson-Pitts}},\ }\href {\doibase 10.1038/nchem.1717} {\bibfield
  {journal} {\bibinfo  {journal} {Nat. Chem.}\ }\textbf {\bibinfo {volume}
  {5}},\ \bibinfo {pages} {724} (\bibinfo {year} {2013})}\BibitemShut {NoStop}%
\bibitem [{\citenamefont {Pillar-Little}, \citenamefont {Guzman},\ and\
  \citenamefont {Rodriguez}(2013)}]{PillarLittle2013}%
  \BibitemOpen
  \bibfield  {author} {\bibinfo {author} {\bibfnamefont {E.~A.}\ \bibnamefont
  {Pillar-Little}}, \bibinfo {author} {\bibfnamefont {M.~I.}\ \bibnamefont
  {Guzman}}, \ and\ \bibinfo {author} {\bibfnamefont {J.~M.}\ \bibnamefont
  {Rodriguez}},\ }\href {\doibase 10.1021/es401700h} {\bibfield  {journal}
  {\bibinfo  {journal} {Environ. Sci. Technol.}\ }\textbf {\bibinfo {volume}
  {47}},\ \bibinfo {pages} {10971} (\bibinfo {year} {2013})}\BibitemShut
  {NoStop}%
\bibitem [{\citenamefont {Simpson}\ \emph {et~al.}(2015)\citenamefont
  {Simpson}, \citenamefont {Brown}, \citenamefont {Saiz-Lopez}, \citenamefont
  {Thornton},\ and\ \citenamefont {von Glasow}}]{Simpson2015}%
  \BibitemOpen
  \bibfield  {author} {\bibinfo {author} {\bibfnamefont {W.~R.}\ \bibnamefont
  {Simpson}}, \bibinfo {author} {\bibfnamefont {S.~S.}\ \bibnamefont {Brown}},
  \bibinfo {author} {\bibfnamefont {A.}~\bibnamefont {Saiz-Lopez}}, \bibinfo
  {author} {\bibfnamefont {J.~A.}\ \bibnamefont {Thornton}}, \ and\ \bibinfo
  {author} {\bibfnamefont {R.}~\bibnamefont {von Glasow}},\ }\href {\doibase
  10.1021/cr5006638} {\bibfield  {journal} {\bibinfo  {journal} {Chem. Rev.}\
  }\textbf {\bibinfo {volume} {115}},\ \bibinfo {pages} {4035} (\bibinfo {year}
  {2015})}\BibitemShut {NoStop}%
\bibitem [{\citenamefont {Kong}\ \emph {et~al.}(2017)\citenamefont {Kong},
  \citenamefont {Waldner}, \citenamefont {Orlando}, \citenamefont {Artiglia},
  \citenamefont {Huthwelker}, \citenamefont {Ammann},\ and\ \citenamefont
  {Bartels-Rausch}}]{Kong2017}%
  \BibitemOpen
  \bibfield  {author} {\bibinfo {author} {\bibfnamefont {X.}~\bibnamefont
  {Kong}}, \bibinfo {author} {\bibfnamefont {A.}~\bibnamefont {Waldner}},
  \bibinfo {author} {\bibfnamefont {F.}~\bibnamefont {Orlando}}, \bibinfo
  {author} {\bibfnamefont {L.}~\bibnamefont {Artiglia}}, \bibinfo {author}
  {\bibfnamefont {T.}~\bibnamefont {Huthwelker}}, \bibinfo {author}
  {\bibfnamefont {M.}~\bibnamefont {Ammann}}, \ and\ \bibinfo {author}
  {\bibfnamefont {T.}~\bibnamefont {Bartels-Rausch}},\ }\href {\doibase
  10.1021/acs.jpclett.7b01573} {\bibfield  {journal} {\bibinfo  {journal} {J.
  Phys. Chem. Lett.}\ }\textbf {\bibinfo {volume} {8}},\ \bibinfo {pages}
  {4757} (\bibinfo {year} {2017})}\BibitemShut {NoStop}%
\bibitem [{\citenamefont {Finlayson-Pitts}(2019)}]{FinlaysonPitts2019}%
  \BibitemOpen
  \bibfield  {author} {\bibinfo {author} {\bibfnamefont {B.~J.}\ \bibnamefont
  {Finlayson-Pitts}},\ }\href {\doibase 10.1002/kin.21305} {\bibfield
  {journal} {\bibinfo  {journal} {Int. J. Chem. Kinet.}\ }\textbf {\bibinfo
  {volume} {51}},\ \bibinfo {pages} {736} (\bibinfo {year} {2019})}\BibitemShut
  {NoStop}%
\bibitem [{\citenamefont {Yu}\ and\ \citenamefont {Li}(2021)}]{Yu2021}%
  \BibitemOpen
  \bibfield  {author} {\bibinfo {author} {\bibfnamefont {Z.}~\bibnamefont
  {Yu}}\ and\ \bibinfo {author} {\bibfnamefont {Y.}~\bibnamefont {Li}},\ }\href
  {\doibase 10.1016/j.scitotenv.2021.145054} {\bibfield  {journal} {\bibinfo
  {journal} {Sci. Total Environ.}\ }\textbf {\bibinfo {volume} {768}},\
  \bibinfo {pages} {145054} (\bibinfo {year} {2021})}\BibitemShut {NoStop}%
\bibitem [{\citenamefont {Bartels-Rausch}\ \emph {et~al.}(2021)\citenamefont
  {Bartels-Rausch}, \citenamefont {Kong}, \citenamefont {Orlando},
  \citenamefont {Artiglia}, \citenamefont {Waldner}, \citenamefont
  {Huthwelker},\ and\ \citenamefont {Ammann}}]{BartelsRausch2021}%
  \BibitemOpen
  \bibfield  {author} {\bibinfo {author} {\bibfnamefont {T.}~\bibnamefont
  {Bartels-Rausch}}, \bibinfo {author} {\bibfnamefont {X.}~\bibnamefont
  {Kong}}, \bibinfo {author} {\bibfnamefont {F.}~\bibnamefont {Orlando}},
  \bibinfo {author} {\bibfnamefont {L.}~\bibnamefont {Artiglia}}, \bibinfo
  {author} {\bibfnamefont {A.}~\bibnamefont {Waldner}}, \bibinfo {author}
  {\bibfnamefont {T.}~\bibnamefont {Huthwelker}}, \ and\ \bibinfo {author}
  {\bibfnamefont {M.}~\bibnamefont {Ammann}},\ }\href {\doibase
  10.5194/tc-15-2001-2021} {\bibfield  {journal} {\bibinfo  {journal} {The
  Cryosphere}\ }\textbf {\bibinfo {volume} {15}},\ \bibinfo {pages} {2001}
  (\bibinfo {year} {2021})}\BibitemShut {NoStop}%
\bibitem [{\citenamefont {Wesolowski}\ and\ \citenamefont
  {Warshel}(1993)}]{wesolowski93}%
  \BibitemOpen
  \bibfield  {author} {\bibinfo {author} {\bibfnamefont {T.~A.}\ \bibnamefont
  {Wesolowski}}\ and\ \bibinfo {author} {\bibfnamefont {A.}~\bibnamefont
  {Warshel}},\ }\href {\doibase 10.1021/j100132a040} {\bibfield  {journal}
  {\bibinfo  {journal} {J. Phys. Chem.}\ }\textbf {\bibinfo {volume} {97}},\
  \bibinfo {pages} {8050} (\bibinfo {year} {1993})},\ \Eprint
  {http://arxiv.org/abs/https://doi.org/10.1021/j100132a040}
  {https://doi.org/10.1021/j100132a040} \BibitemShut {NoStop}%
\bibitem [{\citenamefont {Lopata}\ and\ \citenamefont {Govind}(2011)}]{nwchem}%
  \BibitemOpen
  \bibfield  {author} {\bibinfo {author} {\bibfnamefont {K.}~\bibnamefont
  {Lopata}}\ and\ \bibinfo {author} {\bibfnamefont {N.}~\bibnamefont
  {Govind}},\ }\href {\doibase 10.1021/ct200137z} {\bibfield  {journal}
  {\bibinfo  {journal} {J. Chem. Theory Comput.}\ }\textbf {\bibinfo {volume}
  {7}},\ \bibinfo {pages} {1344} (\bibinfo {year} {2011})},\ \Eprint
  {http://arxiv.org/abs/https://doi.org/10.1021/ct200137z}
  {https://doi.org/10.1021/ct200137z} \BibitemShut {NoStop}%
\bibitem [{\citenamefont {Castro}, \citenamefont {Marques},\ and\ \citenamefont
  {Rubio}(2004)}]{castro}%
  \BibitemOpen
  \bibfield  {author} {\bibinfo {author} {\bibfnamefont {A.}~\bibnamefont
  {Castro}}, \bibinfo {author} {\bibfnamefont {M.~A.~L.}\ \bibnamefont
  {Marques}}, \ and\ \bibinfo {author} {\bibfnamefont {A.}~\bibnamefont
  {Rubio}},\ }\href {\doibase 10.1063/1.1774980} {\bibfield  {journal}
  {\bibinfo  {journal} {J. Comp. Phys.}\ }\textbf {\bibinfo {volume} {121}},\
  \bibinfo {pages} {3425} (\bibinfo {year} {2004})},\ \Eprint
  {http://arxiv.org/abs/https://doi.org/10.1063/1.1774980}
  {https://doi.org/10.1063/1.1774980} \BibitemShut {NoStop}%
\bibitem [{\citenamefont {Thomas}(1927)}]{thomas_1927}%
  \BibitemOpen
  \bibfield  {author} {\bibinfo {author} {\bibfnamefont {L.~H.}\ \bibnamefont
  {Thomas}},\ }\href {\doibase 10.1017/S0305004100011683} {\bibfield  {journal}
  {\bibinfo  {journal} {Mathematical Proceedings of the Cambridge Philosophical
  Society}\ }\textbf {\bibinfo {volume} {23}},\ \bibinfo {pages} {542}
  (\bibinfo {year} {1927})}\BibitemShut {NoStop}%
\bibitem [{\citenamefont {Lembarki}\ and\ \citenamefont
  {Chermette}(1994)}]{PhysRevA.50.5328}%
  \BibitemOpen
  \bibfield  {author} {\bibinfo {author} {\bibfnamefont {A.}~\bibnamefont
  {Lembarki}}\ and\ \bibinfo {author} {\bibfnamefont {H.}~\bibnamefont
  {Chermette}},\ }\href {\doibase 10.1103/PhysRevA.50.5328} {\bibfield
  {journal} {\bibinfo  {journal} {Phys. Rev. A}\ }\textbf {\bibinfo {volume}
  {50}},\ \bibinfo {pages} {5328} (\bibinfo {year} {1994})}\BibitemShut
  {NoStop}%
\bibitem [{\citenamefont {te~Velde}\ \emph {et~al.}(2001)\citenamefont
  {te~Velde}, \citenamefont {Bickelhaupt}, \citenamefont {Baerends},
  \citenamefont {Fonseca~Guerra}, \citenamefont {van Gisbergen}, \citenamefont
  {Snijders},\ and\ \citenamefont {Ziegler}}]{ADF2001}%
  \BibitemOpen
  \bibfield  {author} {\bibinfo {author} {\bibfnamefont {G.}~\bibnamefont
  {te~Velde}}, \bibinfo {author} {\bibfnamefont {F.~M.}\ \bibnamefont
  {Bickelhaupt}}, \bibinfo {author} {\bibfnamefont {E.~J.}\ \bibnamefont
  {Baerends}}, \bibinfo {author} {\bibfnamefont {C.}~\bibnamefont
  {Fonseca~Guerra}}, \bibinfo {author} {\bibfnamefont {S.~J.~A.}\ \bibnamefont
  {van Gisbergen}}, \bibinfo {author} {\bibfnamefont {J.~G.}\ \bibnamefont
  {Snijders}}, \ and\ \bibinfo {author} {\bibfnamefont {T.}~\bibnamefont
  {Ziegler}},\ }\href {\doibase 10.1002/jcc.1056} {\bibfield  {journal}
  {\bibinfo  {journal} {J. Comput. Chem.}\ }\textbf {\bibinfo {volume} {22}},\
  \bibinfo {pages} {931} (\bibinfo {year} {2001})}\BibitemShut {NoStop}%
\bibitem [{ADF(2019)}]{ADF2019.3}%
  \BibitemOpen
  \href@noop {} {} (\bibinfo {year} {2019}),\ \bibinfo {note} {aDF 2019.3, SCM,
  Theoretical Chemistry, Vrije Universiteit, Amsterdam, The Netherlands,
  http://www.scm.com. Optionally, you may add the following list of authors and
  contributors: E.J. Baerends, T. Ziegler, A.J. Atkins, J. Autschbach, O.
  Baseggio, D. Bashford, A. Bérces, F.M. Bickelhaupt, C. Bo, P.M. Boerrigter,
  L. Cavallo, C. Daul, D.P. Chong, D.V. Chulhai, L. Deng, R.M. Dickson, J.M.
  Dieterich, D.E. Ellis, M. van Faassen, L. Fan, T.H. Fischer, A. Förster, C.
  Fonseca Guerra, M. Franchini, A. Ghysels, A. Giammona, S.J.A. van Gisbergen,
  A. Goez, A.W. Götz, J.A. Groeneveld, O.V. Gritsenko, M. Grüning, S.
  Gusarov, F.E. Harris, P. van den Hoek, Z. Hu, C.R. Jacob, H. Jacobsen, L.
  Jensen, L. Joubert, J.W. Kaminski, G. van Kessel, C. König, F. Kootstra, A.
  Kovalenko, M.V. Krykunov, E. van Lenthe, D.A. McCormack, A. Michalak, M.
  Mitoraj, S.M. Morton, J. Neugebauer, V.P. Nicu, L. Noodleman, V.P. Osinga, S.
  Patchkovskii, M. Pavanello, C.A. Peeples, P.H.T. Philipsen, D. Post, C.C.
  Pye, H. Ramanantoanina, P. Ramos, W. Ravenek, J.I. Rodríguez, P. Ros, R.
  Rüger, P.R.T. Schipper, D. Schlüns, H. van Schoot, G. Schreckenbach, J.S.
  Seldenthuis, M. Seth, J.G. Snijders, M. Solà, M. Stener, M. Swart, D.
  Swerhone, V. Tognetti, G. te Velde, P. Vernooijs, L. Versluis, L. Visscher,
  O. Visser, F. Wang, T.A. Wesolowski, E.M. van Wezenbeek, G. Wiesenekker, S.K.
  Wolff, T.K. Woo, A.L. Yakovlev}\BibitemShut {NoStop}%
\bibitem [{\citenamefont {Van~Lenthe}\ and\ \citenamefont
  {Baerends}(2003)}]{basis-Van-Lenthe-JCC2003-24-1142}%
  \BibitemOpen
  \bibfield  {author} {\bibinfo {author} {\bibfnamefont {E.}~\bibnamefont
  {Van~Lenthe}}\ and\ \bibinfo {author} {\bibfnamefont {E.~J.}\ \bibnamefont
  {Baerends}},\ }\href {\doibase 10.1002/jcc.10255} {\bibfield  {journal}
  {\bibinfo  {journal} {J. Comput. Chem.}\ }\textbf {\bibinfo {volume} {24}},\
  \bibinfo {pages} {1142} (\bibinfo {year} {2003})}\BibitemShut {NoStop}%
\bibitem [{\citenamefont {Jacob}\ \emph {et~al.}(2011)\citenamefont {Jacob},
  \citenamefont {Beyhan}, \citenamefont {Bulo}, \citenamefont {Gomes},
  \citenamefont {G\"{o}tz}, \citenamefont {Kiewisch}, \citenamefont {Sikkema},\
  and\ \citenamefont {Visscher}}]{Jacob2011}%
  \BibitemOpen
  \bibfield  {author} {\bibinfo {author} {\bibfnamefont {C.~R.}\ \bibnamefont
  {Jacob}}, \bibinfo {author} {\bibfnamefont {S.~M.}\ \bibnamefont {Beyhan}},
  \bibinfo {author} {\bibfnamefont {R.~E.}\ \bibnamefont {Bulo}}, \bibinfo
  {author} {\bibfnamefont {A.~S.~P.}\ \bibnamefont {Gomes}}, \bibinfo {author}
  {\bibfnamefont {A.~W.}\ \bibnamefont {G\"{o}tz}}, \bibinfo {author}
  {\bibfnamefont {K.}~\bibnamefont {Kiewisch}}, \bibinfo {author}
  {\bibfnamefont {J.}~\bibnamefont {Sikkema}}, \ and\ \bibinfo {author}
  {\bibfnamefont {L.}~\bibnamefont {Visscher}},\ }\href {\doibase
  10.1002/jcc.21810} {\bibfield  {journal} {\bibinfo  {journal} {Journal of
  Computational Chemistry}\ }\textbf {\bibinfo {volume} {32}},\ \bibinfo
  {pages} {2328} (\bibinfo {year} {2011})}\BibitemShut {NoStop}%
\bibitem [{\citenamefont {Jacob}, \citenamefont {Neugebauer},\ and\
  \citenamefont {Visscher}(2008)}]{Jacob2008}%
  \BibitemOpen
  \bibfield  {author} {\bibinfo {author} {\bibfnamefont {C.~R.}\ \bibnamefont
  {Jacob}}, \bibinfo {author} {\bibfnamefont {J.}~\bibnamefont {Neugebauer}}, \
  and\ \bibinfo {author} {\bibfnamefont {L.}~\bibnamefont {Visscher}},\ }\href
  {\doibase 10.1002/jcc.20861} {\bibfield  {journal} {\bibinfo  {journal}
  {Journal of Computational Chemistry}\ }\textbf {\bibinfo {volume} {29}},\
  \bibinfo {pages} {1011} (\bibinfo {year} {2008})}\BibitemShut {NoStop}%
\bibitem [{\citenamefont {Santis}, \citenamefont {Vallet},\ and\ \citenamefont
  {Gomes}(2021)}]{DeSantis-dataset-xas}%
  \BibitemOpen
  \bibfield  {author} {\bibinfo {author} {\bibfnamefont {M.~D.}\ \bibnamefont
  {Santis}}, \bibinfo {author} {\bibfnamefont {V.}~\bibnamefont {Vallet}}, \
  and\ \bibinfo {author} {\bibfnamefont {A.~S.~P.}\ \bibnamefont {Gomes}},\
  }\href {\doibase 10.5281/zenodo.5729961} {\enquote {\bibinfo {title}
  {Dataset: Environment effects on x-ray absorption spectra with quantum
  embedded real-time time-dependent density functional theory approaches},}\ }
  (\bibinfo {year} {2021})\BibitemShut {NoStop}%
\bibitem [{\citenamefont {Smith}\ \emph {et~al.}(2018)\citenamefont {Smith},
  \citenamefont {Burns}, \citenamefont {Sirianni}, \citenamefont {Nascimento},
  \citenamefont {Kumar}, \citenamefont {James}, \citenamefont {Schriber},
  \citenamefont {Zhang}, \citenamefont {Zhang}, \citenamefont {Abbott},
  \citenamefont {Berquist}, \citenamefont {Lechner}, \citenamefont {Cunha},
  \citenamefont {Heide}, \citenamefont {Waldrop}, \citenamefont {Takeshita},
  \citenamefont {Alenaizan}, \citenamefont {Neuhauser}, \citenamefont {King},
  \citenamefont {Simmonett}, \citenamefont {Turney}, \citenamefont {Schaefer},
  \citenamefont {Evangelista}, \citenamefont {DePrince}, \citenamefont
  {Crawford}, \citenamefont {Patkowski},\ and\ \citenamefont
  {Sherrill}}]{Smith2018}%
  \BibitemOpen
  \bibfield  {author} {\bibinfo {author} {\bibfnamefont {D.~G.~A.}\
  \bibnamefont {Smith}}, \bibinfo {author} {\bibfnamefont {L.~A.}\ \bibnamefont
  {Burns}}, \bibinfo {author} {\bibfnamefont {D.~A.}\ \bibnamefont {Sirianni}},
  \bibinfo {author} {\bibfnamefont {D.~R.}\ \bibnamefont {Nascimento}},
  \bibinfo {author} {\bibfnamefont {A.}~\bibnamefont {Kumar}}, \bibinfo
  {author} {\bibfnamefont {A.~M.}\ \bibnamefont {James}}, \bibinfo {author}
  {\bibfnamefont {J.~B.}\ \bibnamefont {Schriber}}, \bibinfo {author}
  {\bibfnamefont {T.}~\bibnamefont {Zhang}}, \bibinfo {author} {\bibfnamefont
  {B.}~\bibnamefont {Zhang}}, \bibinfo {author} {\bibfnamefont {A.~S.}\
  \bibnamefont {Abbott}}, \bibinfo {author} {\bibfnamefont {E.~J.}\
  \bibnamefont {Berquist}}, \bibinfo {author} {\bibfnamefont {M.~H.}\
  \bibnamefont {Lechner}}, \bibinfo {author} {\bibfnamefont {L.~A.}\
  \bibnamefont {Cunha}}, \bibinfo {author} {\bibfnamefont {A.~G.}\ \bibnamefont
  {Heide}}, \bibinfo {author} {\bibfnamefont {J.~M.}\ \bibnamefont {Waldrop}},
  \bibinfo {author} {\bibfnamefont {T.~Y.}\ \bibnamefont {Takeshita}}, \bibinfo
  {author} {\bibfnamefont {A.}~\bibnamefont {Alenaizan}}, \bibinfo {author}
  {\bibfnamefont {D.}~\bibnamefont {Neuhauser}}, \bibinfo {author}
  {\bibfnamefont {R.~A.}\ \bibnamefont {King}}, \bibinfo {author}
  {\bibfnamefont {A.~C.}\ \bibnamefont {Simmonett}}, \bibinfo {author}
  {\bibfnamefont {J.~M.}\ \bibnamefont {Turney}}, \bibinfo {author}
  {\bibfnamefont {H.~F.}\ \bibnamefont {Schaefer}}, \bibinfo {author}
  {\bibfnamefont {F.~A.}\ \bibnamefont {Evangelista}}, \bibinfo {author}
  {\bibfnamefont {A.~E.}\ \bibnamefont {DePrince}}, \bibinfo {author}
  {\bibfnamefont {T.~D.}\ \bibnamefont {Crawford}}, \bibinfo {author}
  {\bibfnamefont {K.}~\bibnamefont {Patkowski}}, \ and\ \bibinfo {author}
  {\bibfnamefont {C.~D.}\ \bibnamefont {Sherrill}},\ }\href {\doibase
  10.1021/acs.jctc.8b00286} {\bibfield  {journal} {\bibinfo  {journal} {J.
  Chem. Theory Comput.}\ }\textbf {\bibinfo {volume} {14}},\ \bibinfo {pages}
  {3504} (\bibinfo {year} {2018})}\BibitemShut {NoStop}%
\bibitem [{\citenamefont {Smith}\ \emph {et~al.}(2020)\citenamefont {Smith},
  \citenamefont {Burns}, \citenamefont {Simmonett}, \citenamefont {Parrish},
  \citenamefont {Schieber}, \citenamefont {Galvelis}, \citenamefont {Kraus},
  \citenamefont {Kruse}, \citenamefont {Remigio}, \citenamefont {Alenaizan},
  \citenamefont {James}, \citenamefont {Lehtola}, \citenamefont {Misiewicz},
  \citenamefont {Scheurer}, \citenamefont {Shaw}, \citenamefont {Schriber},
  \citenamefont {Xie}, \citenamefont {Glick}, \citenamefont {Sirianni},
  \citenamefont {O'Brien}, \citenamefont {Waldrop}, \citenamefont {Kumar},
  \citenamefont {Hohenstein}, \citenamefont {Pritchard}, \citenamefont
  {Brooks}, \citenamefont {Schaefer}, \citenamefont {Sokolov}, \citenamefont
  {Patkowski}, \citenamefont {DePrince}, \citenamefont {Bozkaya}, \citenamefont
  {King}, \citenamefont {Evangelista}, \citenamefont {Turney}, \citenamefont
  {Crawford},\ and\ \citenamefont {Sherrill}}]{Smith2020}%
  \BibitemOpen
  \bibfield  {author} {\bibinfo {author} {\bibfnamefont {D.~G.~A.}\
  \bibnamefont {Smith}}, \bibinfo {author} {\bibfnamefont {L.~A.}\ \bibnamefont
  {Burns}}, \bibinfo {author} {\bibfnamefont {A.~C.}\ \bibnamefont
  {Simmonett}}, \bibinfo {author} {\bibfnamefont {R.~M.}\ \bibnamefont
  {Parrish}}, \bibinfo {author} {\bibfnamefont {M.~C.}\ \bibnamefont
  {Schieber}}, \bibinfo {author} {\bibfnamefont {R.}~\bibnamefont {Galvelis}},
  \bibinfo {author} {\bibfnamefont {P.}~\bibnamefont {Kraus}}, \bibinfo
  {author} {\bibfnamefont {H.}~\bibnamefont {Kruse}}, \bibinfo {author}
  {\bibfnamefont {R.~D.}\ \bibnamefont {Remigio}}, \bibinfo {author}
  {\bibfnamefont {A.}~\bibnamefont {Alenaizan}}, \bibinfo {author}
  {\bibfnamefont {A.~M.}\ \bibnamefont {James}}, \bibinfo {author}
  {\bibfnamefont {S.}~\bibnamefont {Lehtola}}, \bibinfo {author} {\bibfnamefont
  {J.~P.}\ \bibnamefont {Misiewicz}}, \bibinfo {author} {\bibfnamefont
  {M.}~\bibnamefont {Scheurer}}, \bibinfo {author} {\bibfnamefont {R.~A.}\
  \bibnamefont {Shaw}}, \bibinfo {author} {\bibfnamefont {J.~B.}\ \bibnamefont
  {Schriber}}, \bibinfo {author} {\bibfnamefont {Y.}~\bibnamefont {Xie}},
  \bibinfo {author} {\bibfnamefont {Z.~L.}\ \bibnamefont {Glick}}, \bibinfo
  {author} {\bibfnamefont {D.~A.}\ \bibnamefont {Sirianni}}, \bibinfo {author}
  {\bibfnamefont {J.~S.}\ \bibnamefont {O'Brien}}, \bibinfo {author}
  {\bibfnamefont {J.~M.}\ \bibnamefont {Waldrop}}, \bibinfo {author}
  {\bibfnamefont {A.}~\bibnamefont {Kumar}}, \bibinfo {author} {\bibfnamefont
  {E.~G.}\ \bibnamefont {Hohenstein}}, \bibinfo {author} {\bibfnamefont
  {B.~P.}\ \bibnamefont {Pritchard}}, \bibinfo {author} {\bibfnamefont {B.~R.}\
  \bibnamefont {Brooks}}, \bibinfo {author} {\bibfnamefont {H.~F.}\
  \bibnamefont {Schaefer}}, \bibinfo {author} {\bibfnamefont {A.~Y.}\
  \bibnamefont {Sokolov}}, \bibinfo {author} {\bibfnamefont {K.}~\bibnamefont
  {Patkowski}}, \bibinfo {author} {\bibfnamefont {A.~E.}\ \bibnamefont
  {DePrince}}, \bibinfo {author} {\bibfnamefont {U.}~\bibnamefont {Bozkaya}},
  \bibinfo {author} {\bibfnamefont {R.~A.}\ \bibnamefont {King}}, \bibinfo
  {author} {\bibfnamefont {F.~A.}\ \bibnamefont {Evangelista}}, \bibinfo
  {author} {\bibfnamefont {J.~M.}\ \bibnamefont {Turney}}, \bibinfo {author}
  {\bibfnamefont {T.~D.}\ \bibnamefont {Crawford}}, \ and\ \bibinfo {author}
  {\bibfnamefont {C.~D.}\ \bibnamefont {Sherrill}},\ }\href {\doibase
  10.1063/5.0006002} {\bibfield  {journal} {\bibinfo  {journal} {The Journal of
  Chemical Physics}\ }\textbf {\bibinfo {volume} {152}},\ \bibinfo {pages}
  {184108} (\bibinfo {year} {2020})}\BibitemShut {NoStop}%
\bibitem [{\citenamefont {Dunning}(1989)}]{basis-Dunning-JCP1989-90-1007}%
  \BibitemOpen
  \bibfield  {author} {\bibinfo {author} {\bibfnamefont {T.~H.}\ \bibnamefont
  {Dunning}, \bibfnamefont {Jr.}},\ }\href {\doibase 10.1063/1.456153}
  {\bibfield  {journal} {\bibinfo  {journal} {J. Chem. Phys.}\ }\textbf
  {\bibinfo {volume} {90}},\ \bibinfo {pages} {1007} (\bibinfo {year}
  {1989})}\BibitemShut {NoStop}%
\bibitem [{\citenamefont {Kendall}, \citenamefont {Dunning},\ and\
  \citenamefont {Harrison}(1992)}]{basis-Kendall-JCP1992-96-6796}%
  \BibitemOpen
  \bibfield  {author} {\bibinfo {author} {\bibfnamefont {R.~A.}\ \bibnamefont
  {Kendall}}, \bibinfo {author} {\bibfnamefont {T.~H.}\ \bibnamefont {Dunning},
  \bibfnamefont {Jr.}}, \ and\ \bibinfo {author} {\bibfnamefont {R.~J.}\
  \bibnamefont {Harrison}},\ }\href {\doibase 10.1063/1.462569} {\bibfield
  {journal} {\bibinfo  {journal} {J. Chem. Phys.}\ }\textbf {\bibinfo {volume}
  {96}},\ \bibinfo {pages} {6796} (\bibinfo {year} {1992})}\BibitemShut
  {NoStop}%
\bibitem [{\citenamefont {Woon}\ and\ \citenamefont
  {Dunning}(1994)}]{basis-Woon-JCP1994-100-2975}%
  \BibitemOpen
  \bibfield  {author} {\bibinfo {author} {\bibfnamefont {D.~E.}\ \bibnamefont
  {Woon}}\ and\ \bibinfo {author} {\bibfnamefont {T.~H.}\ \bibnamefont
  {Dunning}, \bibfnamefont {Jr.}},\ }\href {\doibase 10.1063/1.466439}
  {\bibfield  {journal} {\bibinfo  {journal} {J. Chem. Phys.}\ }\textbf
  {\bibinfo {volume} {100}},\ \bibinfo {pages} {2975} (\bibinfo {year}
  {1994})}\BibitemShut {NoStop}%
\bibitem [{\citenamefont {Hehre}, \citenamefont {Stewart},\ and\ \citenamefont
  {Pople}(1969)}]{basis-Hehre-JCP1969-51-2657}%
  \BibitemOpen
  \bibfield  {author} {\bibinfo {author} {\bibfnamefont {W.~J.}\ \bibnamefont
  {Hehre}}, \bibinfo {author} {\bibfnamefont {R.~F.}\ \bibnamefont {Stewart}},
  \ and\ \bibinfo {author} {\bibfnamefont {J.~A.}\ \bibnamefont {Pople}},\
  }\href {\doibase 10.1063/1.1672392} {\bibfield  {journal} {\bibinfo
  {journal} {J. Chem. Phys.}\ }\textbf {\bibinfo {volume} {51}},\ \bibinfo
  {pages} {2657} (\bibinfo {year} {1969})}\BibitemShut {NoStop}%
\bibitem [{\citenamefont {De~Santis}\ \emph
  {et~al.}(2020{\natexlab{b}})\citenamefont {De~Santis}, \citenamefont
  {Storchi}, \citenamefont {Belpassi}, \citenamefont {Quiney},\ and\
  \citenamefont {Tarantelli}}]{rt_pybertha2020}%
  \BibitemOpen
  \bibfield  {author} {\bibinfo {author} {\bibfnamefont {M.}~\bibnamefont
  {De~Santis}}, \bibinfo {author} {\bibfnamefont {L.}~\bibnamefont {Storchi}},
  \bibinfo {author} {\bibfnamefont {L.}~\bibnamefont {Belpassi}}, \bibinfo
  {author} {\bibfnamefont {H.~M.}\ \bibnamefont {Quiney}}, \ and\ \bibinfo
  {author} {\bibfnamefont {F.}~\bibnamefont {Tarantelli}},\ }\href {\doibase
  10.1021/acs.jctc.0c00053} {\bibfield  {journal} {\bibinfo  {journal} {J.
  Chem. Theory Comput.}\ }\textbf {\bibinfo {volume} {16}},\ \bibinfo {pages}
  {2410} (\bibinfo {year} {2020}{\natexlab{b}})},\ \Eprint
  {http://arxiv.org/abs/https://doi.org/10.1021/acs.jctc.0c00053}
  {https://doi.org/10.1021/acs.jctc.0c00053} \BibitemShut {NoStop}%
\bibitem [{\citenamefont {M.~De~Santis}\ and\ \citenamefont
  {Storchi}(2021)}]{pybertha-github}%
  \BibitemOpen
  \bibfield  {author} {\bibinfo {author} {\bibfnamefont {L.~B.}\ \bibnamefont
  {M.~De~Santis}}\ and\ \bibinfo {author} {\bibfnamefont {L.}~\bibnamefont
  {Storchi}},\ }\href@noop {} {\enquote {\bibinfo {title} {Pyberthaembedrt, a
  code to perform real-time frozen density embedding calculations},}\ }\bibinfo
  {howpublished}
  {\url{https://github.com/BERTHA-4c-DKS/pybertha/tree/master/pyberthaembedrt}}
  (\bibinfo {year} {2021})\BibitemShut {NoStop}%
\bibitem [{\citenamefont {Yassine}\ \emph {et~al.}(2018)\citenamefont
  {Yassine}, \citenamefont {Shee}, \citenamefont {Real}, \citenamefont
  {Vallet},\ and\ \citenamefont
  {Gomes}}]{halides-water-Bouchafra-PRL2018-121-266001-Zenodo}%
  \BibitemOpen
  \bibfield  {author} {\bibinfo {author} {\bibnamefont {Yassine}}, \bibinfo
  {author} {\bibfnamefont {A.}~\bibnamefont {Shee}}, \bibinfo {author}
  {\bibfnamefont {F.}~\bibnamefont {Real}}, \bibinfo {author} {\bibfnamefont
  {V.}~\bibnamefont {Vallet}}, \ and\ \bibinfo {author} {\bibfnamefont
  {A.~S.~P.}\ \bibnamefont {Gomes}},\ }\href {\doibase 10.5281/zenodo.1477004}
  {\enquote {\bibinfo {title} {{Predictive simulations of ionization energies
  of solvated halide ions with relativistic embedded Equation of Motion
  Coupled-Cluster Theory: Dataset}},}\ } (\bibinfo {year} {2018})\BibitemShut
  {NoStop}%
\bibitem [{\citenamefont {Repisky}\ \emph {et~al.}(2015)\citenamefont
  {Repisky}, \citenamefont {Konecny}, \citenamefont {Kadek}, \citenamefont
  {Komorovsky}, \citenamefont {Malkin}, \citenamefont {Malkin},\ and\
  \citenamefont {Ruud}}]{repisky2015excitation}%
  \BibitemOpen
  \bibfield  {author} {\bibinfo {author} {\bibfnamefont {M.}~\bibnamefont
  {Repisky}}, \bibinfo {author} {\bibfnamefont {L.}~\bibnamefont {Konecny}},
  \bibinfo {author} {\bibfnamefont {M.}~\bibnamefont {Kadek}}, \bibinfo
  {author} {\bibfnamefont {S.}~\bibnamefont {Komorovsky}}, \bibinfo {author}
  {\bibfnamefont {O.~L.}\ \bibnamefont {Malkin}}, \bibinfo {author}
  {\bibfnamefont {V.~G.}\ \bibnamefont {Malkin}}, \ and\ \bibinfo {author}
  {\bibfnamefont {K.}~\bibnamefont {Ruud}},\ }\href {\doibase
  10.1021/ct501078d} {\bibfield  {journal} {\bibinfo  {journal} {J. Chem.
  Theory Comput.}\ }\textbf {\bibinfo {volume} {11}},\ \bibinfo {pages} {980}
  (\bibinfo {year} {2015})}\BibitemShut {NoStop}%
\bibitem [{\citenamefont {Repisky}\ \emph {et~al.}(2020)\citenamefont
  {Repisky}, \citenamefont {Komorovsky}, \citenamefont {Kadek}, \citenamefont
  {Konecny}, \citenamefont {Ekstr\"{o}m}, \citenamefont {Malkin}, \citenamefont
  {Kaupp}, \citenamefont {Ruud}, \citenamefont {Malkina},\ and\ \citenamefont
  {Malkin}}]{Repisky2020}%
  \BibitemOpen
  \bibfield  {author} {\bibinfo {author} {\bibfnamefont {M.}~\bibnamefont
  {Repisky}}, \bibinfo {author} {\bibfnamefont {S.}~\bibnamefont {Komorovsky}},
  \bibinfo {author} {\bibfnamefont {M.}~\bibnamefont {Kadek}}, \bibinfo
  {author} {\bibfnamefont {L.}~\bibnamefont {Konecny}}, \bibinfo {author}
  {\bibfnamefont {U.}~\bibnamefont {Ekstr\"{o}m}}, \bibinfo {author}
  {\bibfnamefont {E.}~\bibnamefont {Malkin}}, \bibinfo {author} {\bibfnamefont
  {M.}~\bibnamefont {Kaupp}}, \bibinfo {author} {\bibfnamefont
  {K.}~\bibnamefont {Ruud}}, \bibinfo {author} {\bibfnamefont {O.~L.}\
  \bibnamefont {Malkina}}, \ and\ \bibinfo {author} {\bibfnamefont {V.~G.}\
  \bibnamefont {Malkin}},\ }\href {\doibase 10.1063/5.0005094} {\bibfield
  {journal} {\bibinfo  {journal} {J. Chem. Phys.}\ }\textbf {\bibinfo {volume}
  {152}},\ \bibinfo {pages} {184101} (\bibinfo {year} {2020})}\BibitemShut
  {NoStop}%
\bibitem [{\citenamefont {Bruner}, \citenamefont {LaMaster},\ and\
  \citenamefont {Lopata}(2016)}]{bruner2016accelerated}%
  \BibitemOpen
  \bibfield  {author} {\bibinfo {author} {\bibfnamefont {A.}~\bibnamefont
  {Bruner}}, \bibinfo {author} {\bibfnamefont {D.}~\bibnamefont {LaMaster}}, \
  and\ \bibinfo {author} {\bibfnamefont {K.}~\bibnamefont {Lopata}},\
  }\href@noop {} {\bibfield  {journal} {\bibinfo  {journal} {J. Chem. Theory
  Comput.}\ }\textbf {\bibinfo {volume} {12}},\ \bibinfo {pages} {3741}
  (\bibinfo {year} {2016})}\BibitemShut {NoStop}%
\bibitem [{\citenamefont {Hofmann}\ and\ \citenamefont
  {Kümmel}(2012)}]{kuemmel2012}%
  \BibitemOpen
  \bibfield  {author} {\bibinfo {author} {\bibfnamefont {D.}~\bibnamefont
  {Hofmann}}\ and\ \bibinfo {author} {\bibfnamefont {S.}~\bibnamefont
  {Kümmel}},\ }\href {\doibase 10.1063/1.4742763} {\bibfield  {journal}
  {\bibinfo  {journal} {J. Comp. Phys.}\ }\textbf {\bibinfo {volume} {137}},\
  \bibinfo {pages} {064117} (\bibinfo {year} {2012})},\ \Eprint
  {http://arxiv.org/abs/https://doi.org/10.1063/1.4742763}
  {https://doi.org/10.1063/1.4742763} \BibitemShut {NoStop}%
\bibitem [{\citenamefont {Rossi}\ \emph {et~al.}(2017)\citenamefont {Rossi},
  \citenamefont {Kuisma}, \citenamefont {Puska}, \citenamefont {Nieminen},\
  and\ \citenamefont {Erhart}}]{erhart2017}%
  \BibitemOpen
  \bibfield  {author} {\bibinfo {author} {\bibfnamefont {T.~P.}\ \bibnamefont
  {Rossi}}, \bibinfo {author} {\bibfnamefont {M.}~\bibnamefont {Kuisma}},
  \bibinfo {author} {\bibfnamefont {M.~J.}\ \bibnamefont {Puska}}, \bibinfo
  {author} {\bibfnamefont {R.~M.}\ \bibnamefont {Nieminen}}, \ and\ \bibinfo
  {author} {\bibfnamefont {P.}~\bibnamefont {Erhart}},\ }\href {\doibase
  10.1021/acs.jctc.7b00589} {\bibfield  {journal} {\bibinfo  {journal} {J.
  Chem. Theory Comput.}\ }\textbf {\bibinfo {volume} {13}},\ \bibinfo {pages}
  {4779} (\bibinfo {year} {2017})},\ \Eprint
  {http://arxiv.org/abs/https://doi.org/10.1021/acs.jctc.7b00589}
  {https://doi.org/10.1021/acs.jctc.7b00589} \BibitemShut {NoStop}%
\bibitem [{\citenamefont {Sinha-Roy}\ \emph {et~al.}(2018)\citenamefont
  {Sinha-Roy}, \citenamefont {García-González}, \citenamefont
  {López~Lozano}, \citenamefont {Whetten},\ and\ \citenamefont
  {Weissker}}]{weissker2018}%
  \BibitemOpen
  \bibfield  {author} {\bibinfo {author} {\bibfnamefont {R.}~\bibnamefont
  {Sinha-Roy}}, \bibinfo {author} {\bibfnamefont {P.}~\bibnamefont
  {García-González}}, \bibinfo {author} {\bibfnamefont {X.}~\bibnamefont
  {López~Lozano}}, \bibinfo {author} {\bibfnamefont {R.~L.}\ \bibnamefont
  {Whetten}}, \ and\ \bibinfo {author} {\bibfnamefont {H.-C.}\ \bibnamefont
  {Weissker}},\ }\href {\doibase 10.1021/acs.jctc.8b00750} {\bibfield
  {journal} {\bibinfo  {journal} {J. Chem. Theory Comput.}\ }\textbf {\bibinfo
  {volume} {14}},\ \bibinfo {pages} {6417} (\bibinfo {year} {2018})},\ \Eprint
  {http://arxiv.org/abs/https://doi.org/10.1021/acs.jctc.8b00750}
  {https://doi.org/10.1021/acs.jctc.8b00750} \BibitemShut {NoStop}%
\bibitem [{\citenamefont {Schelter}\ and\ \citenamefont
  {K\"ummel}(2018)}]{kuemmel2018}%
  \BibitemOpen
  \bibfield  {author} {\bibinfo {author} {\bibfnamefont {I.}~\bibnamefont
  {Schelter}}\ and\ \bibinfo {author} {\bibfnamefont {S.}~\bibnamefont
  {K\"ummel}},\ }\href {\doibase 10.1021/acs.jctc.7b01013} {\bibfield
  {journal} {\bibinfo  {journal} {J. Chem. Theory Comput.}\ }\textbf {\bibinfo
  {volume} {14}},\ \bibinfo {pages} {1910} (\bibinfo {year} {2018})},\ \Eprint
  {http://arxiv.org/abs/https://doi.org/10.1021/acs.jctc.7b01013}
  {https://doi.org/10.1021/acs.jctc.7b01013} \BibitemShut {NoStop}%
\bibitem [{\citenamefont {Jornet-Somoza}\ and\ \citenamefont
  {Lebedeva}(2019)}]{samoza2019}%
  \BibitemOpen
  \bibfield  {author} {\bibinfo {author} {\bibfnamefont {J.}~\bibnamefont
  {Jornet-Somoza}}\ and\ \bibinfo {author} {\bibfnamefont {I.}~\bibnamefont
  {Lebedeva}},\ }\href {\doibase 10.1021/acs.jctc.9b00209} {\bibfield
  {journal} {\bibinfo  {journal} {J. Chem. Theory Comput.}\ }\textbf {\bibinfo
  {volume} {15}},\ \bibinfo {pages} {3743} (\bibinfo {year} {2019})},\ \Eprint
  {http://arxiv.org/abs/https://doi.org/10.1021/acs.jctc.9b00209}
  {https://doi.org/10.1021/acs.jctc.9b00209} \BibitemShut {NoStop}%
\bibitem [{\citenamefont {Eaton}\ \emph {et~al.}(2020)\citenamefont {Eaton},
  \citenamefont {Bateman}, \citenamefont {Hauberg},\ and\ \citenamefont
  {Wehbring}}]{octave}%
  \BibitemOpen
  \bibfield  {author} {\bibinfo {author} {\bibfnamefont {J.~W.}\ \bibnamefont
  {Eaton}}, \bibinfo {author} {\bibfnamefont {D.}~\bibnamefont {Bateman}},
  \bibinfo {author} {\bibfnamefont {S.}~\bibnamefont {Hauberg}}, \ and\
  \bibinfo {author} {\bibfnamefont {R.}~\bibnamefont {Wehbring}},\ }\href
  {https://www.gnu.org/software/octave/doc/v5.2.0/} {\emph {\bibinfo {title}
  {{GNU Octave} version 5.2.0 manual: a high-level interactive language for
  numerical computations}}} (\bibinfo {year} {2020})\BibitemShut {NoStop}%
\bibitem [{\citenamefont {Valiev}\ \emph {et~al.}(2010)\citenamefont {Valiev},
  \citenamefont {Bylaska}, \citenamefont {Govind}, \citenamefont {Kowalski},
  \citenamefont {Straatsma}, \citenamefont {Dam}, \citenamefont {Wang},
  \citenamefont {Nieplocha}, \citenamefont {Apra}, \citenamefont {Windus},\
  and\ \citenamefont {de~Jong}}]{Valiev2010}%
  \BibitemOpen
  \bibfield  {author} {\bibinfo {author} {\bibfnamefont {M.}~\bibnamefont
  {Valiev}}, \bibinfo {author} {\bibfnamefont {E.}~\bibnamefont {Bylaska}},
  \bibinfo {author} {\bibfnamefont {N.}~\bibnamefont {Govind}}, \bibinfo
  {author} {\bibfnamefont {K.}~\bibnamefont {Kowalski}}, \bibinfo {author}
  {\bibfnamefont {T.}~\bibnamefont {Straatsma}}, \bibinfo {author}
  {\bibfnamefont {H.~V.}\ \bibnamefont {Dam}}, \bibinfo {author} {\bibfnamefont
  {D.}~\bibnamefont {Wang}}, \bibinfo {author} {\bibfnamefont {J.}~\bibnamefont
  {Nieplocha}}, \bibinfo {author} {\bibfnamefont {E.}~\bibnamefont {Apra}},
  \bibinfo {author} {\bibfnamefont {T.}~\bibnamefont {Windus}}, \ and\ \bibinfo
  {author} {\bibfnamefont {W.}~\bibnamefont {de~Jong}},\ }\href {\doibase
  10.1016/j.cpc.2010.04.018} {\bibfield  {journal} {\bibinfo  {journal}
  {Comput. Phys. Commun.}\ }\textbf {\bibinfo {volume} {181}},\ \bibinfo
  {pages} {1477} (\bibinfo {year} {2010})}\BibitemShut {NoStop}%
\end{thebibliography}%
\end{document}


\preprint{*}

\title{Supplemental material for: Environment effects on X-ray absorption spectra with quantum embedded real-time Time-dependent density functional theory approaches} 

\author{Matteo De Santis}
\author{Val\'erie Vallet}
\author{Andr\'e Severo Pereira Gomes}
\affiliation{Univ. Lille, CNRS, UMR 8523 -- PhLAM -- Physique des Lasers, Atomes et Molécules, F-59000 Lille,  France}
\email{andre.gomes@univ-lille.fr}

\date{\today}

\maketitle

\onecolumngrid

\section{Supplementary Data, tables and figures}

\subsection{MO-bases analysis of excitations}
\begin{table}[!ht]
\caption{We report for the MO-based decomposition of the transition at $\omega =$ 668.634 eV from a supermolecular calculation of \ce{[F(H2O)8]-} at B3LYP/aug-cc-pvtz level of theory. The elements of the CI-vector, X$_{ia}$, corresponding to the occupied-virtual MO pair are also listed.}
\begin{tabular}{lll}
\multicolumn{3}{c}{Excitation  energy (eV)} \\
\hline
    & $\omega = $ 668.634 & \\
\hline
	occ. & virt. & X$_{ia}$ \\
	\hline
	1 &   46 &  -0.99431  \\
	1 &  47 & -0.09308 \\
	
\hline
\end{tabular}
\end{table}
\begin{verbatim}
Vector   46  Occ=0.000000D+00  E= 2.882190D-01
   Bfn.  Coefficient  Atom+Function         Bfn.  Coefficient  Atom+Function  
  ----- ------------  ---------------      ----- ------------  ---------------
     5      1.110236   1 F  s                 4     -0.642763   1 F  s         
    17     -0.354755   1 F  pz               16      0.291654   1 F  py        
    63      0.253890   8 O  px               76     -0.246397  14 O  s         
    78      0.238905  14 O  py               71     -0.234837  11 O  py        
    77     -0.202945  14 O  px               58      0.201848   5 O  pz  
\end{verbatim}
     
\begin{verbatim}
 Vector   47  Occ=0.000000D+00  E= 3.017870D-01
   Bfn.  Coefficient  Atom+Function         Bfn.  Coefficient  Atom+Function  
  ----- ------------  ---------------      ----- ------------  ---------------
    16      0.694780   1 F  py               17     -0.688708   1 F  pz        
     5     -0.480913   1 F  s                15      0.381587   1 F  px        
    85      0.345763  17 O  py               83      0.296018  17 O  s         
     4      0.277528   1 F  s                13     -0.240524   1 F  py        
    48      0.240055   2 O  s                14      0.233456   1 F  pz     
\end{verbatim}   
\subsection{DSF and Fourier Transfor of TD induced density}


\begin{figure}[htbp]
\begin{center}
\includegraphics[width=0.95\linewidth]{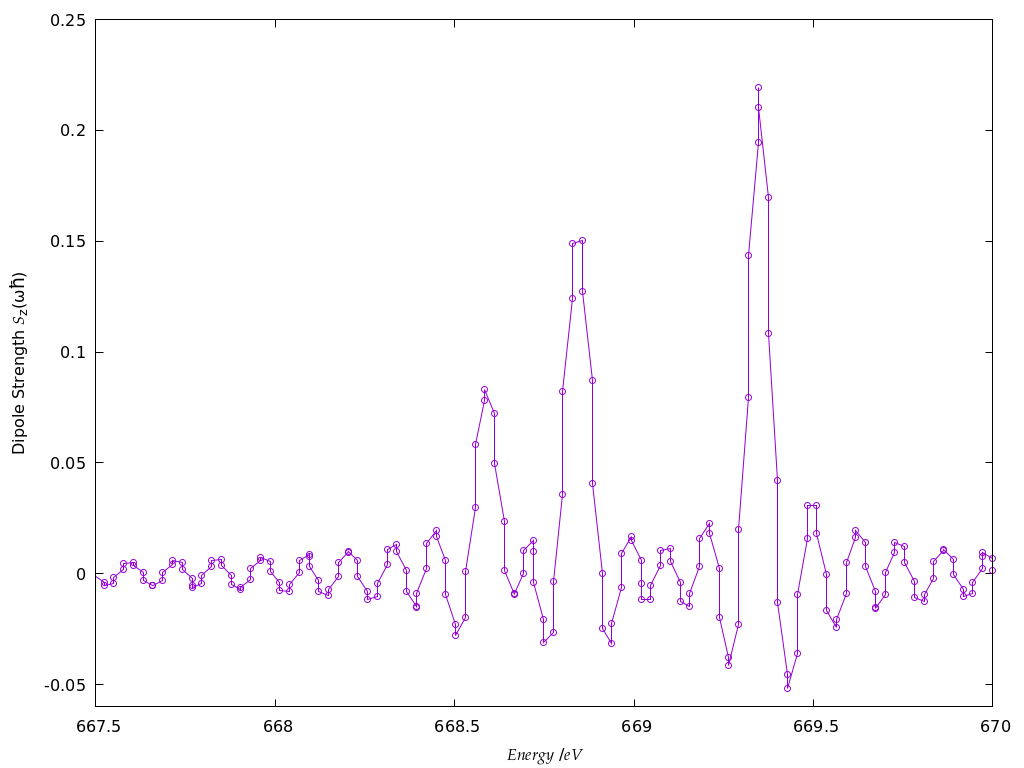}
\end{center}
\caption{We report the $z$-direction contribution to the dipole strength function in th range 667-670 eV for \ce{[F(H2O)8]-}. The time-domain is represented by 56001 sampling points and prior to Fourier transformation was  zero-padded in order to extend it  up to 2$\times 10^{19}$ sampling points. Despite the increased spectral density the DSF features a very low quality  in the energy range of interest.}\label{fig:1}
\end{figure}

\begin{figure}[htbp]
\begin{center}
\includegraphics[width=0.95\linewidth]{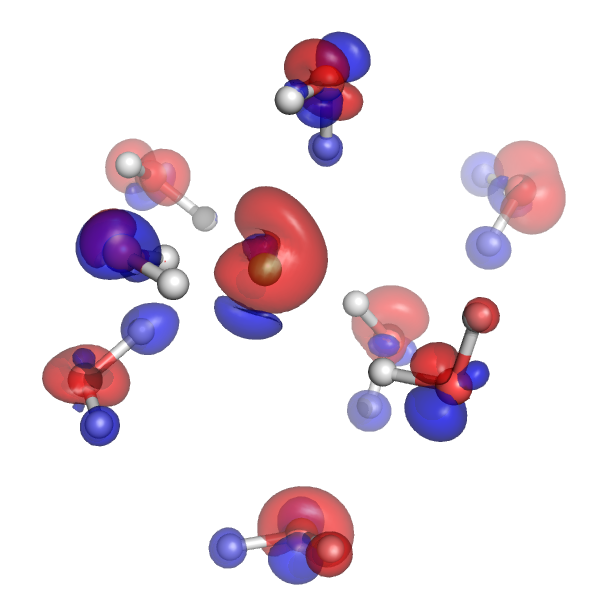}
\end{center}
\caption{Imaginary part of the Fourier Transform of the TD induced density, (Im[$\delta\tilde{\rho}(\bm{r}, \omega)$]) corresponding to the excitation frequency $\omega =$ 668.586 eV for the \ce{[F(H2O)8]-} complex using rt-BOMME.}\label{fig:2}
\end{figure}
\subsection{Spatial component of $S(\omega)$}
\begin{figure}[htbp]
\caption{Details on the three main energy ranges for the cross-section contribution along the $x$-direction ($S_x$) of the K-edge spectra of \ce{F-} and \ce{Cl-}.}
\centering
\begin{minipage}{0.48\linewidth}
\includegraphics[width=0.95\linewidth]{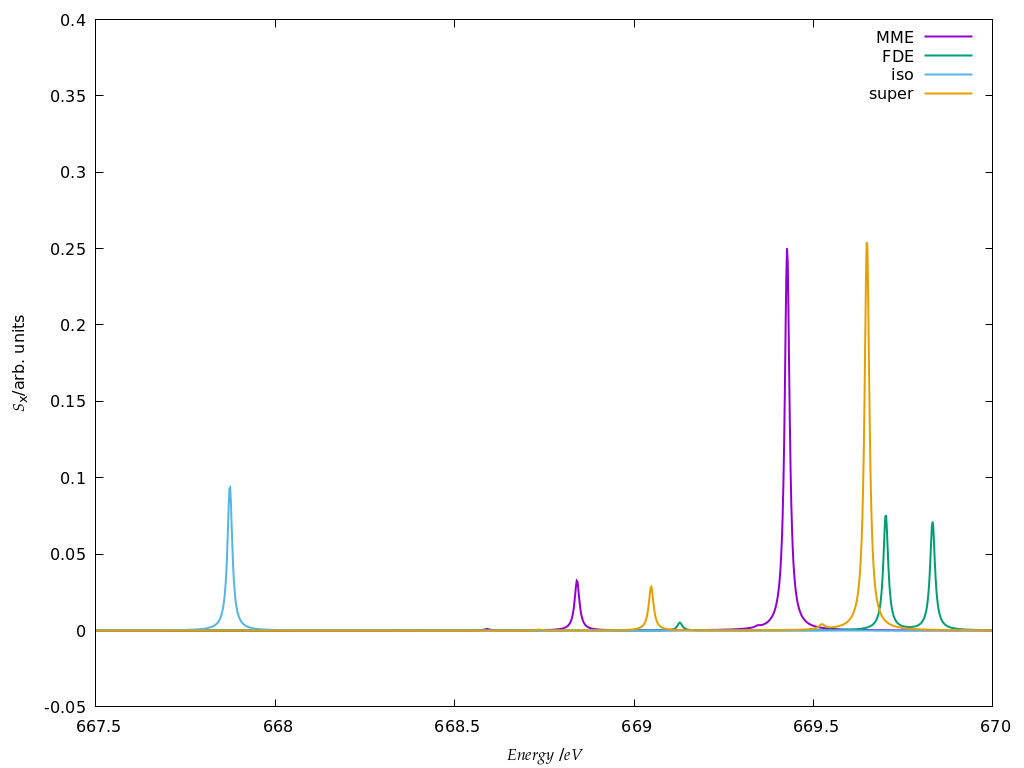}
\end{minipage}
\begin{minipage}{0.48\linewidth}
\includegraphics[width=0.95\linewidth]{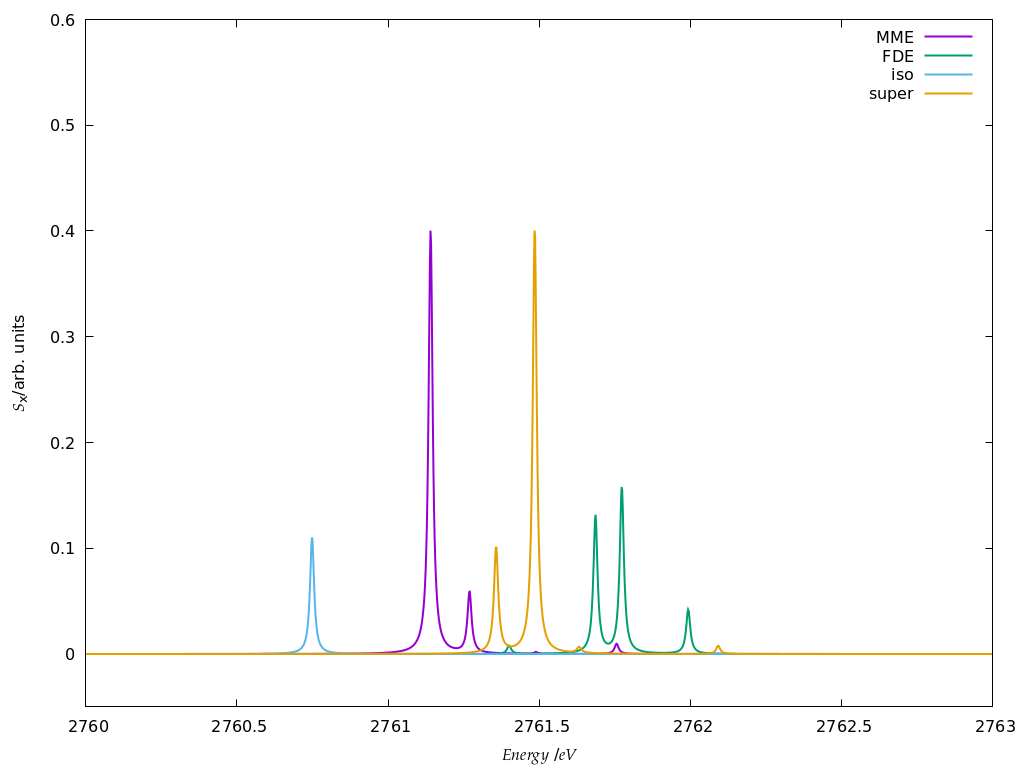}
\end{minipage}
\begin{minipage}{0.48\linewidth}
\includegraphics[width=0.95\linewidth]{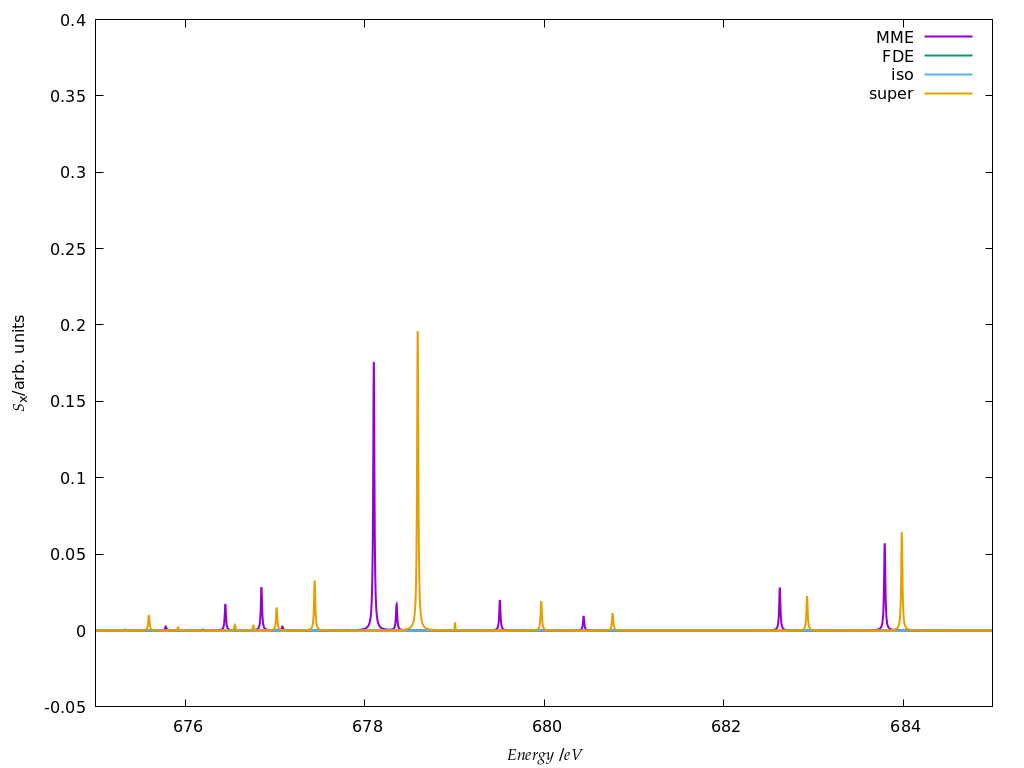}
\end{minipage}
\begin{minipage}{0.48\linewidth}
\includegraphics[width=0.95\linewidth]{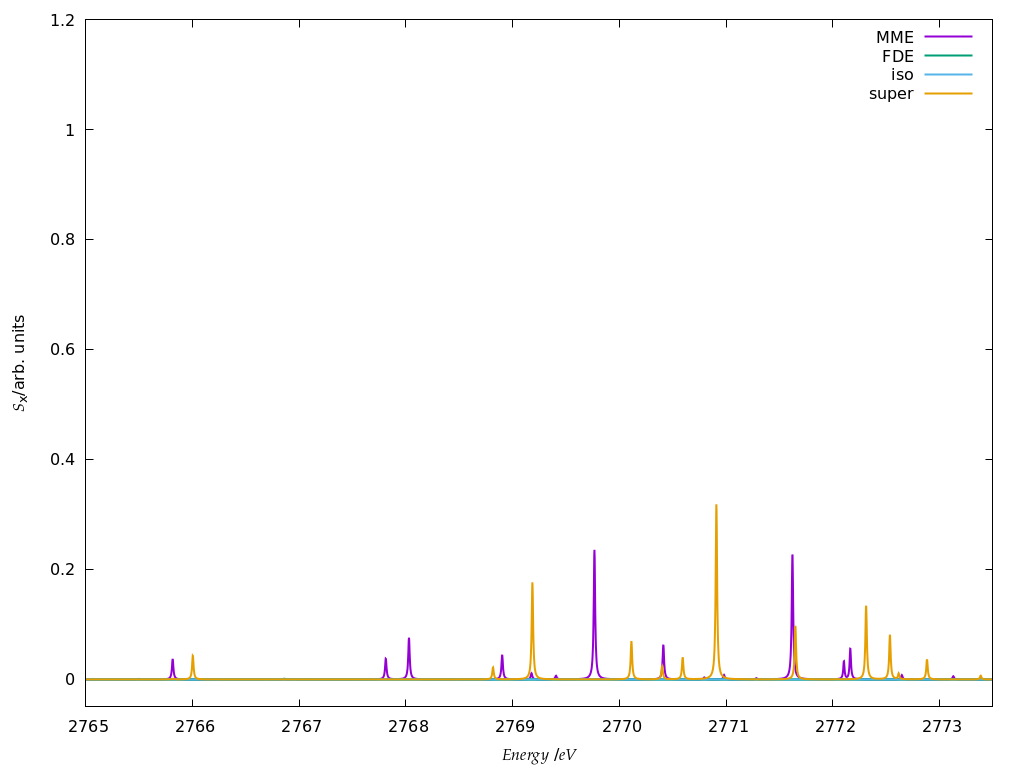}
\end{minipage}
\begin{minipage}{0.48\linewidth}
\includegraphics[width=0.95\linewidth]{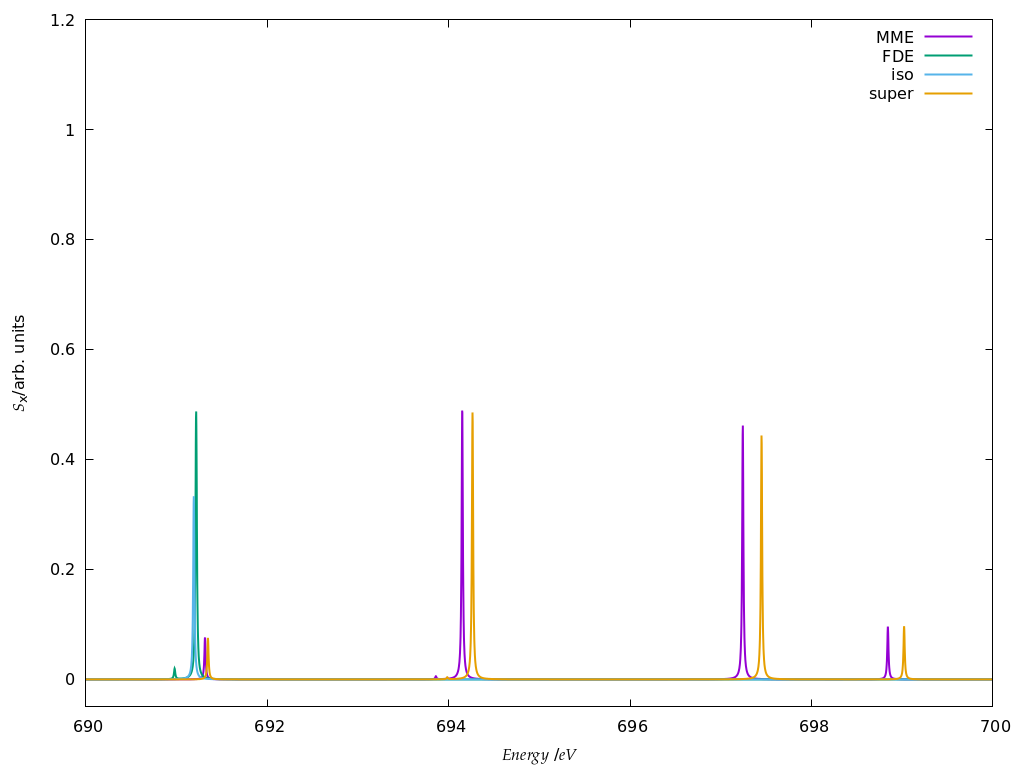}
\end{minipage}
\begin{minipage}{0.48\linewidth}
\includegraphics[width=0.95\linewidth]{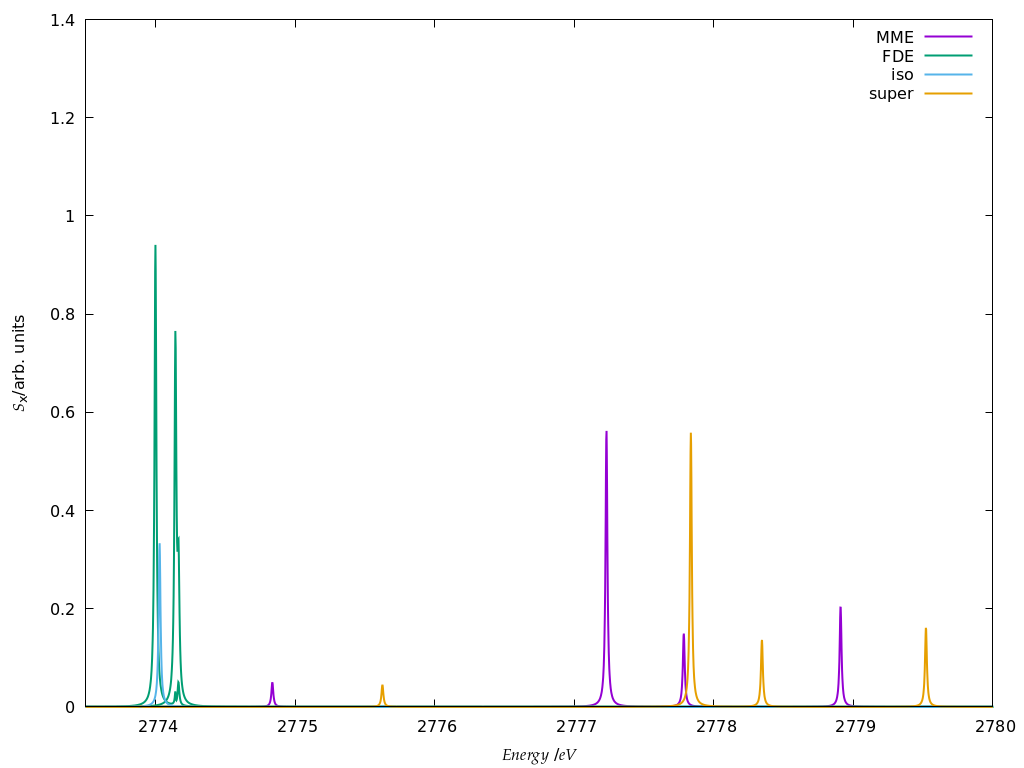}
\end{minipage}
\end{figure}

\newpage
\begin{figure}[htbp]
\caption{Details on the three main energy ranges for the cross-section contribution along the $y$-direction ($S_y$) of the K-edge spectra of \ce{F-} and \ce{Cl-}.}
\begin{minipage}{0.48\linewidth}
\includegraphics[width=0.95\linewidth]{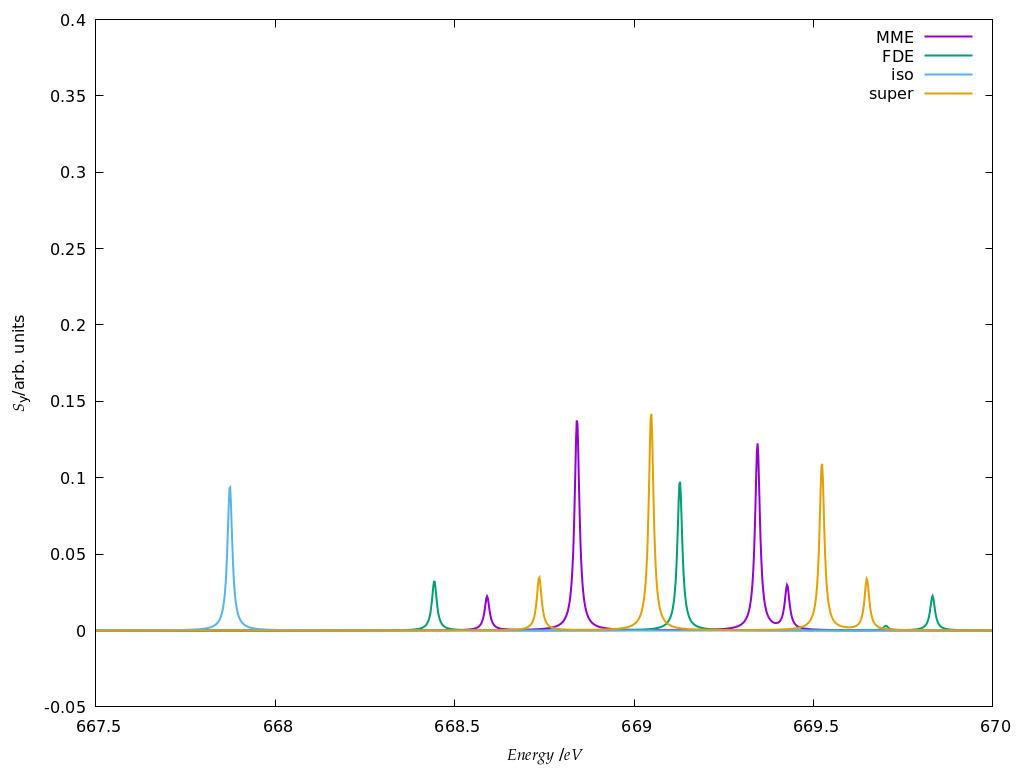}
\end{minipage}
\begin{minipage}{0.48\linewidth}
\includegraphics[width=0.95\linewidth]{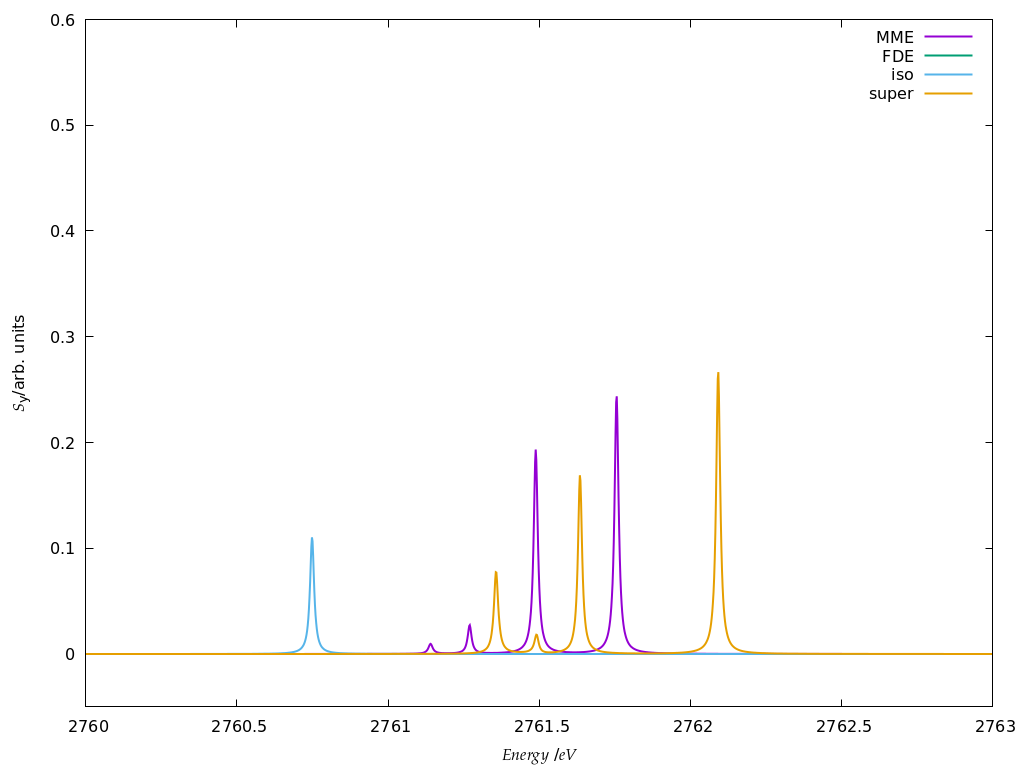}
\end{minipage}
\begin{minipage}{0.48\linewidth}
\includegraphics[width=0.95\linewidth]{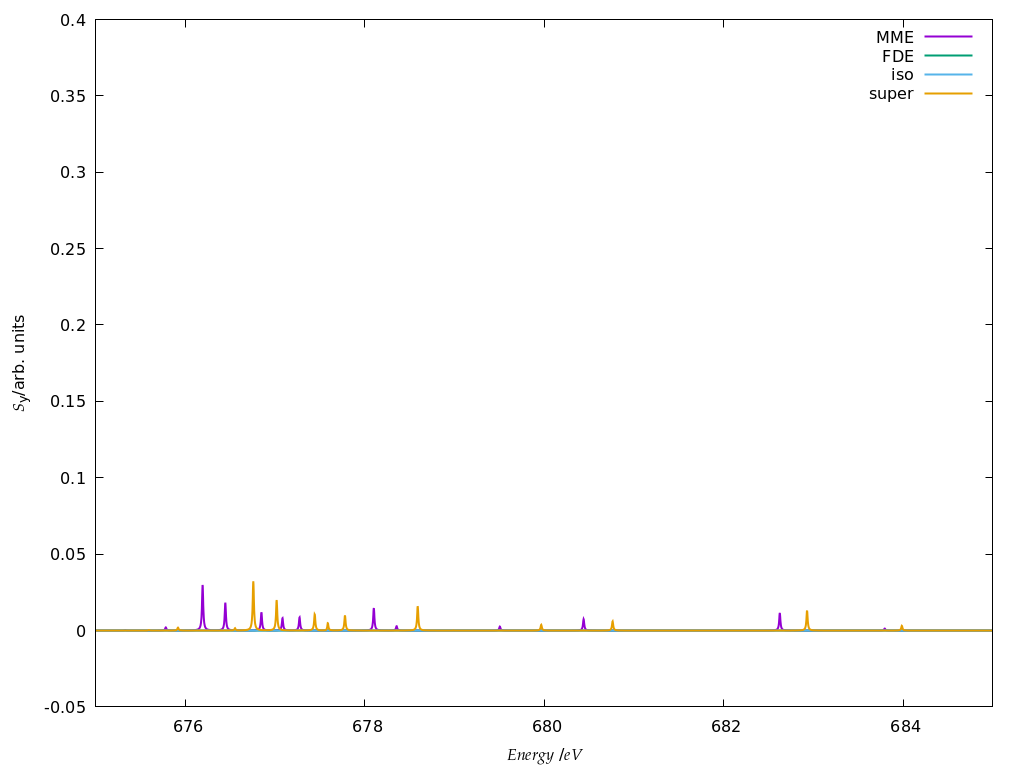}
\end{minipage}
\begin{minipage}{0.48\linewidth}
\includegraphics[width=0.95\linewidth]{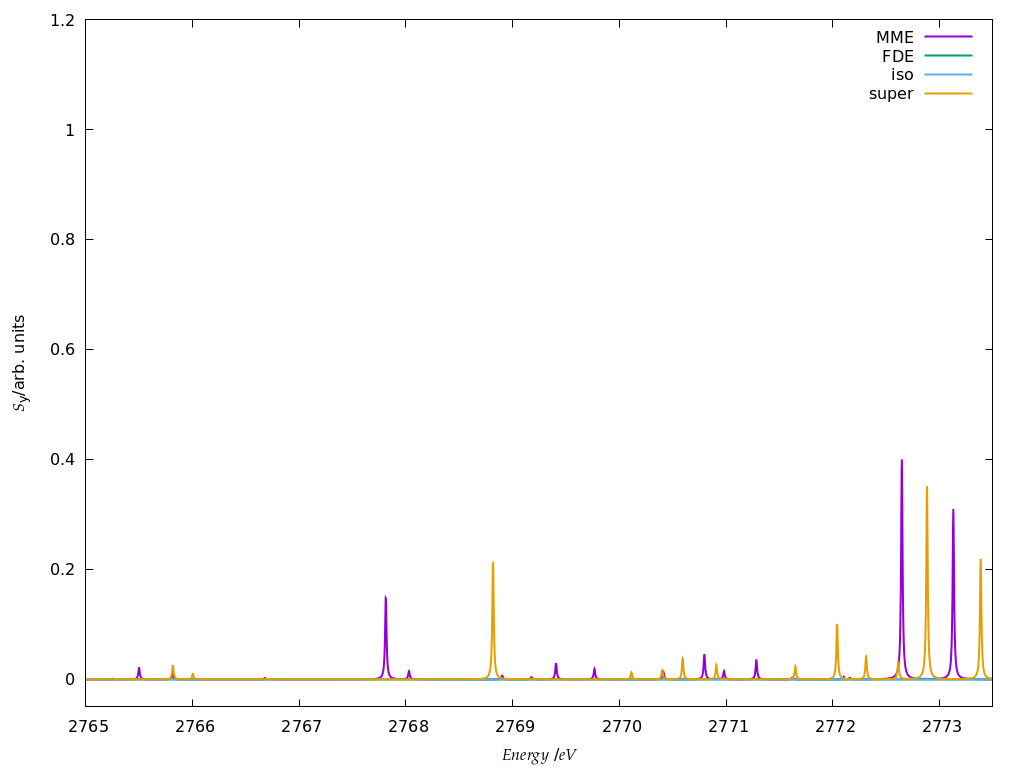}
\end{minipage}
\begin{minipage}{0.48\linewidth}
\includegraphics[width=0.95\linewidth]{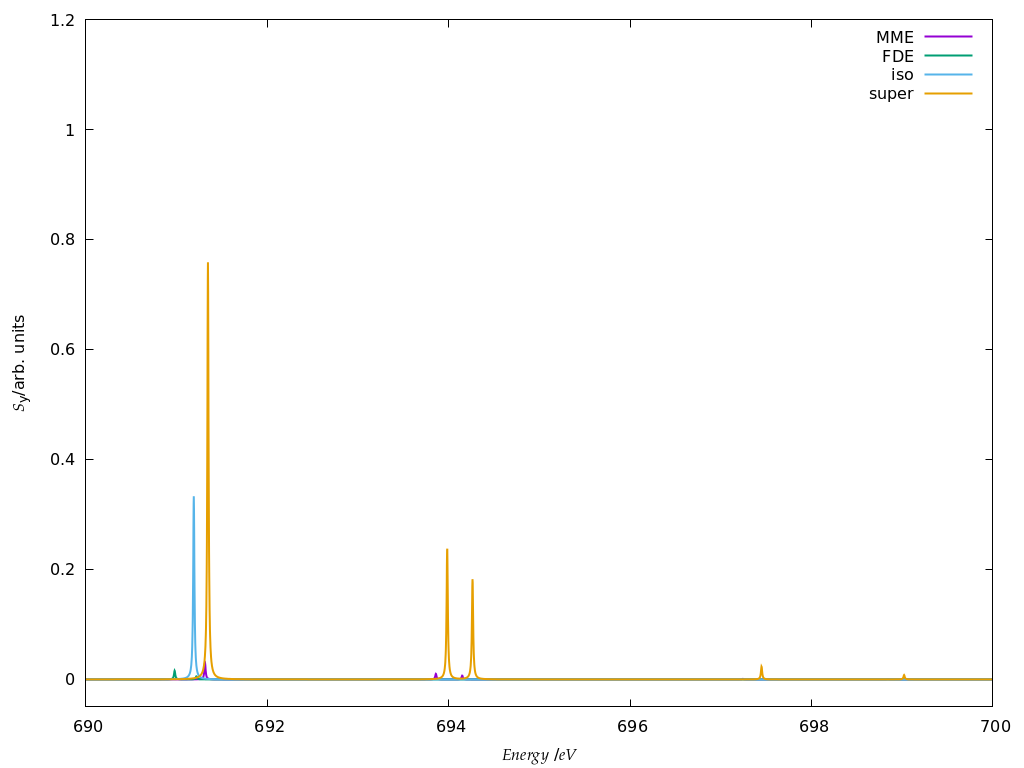}
\end{minipage}
\begin{minipage}{0.48\linewidth}
\includegraphics[width=0.95\linewidth]{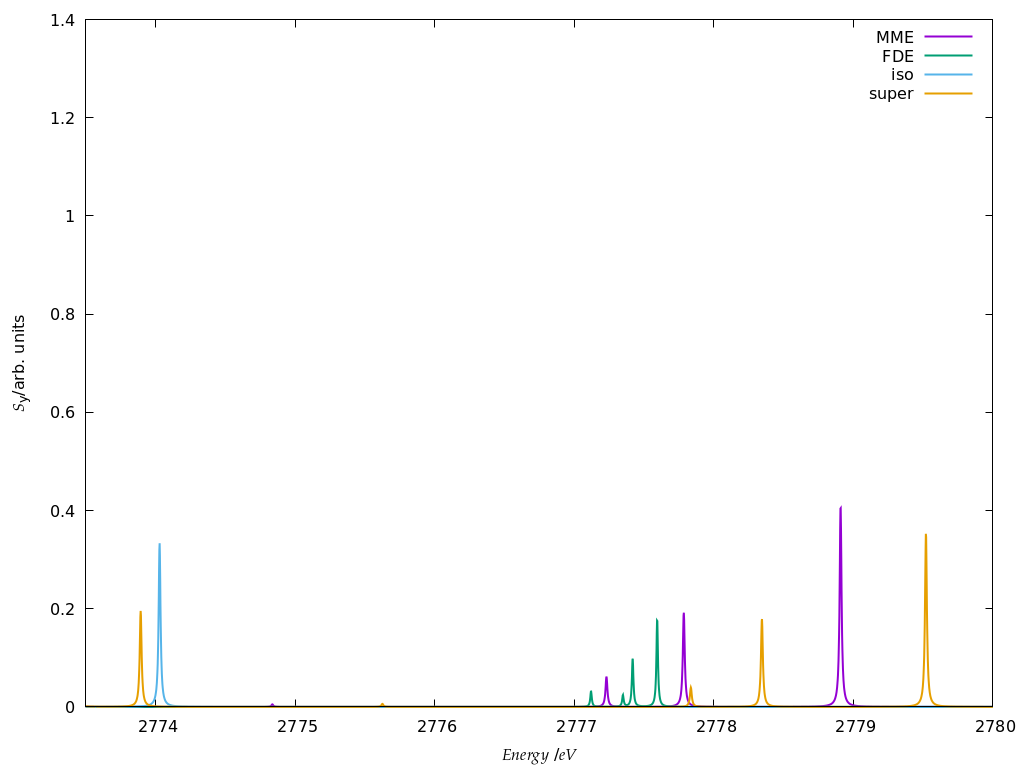}
\end{minipage}
\end{figure}

\newpage
\begin{figure}[htbp]
\caption{Details on the three main energy ranges for the cross-section contribution along the $z$-direction ($S_z$) of the K-edge spectra of \ce{F-} and \ce{Cl-}.}
\centering
\begin{minipage}{0.48\linewidth}
\includegraphics[width=0.95\linewidth]{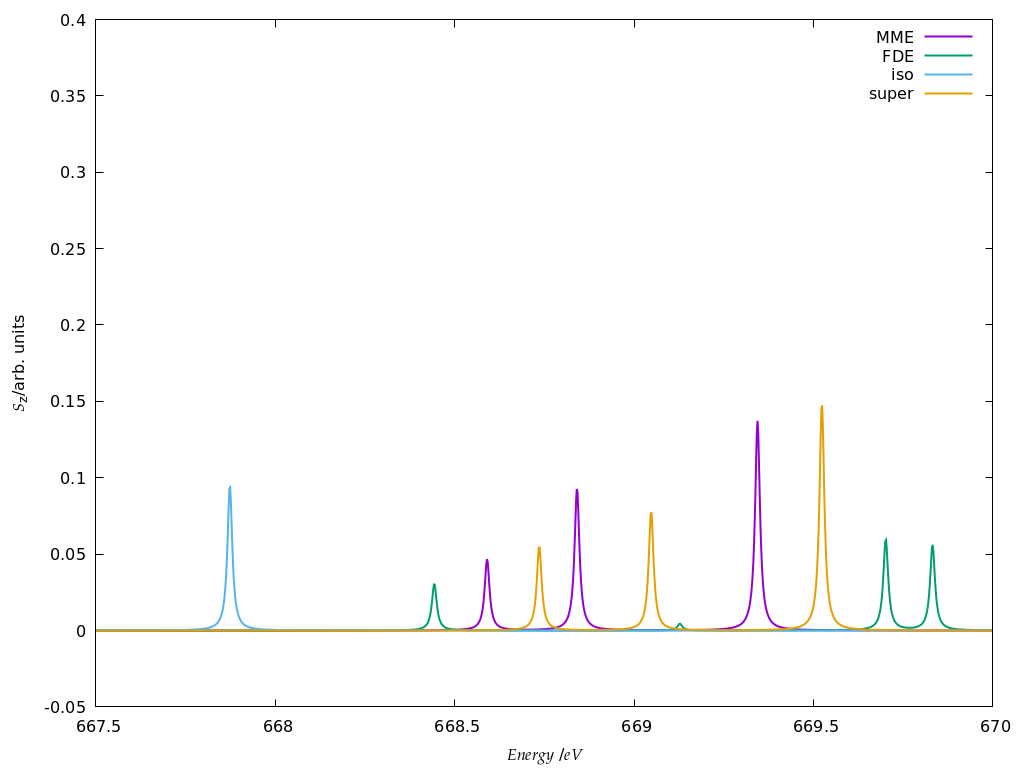}
\end{minipage}
\begin{minipage}{0.48\linewidth}
\includegraphics[width=0.95\linewidth]{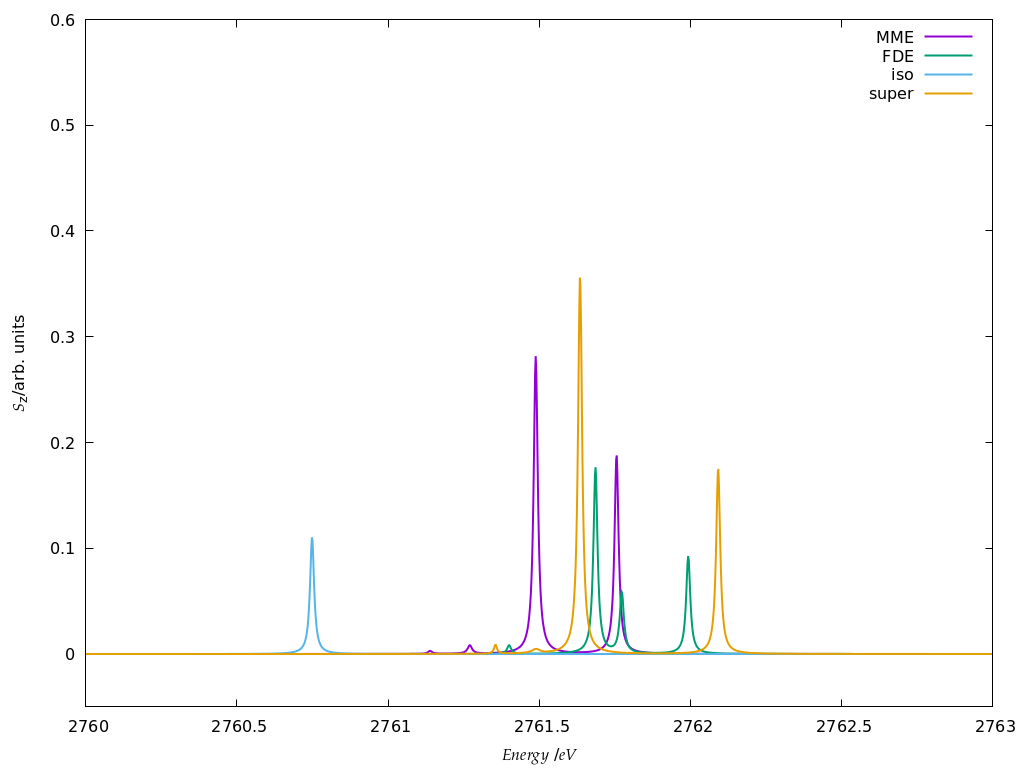}
\end{minipage}
\begin{minipage}{0.48\linewidth}
\includegraphics[width=0.95\linewidth]{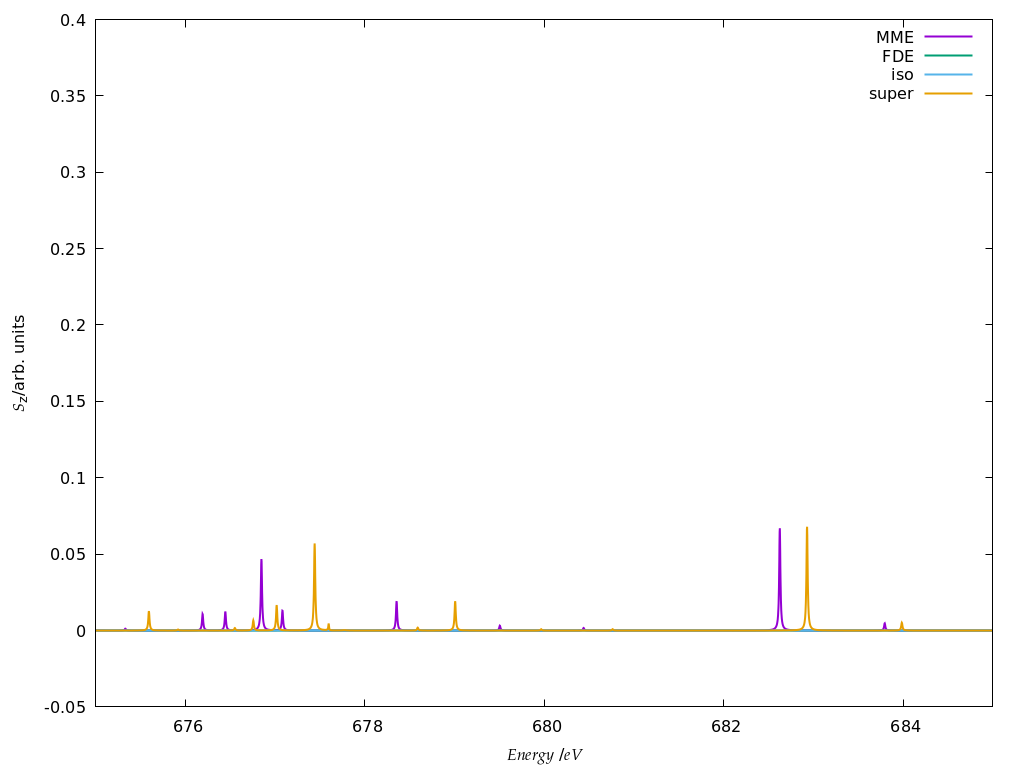}
\end{minipage}
\begin{minipage}{0.48\linewidth}
\includegraphics[width=0.95\linewidth]{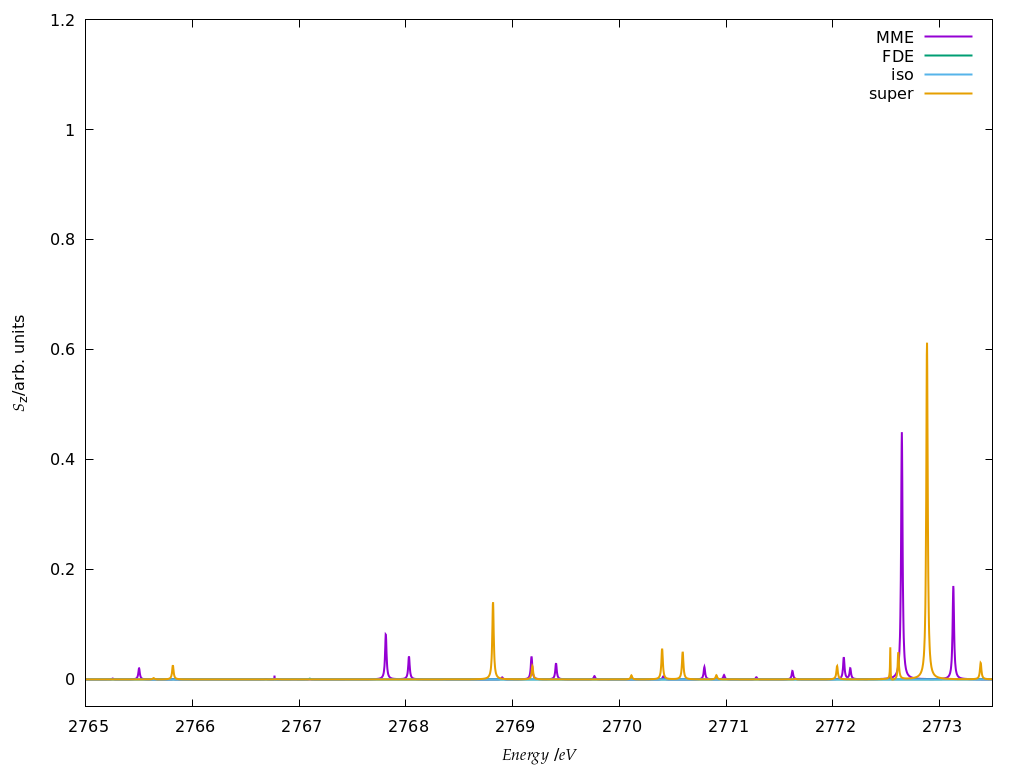}
\end{minipage}
\begin{minipage}{0.48\linewidth}
\includegraphics[width=0.95\linewidth]{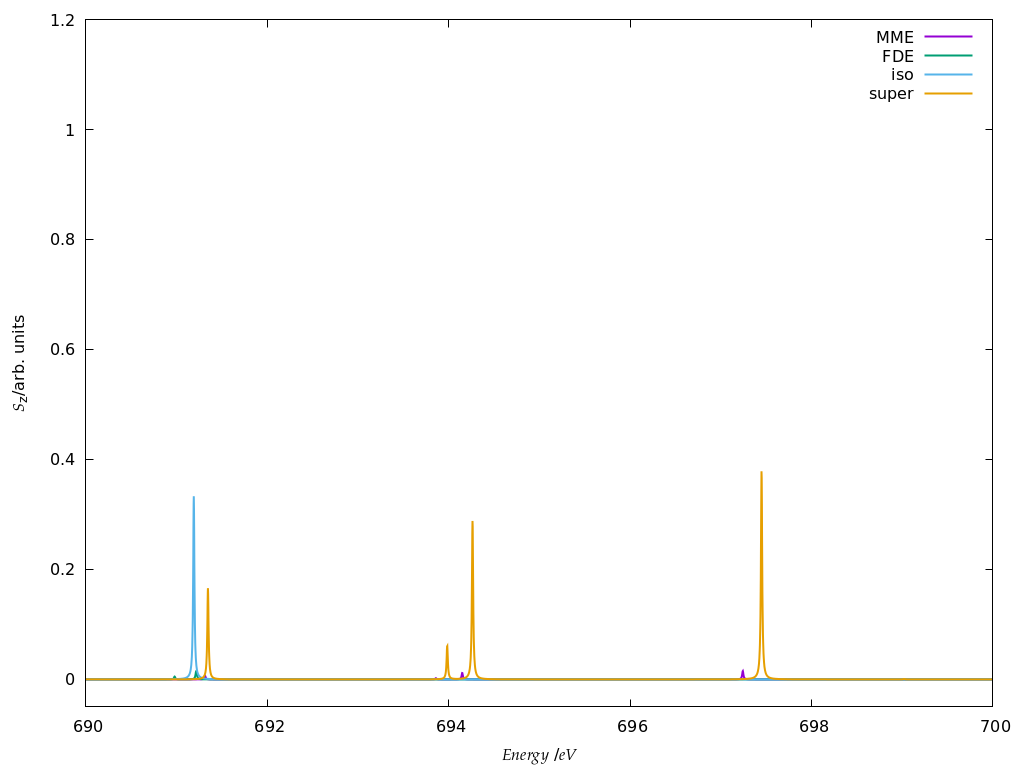}
\end{minipage}
\begin{minipage}{0.48\linewidth}
\includegraphics[width=0.95\linewidth]{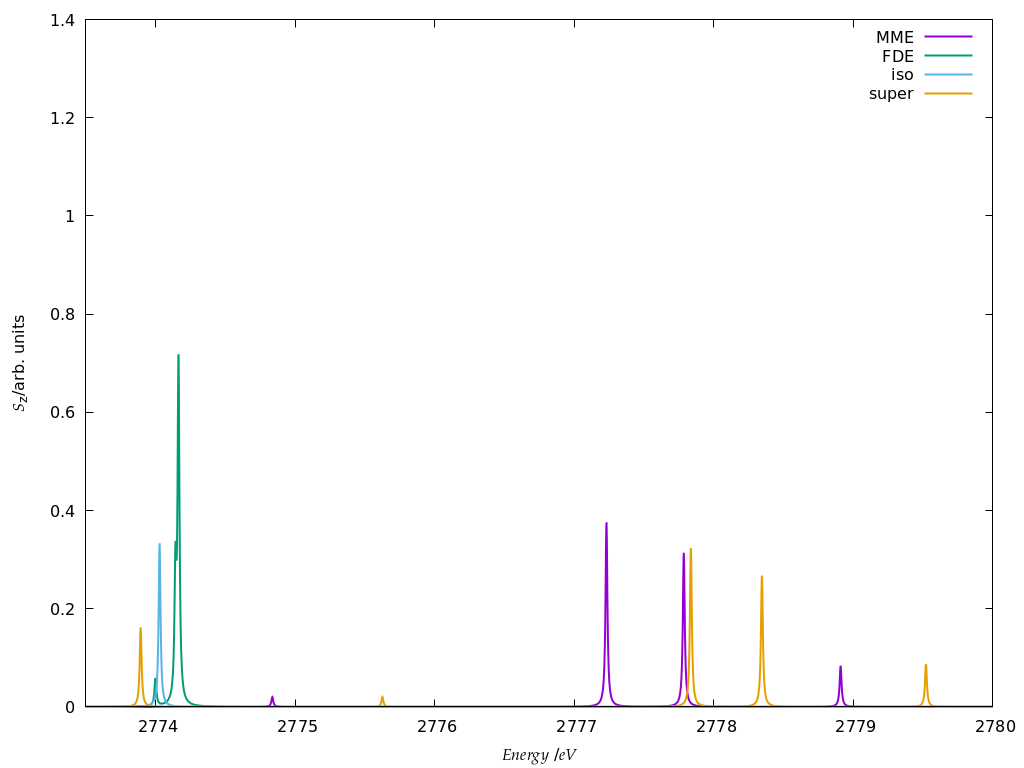}
\end{minipage}
\end{figure}
\newpage
\begin{figure}
\caption{Details of the cross-section contributions along the $x$-,$y$-, and $z$-direction of the \ce{L1}-edge spectrum of \ce{Cl-}.}
\begin{minipage}{0.48\linewidth}
\includegraphics[width=0.95\linewidth]{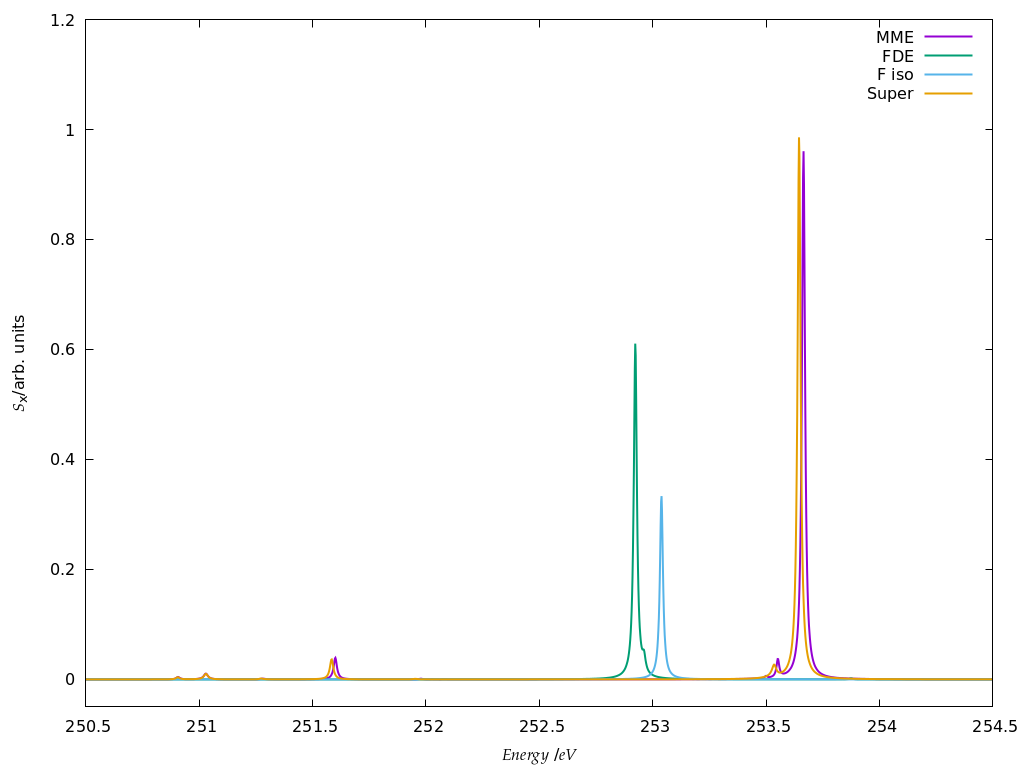}
\end{minipage}
\begin{minipage}{0.48\linewidth}
\includegraphics[width=0.95\linewidth]{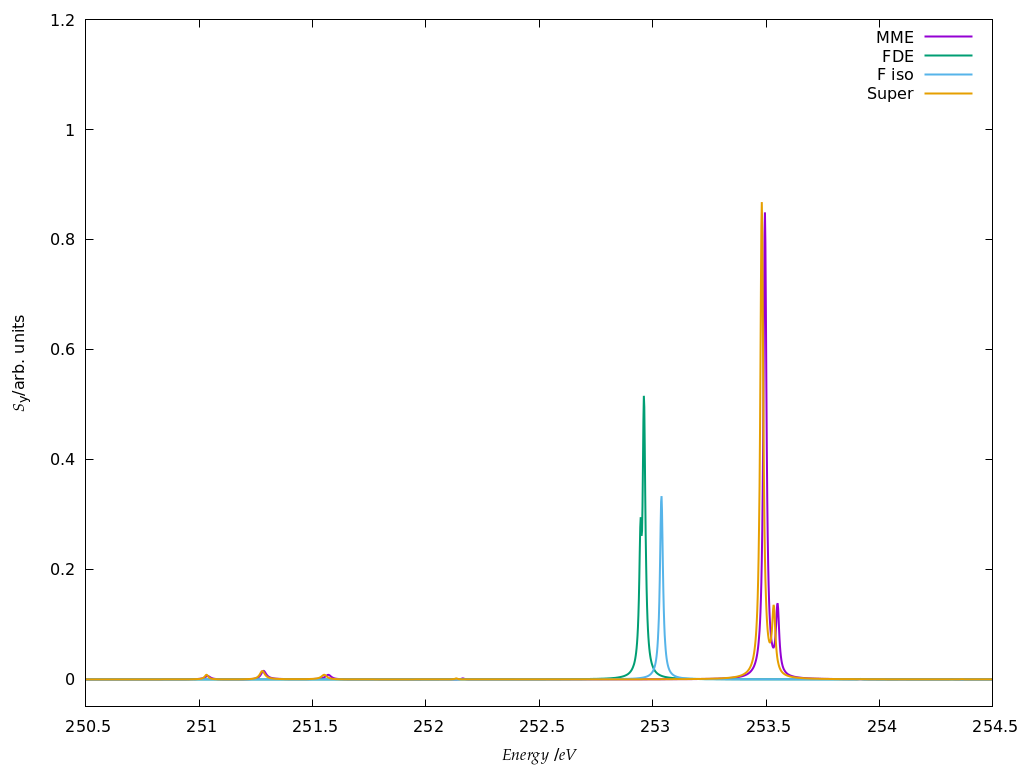}
\end{minipage}
\begin{minipage}{0.48\linewidth}
\includegraphics[width=0.95\linewidth]{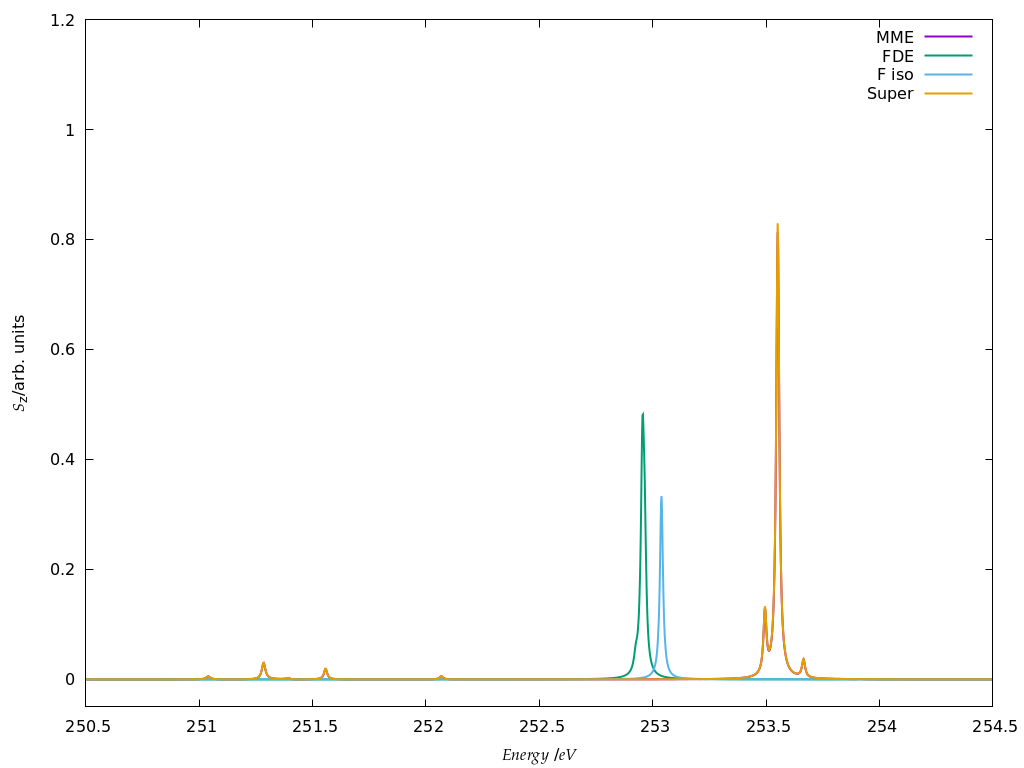}
\end{minipage}
\end{figure}

